\documentclass[aps,prd,groupedaddress,preprintnumbers,
               floatfix, nofootinbib,longbibliography]{revtex4}
\usepackage{textcomp}
\usepackage[latin1]{inputenc}                    
\usepackage{graphicx}                            
\usepackage{epsfig}
\usepackage[export]{adjustbox}
\usepackage{mathrsfs}
\usepackage{mathtools}

\usepackage{epsf}
\usepackage{latexsym}                            
\usepackage{amsfonts}                            
\usepackage{amssymb}                             
\usepackage{amsmath}                             
\usepackage{slashed}
\usepackage[mathscr]{eucal}                      
\usepackage{dcolumn}                             
\usepackage{multirow}                               
\usepackage{bm}                                  
\usepackage{hyperref}                            
\usepackage[usenames]{color}
\usepackage[dvipsnames]{xcolor}

\usepackage{color}

\graphicspath{{.}{Graphics/}}                    

\newcommand{\MSb}{\overline{\mathrm{MS}}}

\newcommand{\be}{\begin{equation}}
\newcommand{\ee}{\end{equation}}
\newcommand{\bea}{\begin{eqnarray}}
\newcommand{\eea}{\end{eqnarray}}

\def \3{\ss }

\newcommand{\beqn}{\begin{eqnarray}}
\newcommand{\eeqn}{\end{eqnarray}}

\newcommand{\UV}{\text{\tiny{UV}}}
\newcommand{\IR}{\text{\tiny{IR}}}

\def\teml{1}

\begin{document}

\begin{titlepage}

  {\vspace{-0.5cm} \normalsize
  \hfill \parbox{70mm}{
}}\\[10mm]
  \begin{center}
    \begin{LARGE} {\textbf{Burkhardt-Cottingham-type sum rules \\[0.3cm] for light-cone and quasi-PDFs}}
    \end{LARGE}
  \end{center}

\vspace*{1cm}

 \vspace{-0.8cm}
  \baselineskip 20pt plus 2pt minus 2pt
  \begin{center}
    \textbf{
      Shohini Bhattacharya$^{(\teml)}$ and 
      Andreas Metz$^{(\teml)}$
      }
\end{center}

  \begin{center}
    \begin{footnotesize}
      \noindent 	
    $^{(\teml)}$ {\it{Department of Physics, SERC, Temple University, Philadelphia, PA 19122, USA}} 
     \vspace{0.2cm}
    \end{footnotesize}
  \end{center}

\centerline{\today}

\begin{abstract}
The Burkhardt-Cottingham (BC) sum rule connects the twist-3 light-cone parton distribution function (PDF) $g_{T}(x)$ to the twist-2 helicity PDF $g_{1}(x)$. 
The chiral-odd counterpart of the BC sum rule relates the twist-3 light-cone PDF $h_{L}(x)$ to the twist-2 transversity PDF $h_{1}(x)$. 
These BC-type sum rules can also be derived for the corresponding quasi-PDFs. 
We perform a perturbative check of the BC-type sum rules in the quark target model and the Yukawa model, by going beyond the ultra-violet (UV) divergent terms. 
We employ dimensional regularization (DR) and cut-off schemes to regulate UV divergences, and show that the BC-type sum rules hold for DR, while they are generally violated when using a cut-off. 
This violation can be traced back to the breaking of rotational invariance.
We find corresponding results for the sum rule relating the mass of the target to the twist-3 PDF $e(x)$.
Moreover, we supplement our analytical results with numerical calculations.
\end{abstract}

\maketitle

\end{titlepage}

\section{Introduction}
\label{s:introduction}
Quarks and gluons, collectively denoted as partons, are the fundamental degrees of freedom of quantum chromodynamics (QCD).
While partons cannot be observed directly, QCD factorization theorems allow one to express physical observables in terms of non-perturbative functions, which contain information about partons inside nucleons~\cite{Collins:1989gx}. 
In this context, parton distribution functions (PDFs) belong to the most important non-perturbative functions~\cite{Collins:1981uw}.
Not only can PDFs be extracted through high-energy scattering experiments, but they can also be computed in models and lattice QCD. 
PDFs can be grouped according to their ``twist", which determines the order in the inverse hard scale at which a PDF contributes to an observable. 
While twist-2 PDFs provide the dominant contribution to physical observables, higher-twist PDFs, such as the twist-3 PDFs, suffer from kinematical suppressions, which precludes an ``easy" experimental extraction. 
At twist-2, a complete (one-dimensional) description of nucleons in terms of quarks can be obtained by means of three PDFs: the unpolarized PDF $f_{1}(x)$, the helicity PDF $g_{1}(x)$, and the transversity PDF $h_{1}(x)$. 
On the other hand, at twist-3, one has the three PDFs: $e(x)$, $g_{T}(x)$ and $h_{L}(x)$. 
Twist-2 PDFs have a probabilistic interpretation of representing momentum distributions of partons inside nucleons. 
On the other hand, twist-3 PDFs do not represent densities, and hence are conceptually intriguing because we are forced to go beyond the simple parton model.
For instance, they can be shown to quantify multi-parton correlations inside nucleons~\cite{Balitsky:1987bk,Kanazawa:2015ajw}. 
Noteworthy is also the semi-classical interpretation of $x^{2}$-moments of $e(x)$ and $g_{T}(x)$ in terms of the average transverse force experienced by quarks in DIS~\cite{Burkardt:2008ps}.  

Lorentz invariance plays a central role in any relativistic quantum field theory such as QCD. 
Certain sum rules are the remarkable consequences of Lorentz invariance.
Several such sum rules are integral relations connecting PDFs of different twists to one another. 
One example, the Burkhardt-Cottingham (BC) sum rule, proposed about 50 years ago, connects the twist-2 $g_{1}(x)$ to the twist-3 $g_{T}(x)$~\cite{Burkhardt:1970ti},
\begin{eqnarray}
\int dx \, g_{1} (x) = \int dx \, g_{T} (x) \, .
\label{e:def_BC_LC}
\end{eqnarray}
The chiral-odd counterpart of the BC sum rule, also known as the h-sum rule, connects the twist-2 $h_{1}(x)$ to the twist-3 $h_{L}(x)$~\cite{Tangerman:1994bb, Burkardt:1995ts},
\begin{eqnarray}
\int dx \, h_{1} (x) = \int dx \, h_{L} (x) \, .
\label{e:def_h_LC}
\end{eqnarray}
These BC-type sum rules have been under scrutiny for decades --- see, for instance, Refs.~\cite{Burkardt:1995ts, Kundu:2001pk, Burkardt:2001iy, Aslan:2018tff}. 
One of the most interesting features of the twist-3 PDFs concerns the possible existence of singular zero-mode contributions~\cite{Burkardt:1995ts, Burkardt:2001iy, Efremov:2002qh, Wakamatsu:2003uu, Aslan:2018zzk, Pasquini:2018oyz, Aslan:2018tff}, that is, terms proportional to $\delta(x)$, and their potential impact on the sum rules. 
Obviously, a $\delta(x)$ contribution would preclude experimental checks of the sum rule as the point $x = 0$ cannot be reached in experiment.
Put differently, measurements would suggest a violation of the sum rule.

The checks of the BC-type sum rules include, in particular, perturbative calculations in models such as the quark target model (QTM)~\cite{Burkardt:1995ts, Kundu:2001pk, Burkardt:2001iy, Aslan:2018tff}.
The final conclusion reached in those studies was that the sum rules hold in the models, provided that one takes into account the zero-mode contributions.
However, in those works, for analytical simplicity, only the UV-divergent contributions were considered, and it was tacitly assumed that this is sufficient.
One may ask if this is really enough, and whether the UV-finite terms satisfy the sum rules as well.
To address those questions is one of the main purposes of the present work.
To this end, we perform a check of the BC-type sum rules by going beyond the UV-divergent terms at one-loop order. 
We exploit two models for this analysis: the QTM and the Yukawa Model (YM). 
We use two regularization schemes for the UV divergences: Dimensional regularization (DR) and a cut-off. 
For the IR divergences, we employ three schemes: non-zero gluon mass $m_{g} \neq 0$, non-zero quark mass $m_{q} \neq 0$, and DR. 
Our work suggests that it is indeed not sufficient to limit the check of the BC-type sum rules to the UV-divergent parts of the PDFs. 
In fact, the sum rules can be expected to be violated for the UV-finite terms in schemes that break rotational invariance.
Specifically, the sum rules are typically violated when using a cut-off, while they hold in DR (which preserves rotational invariance).
As a by-product, we find that working with $m_{g} \neq 0$ as an IR regulator at twist-3 can in general cause problems. 

The second major point of this work is the discussion of BC-type sum rules for parton quasi-distributions (quasi-PDFs), which became key quantities for hadron structure calculations in lattice QCD.
For a long time, lattice-QCD extractions of the full $x$-dependence of the parton distributions were hindered by the explicit time-dependence of the underlying correlation functions. 
The quasi-PDF approach, proposed by Ji in 2013, has sparked a new wave of interest in extracting PDFs from lattice QCD~\cite{Ji:2013dva,Ji:2014gla}.
This approach relies on the extraction of matrix elements for boosted nucleons involving space-like separated fields. 
Over the years, enormous progress has taken place on the extraction of PDFs through such an approach from lattice QCD~\cite{Orginos:2017kos,Ji:2015jwa,Ishikawa:2016znu,Chen:2016fxx,Constantinou:2017sej,Alexandrou:2017huk,Chen:2017mzz,Ji:2017oey,Ishikawa:2017faj,Green:2017xeu,Spanoudes:2018zya,Zhang:2018diq,Li:2018tpe,Constantinou:2019vyb,Lin:2014zya,Alexandrou:2015rja,Chen:2016utp,Alexandrou:2016jqi,Zhang:2017bzy,Lin:2017ani,Bali:2017gfr,Alexandrou:2017dzj,Chen:2017gck,Alexandrou:2018pbm,Chen:2018fwa,Alexandrou:2018eet,Liu:2018uuj,Bali:2018spj,Lin:2018qky,Fan:2018dxu,Bali:2018qat,Sufian:2019bol,Alexandrou:2019lfo,Izubuchi:2019lyk,Cichy:2019ebf,Joo:2019jct,Joo:2019bzr,Alexandrou:2019dax,Chai:2020nxw,Joo:2020spy,Bhat:2020ktg,Sufian:2020vzb}. 
(See Refs.~\cite{Cichy:2018mum, Ji:2020ect, Constantinou:2020pek} for reviews on quasi-PDFs.) 
While for quite some time studies of quasi-PDFs were limited to the twist-2 case, recently first studies appeared that are related to twist-3 quasi-PDFs~\cite{Bhattacharya:2020cen, Bhattacharya:2020xlt, Bhattacharya:2020jfj, Braun:2021aon}.
In this work, we establish the BC-type sum rules for quasi-PDFs, both through a model-independent analysis and through analytical as well as numerical model calculations. 
Like in the case of the BC-type sum rules for light-cone PDFs, the corresponding sum rules for quasi-PDFs are violated in the cut-off scheme which breaks rotational invariance. 

Finally, we calculate the twist-3 PDF $e(x)$ and its corresponding quasi-PDF.
We also explore, in the QTM and the YM, the sum rule which relates this function to the target mass.
The general finding of the model calculations matches our study of the BC-type sum rules.
While the sum rule for $e(x)$ holds trivially for the UV-divergent terms, care must be taken in the case of finite terms.

We organize the manuscript as follows: In Sec.~\ref{s:definitions} we provide definitions of the light-cone PDFs ($g_{1}(x), \, $ $g_{T}(x)$) and ($h_{1}(x), \, $ $h_{L}(x)$), and of the quasi-PDFs ($g_{1, \rm{Q}}(x), \, $ $g_{T, \rm{Q}}(x)$) and ($h_{1, \rm{Q}}(x), \, $ $h_{L, \rm{Q}}(x)$). 
In that section, we give a model-independent discussion of the BC-type sum rules for both light-cone and quasi-PDFs.
In Sec.~\ref{s:main_results_lc} we present the one-loop perturbative results for the light-cone PDFs ($g_{1}(x), \, $ $g_{T}(x)$) and ($h_{1}(x), \, $ $h_{L}(x)$) in the QTM and the YM. With the help of these results, we show analytically that the BC-type sum rules are valid when one uses DR for the UV. Besides, we draw attention to the explicit violation of these sum rules for the UV-finite terms when one employs a cut-off scheme. To the best of our knowledge, such an issue, in the context of BC-type sum rules, has never been reported before.
We argue that the observed problem with a cut-off scheme is the lack of rotational invariance which, as mentioned before, is the key ingredient responsible for the existence of the sum rules in the first place. 
In Sec.~\ref{s:main_results_quasi} we present the one-loop results for the quasi-PDFs ($g_{1, \rm{Q}}(x), \, $ $g_{T, \rm{Q}}(x)$) and ($h_{1, \rm{Q}}(x), \, $ $h_{L, \rm{Q}}(x)$) in the two models.  
Sec.~\ref{s:numerical_results} is dedicated to numerical results for the sum rules for both light-cone and quasi-PDFs. 
In that section, we also clarify when and why the moments of quasi-PDFs should converge, and the impact of a twist-expansion on those moments. In Sec.~\ref{s:results_e_eQ} we calculate the light-cone PDF $e(x)$, and its corresponding quasi-PDF $e_{\rm{Q}}(x)$ in the QTM and the YM, and show that their moments agree. In particular, we consider the relation between $e(x)$ and the mass of the target.
We summarize our work in Sec.~\ref{s:summary}.
The Appendix~\ref{a:interesting_point_DR_IR} contains a discussion about an issue that can arise when applying DR to both UV and IR divergences, which is (also) related to the breaking of rotational invariance.

\section{Definition of PDFs and Overview of BC-type Sum rules}
\label{s:definitions}
\subsection{PDF Definitions}
Light-cone PDFs are defined in terms of the correlation function\footnote{For any generic four-vector $v$, we
define the Minkowski components by $(v^0, v^1, v^2, v^3)$ and
the light-cone components by $(v^+, v^-, \vec{v}_\perp)$, where
$v^+ = \frac{1}{\sqrt{2}} (v^0 + v^3)$, $v^- = \frac{1}{\sqrt{2}}
(v^0 - v^3)$ and $\vec{v}_\perp = (v^1, v^2)$.}
\begin{equation}
\Phi^{[\Gamma]}(x, S) = \frac{1}{2} \int \frac{dz^-}{2\pi} \, e^{i k \cdot z} \, \langle P, S | \bar{\psi}(- \tfrac{z}{2}) \, \Gamma \, {\cal W}(- \tfrac{z}{2}, \tfrac{z}{2}) \,\psi(\tfrac{z}{2})  | P, S \rangle \Big|_{z^+ = 0, \vec{z}_\perp = \vec{0}_\perp} \, ,
\label{e:corr_standard_GPD}
\end{equation}
where, $\Gamma$ is a generic gamma matrix.
Color gauge invariance of this non-local quark-quark correlator is ensured by the Wilson line
\begin{equation}
{\cal W}(- \tfrac{z}{2}, \tfrac{z}{2}) \Big|_{z^+ = 0, \vec{z}_\perp = \vec{0}_\perp}
= {\cal P} \, \textrm{exp} \, \Bigg( - i g_s \int_{-\tfrac{z^-}{2}}^{\tfrac{z^-}{2}} \, dy^- \, A^+(0^+, y^-, \vec{0}_\perp) \Bigg) \, .
\end{equation}
Here, ${\cal P}$ is a path-ordered exponential depending on $A^{+}$, which is the plus-component of the gluon field, and $g_{s}$ denotes the strong coupling constant. The state of the hadron is characterized by the 4-momentum $P^{\mu} = (P^{+}, P^{-}, \vec{0}_{\perp})$ and a covariant spin vector $S$ for which one can write
\begin{equation}
S^{\mu} = (S^+, S^-, \vec{S}_\perp) = \bigg( \lambda \frac{P^+}{M}, - \lambda \frac{M}{2 P^+}, \vec{S}_\perp \bigg) \, .
\end{equation}
Here, $\lambda$  and $M$ denotes the helicity and the mass of the hadron, respectively, and $i$ is an index in the transverse space. By definition, the spin vector satisfies the constraint $P \cdot S = 0$. The twist-2 light-cone PDFs $g_{1}(x)$ and $h_{1}(x)$ are defined as
\begin{eqnarray}
\Phi^{[\gamma^{+} \gamma_{5}]} = \lambda \, g_{1}(x) \, , \qquad
\Phi^{[i \sigma^{i+} \gamma_5]}  = S^{i}_{\perp} \,  h_{1}(x) \,,
\end{eqnarray}
while the twist-3 light-cone PDFs $g_{T}(x)$ and $h_L(x)$ are defined as
\begin{eqnarray}
\Phi^{[\gamma^{i}_{\perp} \gamma_{5}]} = \dfrac{M}{P^{+}} S^{i}_{\perp} \, g_{T}(x) \, , \qquad
\Phi^{[i \sigma^{+-} \gamma_5]}  =\dfrac{M}{P^{+}} \lambda \, h_{L}(x) \, .
\end{eqnarray}
In the above expressions, $\sigma^{\mu\nu}=\frac{i}{2}(\gamma^{\mu}\gamma^{\nu}-\gamma^{\nu}\gamma^{\mu})$ and $\gamma_{5}$ is the matrix which anti-commutes with other Dirac matrices. 
As evident from these expressions, longitudinal target polarization is required to address $g_{1}(x)$ and $h_{L}(x)$, while transverse polarization is needed for $h_{1}(x)$ and $g_{T}(x)$.
The light-cone PDFs depend on $x = k^{+}/P^{+}$ and have support in the region $- 1 \leq x \leq 1$.

Quasi-PDFs are defined through the spatial correlation function~\cite{Ji:2013dva,Ji:2014gla}
\begin{equation}
\Phi_{\rm Q}^{[\Gamma]}(x, S; P^{3}) = \frac{1}{2} \int \frac{dz^3}{2\pi} \, e^{i k \cdot z} \, \langle P, S | \bar{\psi}(- \tfrac{z}{2}) \, \Gamma \, {\cal W}_{\rm Q}(- \tfrac{z}{2}, \tfrac{z}{2}) \,\psi(\tfrac{z}{2})  | P, S \rangle \Big|_{z^0 = 0, \vec{z}_\perp = \vec{0}_\perp} \,,
\label{e:corr_quasi_GPD}
\end{equation}
with the Wilson line
\begin{equation}
{\cal W}_{\rm Q}(- \tfrac{z}{2}, \tfrac{z}{2}) \Big|_{z^0 = 0, \vec{z}_\perp = \vec{0}_\perp}
= {\cal P} \, \textrm{exp} \, \Bigg( - i g_s \int_{-\tfrac{z^3}{2}}^{\tfrac{z^3}{2}} \, dy^3 \, A^3(0, \vec{0}_\perp, y^3) \Bigg) \, .
\end{equation}
In this case, we write the 4-momentum of the hadron as $P^{\mu} = (P^{0}, \vec{0}_{\perp}, P^{3})$, and the spin vector as
\begin{equation}
S^{\mu} = (S^{0}, \vec{S}_\perp, S^{3}) = \bigg( \lambda \frac{P^{3}}{M}, \vec{S}_\perp, \lambda \frac{P^{0}}{M} \bigg) \, .
\end{equation}
The quasi-PDFs $g_{1, \rm{Q}}(x)$ and $h_{1, \rm{Q}}(x)$ are defined according to
\begin{eqnarray}
\Phi_{\rm Q}^{[\gamma^{3}\gamma_{5}]} = \lambda \, \delta_{0} \, g_{1, \rm{Q}}(x; P^{3}) \,, \qquad
\Phi_{\rm Q}^{[i \sigma^{i0} \gamma_5]}  =  S^{i}_{\perp} \, \delta_{0} \, h_{1, {\rm Q}}(x; P^{3}) \, ,
\end{eqnarray} 
where $\delta_{0} = \sqrt{1 + M^{2}/P^{2}_{3}} \,$ \footnote{For convenience of notation, throughout our work we will be using $(P^{3})^{2} \rightarrow P^{2}_{3}$.} 
so that $P^0 = \delta_0 \, P^3$.
The factor $\delta_0$ in the above equations is needed for getting the same lowest moment of the quasi-PDFs and the corresponding light-cone PDFs~\cite{Bhattacharya:2019cme}.
Note that one can choose to work with the gamma matrix $\gamma^{0} \gamma_{5}$ for $g_{1, \rm{Q}} (x)$, and $i \sigma^{i3} \gamma_5$ for $h_{1, \rm{Q}} (x)$~\cite{Bhattacharya:2019cme}. The conclusions of our present work are not affected by these alternative choices. 
The quasi-PDFs $g_{T, \rm{Q}}(x)$ and $h_{L, \rm{Q}}(x)$ are defined as
\begin{eqnarray}
\Phi_{\rm Q}^{[\gamma^{i}_{\perp} \gamma_{5}]} = \dfrac{M}{P^{3}} \, S^{i}_{\perp} \, g_{T, \rm{Q}}(x; P^{3}) \,, \qquad
\Phi_{\rm Q}^{[i \sigma^{3 0} \gamma_5]}  =  \dfrac{M}{P^{3}} \, \lambda \, h_{L, {\rm Q}}(x; P^{3}) \, ,
\end{eqnarray} 
where, $x = k^{3}/P^{3}$. In contrast to the light-cone PDFs, quasi-PDFs have support in the region $- \infty \leq x \leq \infty$. (The variable $x$ for the quasi-PDFs should not be confused with the momentum fraction for light-cone PDFs.) 
Note that the quasi-PDFs have an explicit dependence on $P^{3}$.

\subsection{BC-type sum rules}
\label{ss:model_independent_sum_rule_quasi}
The local axial current and tensor current define the axial charge $g_A$ and the tensor charge $g_T$, respectively, through
\begin{eqnarray}
2 M S^{\mu} g_{A} &=&  \langle P, S | \bar{\psi}(0) \, \gamma^{\mu} \gamma_{5} \, \psi(0)  | P, S \rangle  \nonumber \, , 
\\[0.2cm]
2 (S^{\mu} P^{\nu} - S^{\nu} P^{\mu})  g_{T} &=&  \langle P, S | \bar{\psi}(0) \, i \sigma^{\mu \nu} \gamma_{5} \, \psi(0)  | P, S \rangle  \, .
\label{e:axial_tensor_charge}
\end{eqnarray}
These two equations are a consequence of Lorentz invariance.
It is now straightforward to show that
\begin{eqnarray}
\lambda \int dx \, g_{1} (x) &=& \dfrac{1}{2P^{+}} \langle P, S | \bar{\psi}(0) \, \gamma^{+} \gamma_{5} \, \psi(0)  | P, S \rangle  = \lambda g_{A} \nonumber \, , 
\\[0.2cm]
\dfrac{M}{P^{+}} S^{i}_{\perp}  \int dx \, g_{T} (x) &=&  \dfrac{1}{2P^{+}} \langle P, S | \bar{\psi}(0) \, \gamma^{i}_{\perp} \gamma_{5} \, \psi(0)  | P, S \rangle = \dfrac{M}{P^{+}} S^{i}_{\perp} g_{A} \, ,
\end{eqnarray}
which leads to 
\begin{eqnarray}
\int dx \, g_{1} (x) = \int dx \, g_{T} (x) = g_{A} \, .
\label{e:def_BC_LC}
\end{eqnarray}
Eq.~(\ref{e:def_BC_LC}) is known as the BC sum rule~\cite{Burkhardt:1970ti}. 
Since, according to Eq.~(\ref{e:axial_tensor_charge}), the axial charge appears for both longitudinal and transverse polarization, the BC sum rule can be considered a consequence of rotational invariance.
For the chiral-odd functions $h_{1}$ and $h_{L}$ we get
\begin{eqnarray}
S^{i}_{\perp} \int dx \, h_{1} (x) &=& \dfrac{1}{2P^{+}} \langle P, S | \bar{\psi}(0) \, i \sigma^{i +} \gamma_{5} \, \psi(0)  | P, S \rangle  = S^{i}_{\perp} \, g_{T} \nonumber \, , 
\\[0.2cm]
\dfrac{M}{P^{+}} \lambda  \int dx \, h_{L} (x) &=&  \dfrac{1}{2P^{+}} \langle P, S | \bar{\psi}(0) \, i \sigma^{+ -} \gamma_{5} \, \psi(0)  | P, S \rangle = \dfrac{M}{P^{+}} \lambda \, g_{T}   \, ,
\end{eqnarray}
which leads to 
\begin{eqnarray}
\int dx \, h_{1} (x) = \int dx \, h_{L} (x) = g_{T} \, .
\label{e:def_h_LC}
\end{eqnarray}
Eq.~(\ref{e:def_h_LC}) is known as the h-sum rule~\cite{Tangerman:1994bb, Burkardt:1995ts}.

A corresponding analysis for the quasi-PDFs readily provides the sum rules
\begin{eqnarray}
\int dx \, g_{1, \rm{Q}} (x; P^3) &=& \int dx \, g_{T, \rm{Q}} (x; P^3) = g_{A} \nonumber \, , \\[0.2cm]
\int dx \, h_{1, \rm{Q}} (x; P^3) &=& \int dx \, h_{L, \rm{Q}} (x; P^3) = g_{T} \, ,
\end{eqnarray}
that is, the BC-type sum rules hold for quasi-PDFs as well --- see also the corresponding brief discussion in Ref.~\cite{Bhattacharya:2020cen}.
The lowest moments of quasi-PDFs are $P^{3}$-independent, and they agree with those for the corresponding light-cone PDFs. 
This outcome is in line with a model-independent calculation of moments for the twist-2 PDFs and twist-2 GPDs~\cite{Bhattacharya:2019cme}.

\section{Analytical results for the Light-cone PDFs and the BC-type Sum rules}
\label{s:main_results_lc}
This section focuses on the calculation of the sum rules for the light-cone PDFs ($g_{1}(x), \, $ $g_{T}(x)$), and ($h_{1}(x), \, $ $h_{L}(x)$). 
Explicit calculations to one-loop order are provided in two models: the QTM and the YM.
We use three different schemes to regulate the infra-red (IR) divergences: non-zero gluon mass $m_{g} \neq 0$, non-zero quark mass $m_{q} \neq 0$, and dimensional regularization (DR). For the ultraviolet (UV) divergences, we employ two schemes: DR and cut-off. Since our calculations are at a partonic level, from hereafter we will use $p$ as the momentum label for the target in both models.

\subsection{Results in Quark Target Model}
\begin{figure}[t]
\begin{center}
    \includegraphics[width=15cm]{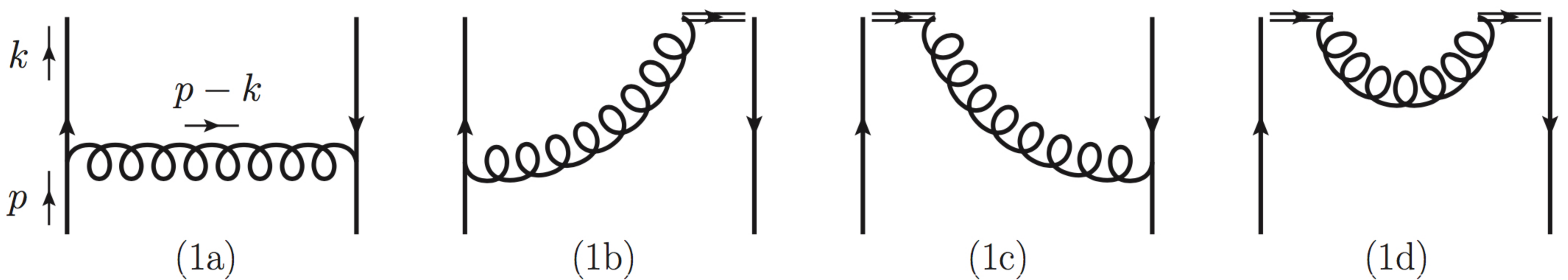}
 	\caption{One-loop real diagrams contributing to the light-cone PDFs and the quasi-PDFs in the QTM.}
 	\label{fig:real}
\vspace{0.5cm}
    \includegraphics[width=14.6cm]{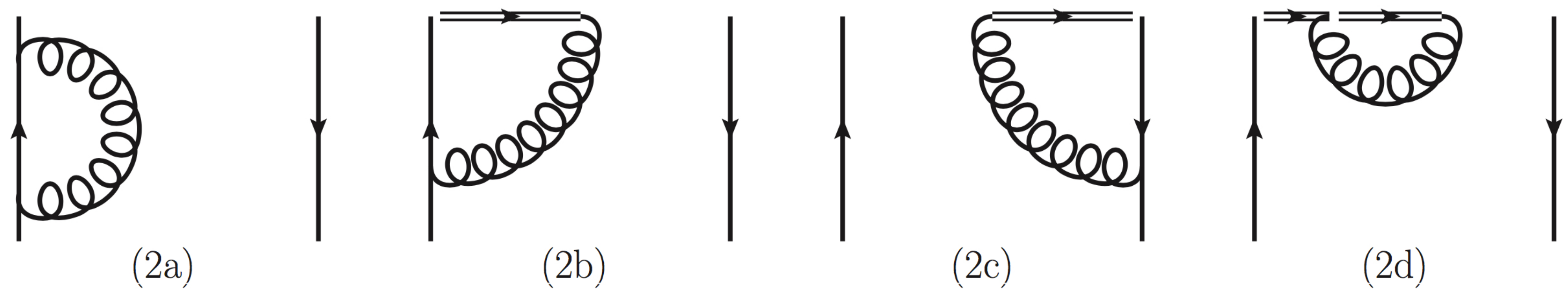}
 	\caption{One-loop virtual diagrams contributing to the light-cone PDFs and the quasi-PDFs in the QTM. The Hermitian conjugate diagrams of (2a) and (2d) have not been shown.}
 	\label{fig:virtual}
\end{center}
\end{figure}
Calculations within QTM can in principle be done in any gauge. Here, we work in the Feynman gauge for which the real and virtual diagrams have been shown in Fig.~\ref{fig:real} and Fig.~\ref{fig:virtual}, respectively. 

\subsubsection{BC sum rule}
We start with the calculation of the real diagrams for $g_{1}(x)$. For the diagram in Fig.~($\rm{1a}$), we obtain the following general expression, before the $k_{\perp}$ integration, in terms of both $m_{g} \neq 0$~\footnote{It is known that a nonzero gluon mass in QCD violates gauge invariance.  However, the calculations in this work do not involve a gluon self-interaction and, therefore, are like QED treatments (modulo a color factor). Generally, in QED a nonzero photon mass can be used to isolate IR singularities.  This feature is sufficient for the purpose of our study.} and $m_{q} \neq 0$, 
\begin{eqnarray}
g^{\rm{(1a)}}_{1} (x) &=& - \dfrac{g^{2}_{s} C_{F} \mu^{2\epsilon}}{2\pi} (1-x) \int \dfrac{d^{n-2}k_{\perp}}{(2\pi)^{n-2}} \, \dfrac{- (1-\epsilon) k^{2}_{\perp} + (1-\epsilon) (1+x^{2}) m^{2}_{q} + 2 \epsilon x m^{2}_{q}}{ \big ( k^{2}_{\perp} + (1-x)^{2} m^{2}_{q} + x m^{2}_{g} \big ) ^{2}} \, .
\label{e:g1_QTM_general}
\end{eqnarray}
Here, $g_{s}$ is the QCD coupling constant, $C_{F} = (N^{2}_{\rm{c}}-1)/2N_{\rm{c}}$ is the color factor and $n=4-2\epsilon$. In Eq.~(\ref{e:g1_QTM_general}), we have applied DR to the transverse momentum integral in order to regulate the UV divergences, and IR divergences present in the case of working with zero partonic masses~\footnote{We have used Kreimer's prescription for the treatment of $\gamma_5$ in $n$-dimensions, that is, before solving for the traces, we have anti-commuted the $\gamma_5$'s and used $(\gamma_5)^2=1$~\cite{Korner:1991sx}. We did not have to use any other property of $\gamma_5$ in $n$ dimensions. In particular, we did not have to evaluate expressions such as ${\rm Tr} (\gamma^{\mu} \gamma^{\nu} \gamma^{\alpha} \gamma^{\beta} \gamma_5)$. We therefore believe that our conclusions are unaffected by the choice of scheme for the treatment of $\gamma_5$ in dimensional regularization.}. For the UV divergences, one must satisfy the condition $\epsilon \rightarrow \epsilon_{\UV} > 0$ (and the corresponding subtraction scale is $\mu \rightarrow \mu_{\UV} > 0$). For the IR divergences, one must ensure the condition $\epsilon \rightarrow \epsilon_{\IR} < 0$ (and $\mu \rightarrow \mu_{\IR} > 0$). For the analytical check of the sum rules, we will not be working with the general expression provided in Eq.~(\ref{e:g1_QTM_general}). Rather, for the sake of simplicity, we will be invoking different IR schemes. However, for the numerical check of the sum rules, we will recourse to the expression in Eq.~(\ref{e:g1_QTM_general}).
After regulating the UV divergences in the DR scheme, we obtain the following results for $g_{1}(x)$ with three different IR regulators, 
\begin{eqnarray}
g^{\rm{(1a)}}_{1} (x) \Big |^{\epsilon_{\UV}}=
\begin{dcases}
& g^{\rm{(1a)}}_{1} (x) \Big |^{\epsilon_{\UV}}_{m_{g}} = \dfrac{\alpha_{s}C_{F}}{2\pi} (1-x) \bigg ( {\cal P}_{\UV} + \ln \dfrac{\mu^{2}_{\UV}}{x m^{2}_{g}} - 2 \bigg )  \, ,
\\[0.2cm]
& g^{\rm{(1a)}}_{1} (x) \Big |^{\epsilon_{\UV}}_{m_q} = \dfrac{\alpha_{s}C_{F}}{2\pi} (1-x) \bigg ( {\cal P}_{\UV} + \ln \dfrac{\mu^{2}_{\UV}}{(1-x)^{2} m^{2}_{q}} - 2 - \dfrac{1+x^{2}}{(1-x)^{2}}\bigg ) \, ,
\\[0.2cm]
& g^{\rm{(1a)}}_{1} (x) \Big |^{\epsilon_{\UV}}_{\epsilon_{\IR}} = \dfrac{\alpha_{s}C_{F}}{2\pi} (1-x) \bigg ( {\cal P}_{\UV} - {\cal P}_{\IR}  + \ln \dfrac{\mu^{2}_{\UV}}{\mu^{2}_{\IR}} \bigg ) \, ,
\end{dcases}
\label{e:g1_DR}
\end{eqnarray}
where
\begin{displaymath}
{\cal P}_{\UV/ \IR} = \frac{1}{\epsilon_{\UV /\IR}} + \ln 4\pi - \gamma_E \, .
\end{displaymath}
On the other hand, if a cut-off is applied on the $k_{\perp}$ integral in Eq.~(\ref{e:g1_QTM_general}), we get
\begin{eqnarray}
g^{\rm{(1a)}}_{1} (x) \Big |^{\Lambda_{\UV}}=
\begin{dcases}
& g^{\rm{(1a)}}_{1} (x) \Big |^{\Lambda_{\UV}}_{m_{g}} = \dfrac{\alpha_{s} C_{F}}{2\pi} (1-x) \bigg ( \ln \dfrac{\Lambda^{2}_{\UV}}{x m^{2}_{g}} - 1 \bigg ) \, ,
\\[0.2cm]
& g^{\rm{(1a)}}_{1} (x) \Big |^{\Lambda_{\UV}}_{m_q} =  \dfrac{\alpha_{s} C_{F}}{2\pi} (1-x) \bigg ( \ln \dfrac{\Lambda^{2}_{\UV}}{(1-x)^{2}m^{2}_{q}} - 1 - \dfrac{1+x^{2}}{(1-x)^{2}} \bigg ) \, ,
\end{dcases}
\label{e:g1_cut-off}
\end{eqnarray}
with $m_{g} \neq 0$ and $m_{q} \neq 0$, respectively. We observe that the coefficient of the UV poles, be it in the DR or in the cut-off scheme, are exactly the same. However, the finite factors are different in the two schemes. 
We will return to this point later towards the end of this section.

It is straightforward to calculate the contribution of the diagram in Fig.~($\rm{1b}$) to $g_{1}(x)$. We obtain the following results when DR is used for the UV,
\begin{eqnarray}
g^{\rm{(1b)}}_{1} (x) \Big |^{\epsilon_{\UV}}=
\begin{dcases}
& g^{\rm{(1b)}}_{1}(x) \Big |^{\epsilon_{\UV}}_{m_{g}}  =  \frac{\alpha_s C_F}{2\pi} \, \frac{x}{1-x} \, \bigg ( {\cal P}_{\UV} + \ln \frac{\mu_{\UV}^2}{x m_g^2} \bigg ) \, , 
\\[0.2cm]
& g^{\rm{(1b)}}_{1}(x) \Big |^{\epsilon_{\UV}}_{m_{q}}  =  \frac{\alpha_s C_F}{2\pi} \, \frac{x}{1-x} \, \bigg ( {\cal P}_{\UV} + \ln \frac{\mu_{\UV}^2}{(1-x)^2 m_q^2} \bigg ) \, ,
\\[0.2cm]
& g^{\rm{(1b)}}_{1}(x) \Big |^{\epsilon_{\UV}}_{\epsilon_{\IR}}  =  \frac{\alpha_s C_F}{2\pi} \, \frac{x}{1-x} \, \bigg ( {\cal P}_{\UV} - {\cal P}_{\IR} + \ln \frac{\mu_{\UV}^2}{\mu_{\IR}^2} \bigg ) \, ,
\end{dcases}
\end{eqnarray}
while in the cut-off scheme we find
\begin{eqnarray}
g^{\rm{(1b)}}_{1} (x) \Big |^{\Lambda_{\UV}}=
\begin{dcases}
& g^{\rm{(1b)}}_{1}(x) \Big |^{\Lambda_{\UV}}_{m_{g}}  =  \frac{\alpha_s C_F}{2\pi} \, \frac{x}{1-x} \, \bigg ( \ln \frac{\Lambda_{\UV}^2}{x m_g^2} \bigg ) \, , 
\\[0.2cm]
& g^{\rm{(1b)}}_{1}(x) \Big |^{\Lambda_{\UV}}_{m_{q}}  =  \frac{\alpha_s C_F}{2\pi} \, \frac{x}{1-x} \, \bigg ( \ln \frac{\Lambda_{\UV}^2}{(1-x)^2 m_q^2} \bigg ) \, .
\end{dcases}
\label{e:g1_1b_DR_QTM}
\end{eqnarray}
The diagram in Fig.~($\rm{1c}$) gives the same result as that of Fig.~($\rm{1b}$). This outcome is due to the relevant trace algebra. In fact, this pattern continues for all the other PDFs. The diagram in Fig.~($\rm{1d}$) does not contribute to the light-cone PDFs because the result is proportional to the square of the unit light-cone vector~\cite{Bhattacharya:2020xlt}.

We now proceed to the calculation of the virtual diagrams. 
All those diagrams exhibit the factor $\delta(1-x)$ which we include below when summing up the terms.
The contribution of the quark self-energy diagram, as shown in Fig~$\rm{(2a)}$, does not depend on the type of PDF under discussion. In Ref.~\cite{Bhattacharya:2020xlt}, we provided the results for this diagram when DR was used for the UV. We quote the results here for the sake of completeness,
\begin{eqnarray}
\dfrac{\partial \Sigma (p)}{\partial \slashed{p}} \Big |^{\epsilon_{\UV}} =
\begin{dcases}
& \dfrac{\partial \Sigma (p)}{\partial \slashed{p}} \Big |^{\epsilon_{\UV}}_{m_{g}} =
-\dfrac{\alpha_{s} C_{F}}{2 \pi} \int_{0}^{1} dy \, y  \bigg ( {\cal P}_{\UV} + \ln \dfrac{\mu_{\UV}^{2}}{y m^{2}_{g} } -1 \bigg ) \,,
\label{e:...}
\\[0.2cm]
& \dfrac{\partial \Sigma (p)}{\partial \slashed{p}} \Big |^{\epsilon_{\UV}}_{m_{q}} =
-\dfrac{\alpha_{s} C_{F}}{2 \pi} \int_{0}^{1} dy \, (1-y)  \bigg ( {\cal P}_{\UV} + \ln \dfrac{\mu_{\UV}^{2}}{(1-y)^2 m_q^{2}} - \dfrac{1+y^2}{(1-y)^{2}} \bigg ) \,,
\\[0.2cm]
& \dfrac{\partial \Sigma (p)}{\partial \slashed{p}} \Big |^{\epsilon_{\UV}}_{\epsilon_{\IR}}  =
-\dfrac{\alpha_{s} C_{F}}{2 \pi} \int_{0}^{1} dy \, y  \bigg (  {\cal P}_{\UV}- {\cal P}_{\IR}  + \ln \dfrac{\mu_{\UV}^{2}}{\mu_{\IR}^{2} } \bigg ) \, .
\end{dcases}
\end{eqnarray}
We obtain the following results in the cut-off scheme,
\begin{eqnarray}
\dfrac{\partial \Sigma (p)}{\partial \slashed{p}} \Big |^{\Lambda_{\UV}} =
\begin{dcases}
& \dfrac{\partial \Sigma (p)}{\partial \slashed{p}} \Big |^{\Lambda_{\UV}}_{m_{g}} =
-\dfrac{\alpha_{s} C_{F}}{2 \pi} \int_{0}^{1} dy  \, \bigg (y \ln \dfrac{\Lambda^{2}_{\UV}}{y m^{2}_{g}} \bigg ) \, ,
\label{e:...}
\\[0.2cm]
& \dfrac{\partial \Sigma (p)}{\partial \slashed{p}} \Big |^{\Lambda_{\UV}}_{m_{q}} =
-\dfrac{\alpha_{s} C_{F}}{2 \pi} \int_{0}^{1} dy \, \bigg ( (1-y) \ln \dfrac{\Lambda^{2}_{\UV}}{(1-y)^{2} m^{2}_{q}} - \dfrac{2y}{1-y} \bigg ) \, .
\end{dcases}
\label{e:2a_DR_QTM}
\end{eqnarray}
The $x$-integrals of the contributions from the diagrams~(2b) and~(2c) are exactly the same as~(1b) and~(1c), except for an overall sign, which is due to the reversed direction for the momentum flow in the Wilson line. Just like Fig.~$\rm{(1d)}$, the contribution from Fig.~$\rm{(2d)}$ drops out.

We now turn our attention to $g_{T}(x)$. In case of the twist-3 PDFs, the result from Fig.~($\rm{1a}$) can be split in two distinct parts: a singular part and a canonical part~\cite{Bhattacharya:2020xlt}, 
\begin{eqnarray}
g^{\rm{(1a)}}_{T} (x)  &=& g^{\rm{(1a)}}_{T \rm{(s)}} (x) + g^{\rm{(1a)}}_{T \rm{(c)}} (x) \, ,
\end{eqnarray}
where the singular parts are related to the zero-mode $\delta(x)$ contributions. The general expressions for the singular and canonical parts, before the $k_{\perp}$ integration, are 
\begin{eqnarray}
g^{\rm{(1a)}}_{T \rm{(s)}} (x) &=& - \dfrac{g^{2}_{s} C_{F} \mu^{2\epsilon}}{2\pi} \, \delta (x) \int \dfrac{d^{n-2}k_{\perp}}{(2\pi)^{n-2}} \, \dfrac{\epsilon}{(k^{2}_{\perp} + m^{2}_{q})} \, , \nonumber \\[0.2cm]
g^{\rm{(1a)}}_{T \rm{(c)}} (x) &=& \dfrac{g^{2}_{s} C_{F} \mu^{2\epsilon}}{2\pi} \int \dfrac{d^{n-2}k_{\perp}}{(2\pi)^{n-2}} \, \dfrac{x k^{2}_{\perp} - (1-x^{2}) m^{2}_{q} + x m^{2}_{g} + \epsilon (1-x) m^{2}_{g}}{ \big ( k^{2}_{\perp} + (1-x)^{2} m^{2}_{q} + x m^{2}_{g} \big ) ^{2} } \, .
\end{eqnarray}
The origin of the delta function is in the integral~\cite{Yan:1973qg,Burkardt:2001iy,Aslan:2018tff,Bhattacharya:2020xlt,Bhattacharya:2020jfj}
\begin{equation}
\int d k^{-} \frac{1}{(k^2 - m_q^2 + i\varepsilon)^{2}} =  \frac{i \pi}{k^2_\perp + m_q^2} \, \frac{\delta (x)}{p^{+}} \, .
\label{e:Dirac_delta}
\end{equation}
After the $k_{\perp}$ integrals are carried out, for $m_{g} \neq 0$, one obtains the following two expressions for the singular parts~\cite{Bhattacharya:2020xlt}
\begin{eqnarray}
g^{\rm{(1a)}}_{T \rm{(s)}} (x) \Big |^{\epsilon_{\UV}} =
\begin{dcases}
& g^{\rm (1a)}_{T \rm{(s)}} (x) \Big |^{\epsilon_{\UV}}_{m_{q}} = - \dfrac{\alpha_{s} C_{F}}{2\pi} \, \delta (x) \, ,
\\[0.2cm]
& g^{\rm{(1a)}}_{T \rm{(s)}} (x) \Big |^{\epsilon_{\UV}}_{\epsilon_{\IR}} = 0 \, .
\end{dcases}
\label{e:gT_sing_DR}
\end{eqnarray}
As evident from Eq.~(\ref{e:Dirac_delta}), the zero-mode contributions originate exclusively from quark propagators. Therefore, to regulate the associated IR divergence in the $k_{\perp}$ integral, one is left with two options only: either work with $m_{q} \neq 0$, or apply DR. In other words, gluon mass never enters the discussion of the zero-mode contributions, because of which the associated IR divergence is left unguarded when $m_{g} \neq 0$. In Ref.~\cite{Bhattacharya:2020xlt}, we suggested that one could in principle keep working with $m_{g} \neq 0$ for the canonical part and for all the other diagrams, provided one uses $m_{q} \neq 0$ or a DR regulation for the zero-mode contributions. 
(Nevertheless, strictly speaking one must conclude that $m_g \neq 0$ is an insufficient IR regulator for twist-3 PDFs.
We will return to this point below.)
In the case of $m_{q} \neq 0$, the UV pole from the $k_{\perp}$ integral allows for a $\delta(x)$ in $g_{T}$. On the other hand, when DR is applied for the IR, both UV and IR poles allow for a $\delta(x)$, but with opposite signs with respect to one another, and hence the $\delta(x)$ contribution drops out~\cite{Bhattacharya:2020xlt}. For the canonical part, we get~\cite{Bhattacharya:2020xlt}
\begin{eqnarray}
g^{\rm{(1a)}}_{T \rm{(c)}} (x) \Big |^{\epsilon_{\UV}}_{m_{g}} &=& \dfrac{\alpha_{s} C_{F}}{2\pi} \bigg ( x \, {\cal P}_{\UV} + x \ln \frac{\mu_{\UV}^2}{x m_g^2} + (1 - x) \bigg ) \, ,
\end{eqnarray}
with $m_{g} \neq 0$. The results for $g_{T}(x)$ with $m_{q} \neq 0$ and DR for the IR are~\cite{Bhattacharya:2020xlt}
\begin{eqnarray}
g^{\rm{(1a)}}_{T} (x) \Big |^{\epsilon_{\UV}}_{m_{q}} &=& g^{\rm{(1a)}}_{T \rm{(s)}} (x) \Big |^{\epsilon_{\UV}}_{m_{q}} + g^{\rm{(1a)}}_{T \rm{(c)}} (x) \Big |^{\epsilon_{\UV}}_{m_{q}} 
\nonumber \\[0.2cm]
&=&  - \dfrac{\alpha_{s} C_{F}}{2\pi} \, \delta (x) + \dfrac{\alpha_{s} C_{F}}{2\pi} \bigg ( x \, {\cal P}_{\UV} +
x \ln \frac{\mu_{\UV}^2}{(1-x)^2 m_q^2} + \frac{x^2-2x-1}{1 - x} \bigg ) \, , \label{e:gT_mq_QTM} 
\\[0.2cm]
g^{\rm{(1a)}}_{T} (x) \Big |^{\epsilon_{\UV}}_{\epsilon_{\IR}}  &=& g^{\rm{(1a)}}_{T \rm{(s)}} (x) \Big |^{\epsilon_{\UV}}_{\epsilon_{\IR}} + g^{\rm{(1a)}}_{T \rm{(c)}} (x) \Big |^{\epsilon_{\UV}}_{\epsilon_{\IR}} \nonumber \\[0.2cm]
&=& 0 \, + \,  \dfrac{\alpha_{s} C_{F}}{2\pi} \, \bigg ( x \,  ({\cal P}_{\UV} - {\cal P}_{\IR} )+
x \ln \frac{\mu_{\UV}^2}{\mu_{\IR}^2}  \bigg ) \, .
\end{eqnarray}
When a cut-off is used for the UV, the zero-mode contribution drops out,
\begin{eqnarray}
g^{\rm (1a)}_{T \rm{(s)}} (x) \Big |^{\Lambda_{\UV}}_{m_{q}} = 0 \, ,
\label{e:gT_sing_cut-off}
\end{eqnarray}
because of the overall proportionality to $\epsilon_\UV$. Therefore, with $m_{g} \neq 0$ the result for $g_{T}(x)$ reads
\begin{eqnarray}
g^{\rm{(1a)}}_{T} (x) \Big |^{\Lambda_{\UV}}_{m_{g}}  &=& g^{\rm{(1a)}}_{T \rm{(s)}} (x) \Big |^{\Lambda_{\UV}} + g^{\rm{(1a)}}_{T \rm{(c)}} (x) \Big |^{\Lambda_{\UV}}_{m_{g}} \nonumber \\[0.2cm]
&=& 0 + \dfrac{\alpha_{s}C_{F}}{2\pi} \bigg ( x \ln \dfrac{\Lambda^{2}_{\UV}}{x m^{2}_{g}} + (1-x) \bigg ) \, .
\end{eqnarray}
Finally, with $m_{q} \neq 0$, we get
\begin{eqnarray}
g^{\rm{(1a)}}_{T} (x) \Big |^{\Lambda_{\UV}}_{m_{q}}  &=& g^{\rm{(1a)}}_{T \rm{(s)}} (x) \Big |^{\Lambda_{\UV}}_{m_{q}} + g^{\rm{(1a)}}_{T \rm{(c)}} (x) \Big |^{\Lambda_{\UV}}_{m_{q}} \nonumber \\[0.2cm]
&=& 0 + \dfrac{\alpha_{s} C_{F}}{2\pi} \bigg ( x \ln \dfrac{\Lambda^{2}_{\UV}}{(1-x)^{2}m^{2}_{q}} + \dfrac{x^{2} -2x -1}{1-x} \bigg )  \, .
\end{eqnarray}
Once again, we observe that the coefficient of the UV poles are exactly the same in the two UV schemes. In contrast to $g_{1}(x)$, the UV-finite pieces for $g_{T}(x)$ in the two UV schemes exactly match. 

For the diagram in Fig.~($\rm{1b}$), the results for $g_{T}(x)$ when DR is used for the UV read~\cite{Bhattacharya:2020xlt},
\begin{eqnarray}
g^{\rm{(1b)}}_{T} (x) \Big |^{\epsilon_{\UV}}=
\begin{dcases}
& g^{\rm{(1b)}}_{T}(x) \Big |^{\epsilon_{\UV}}_{m_{g}}  =  \frac{\alpha_s C_F}{4\pi} \, \frac{1 + x}{1-x} \, \bigg ( {\cal P}_{\UV} + \ln \frac{\mu_{\UV}^2}{x m_g^2} \bigg ) \, , 
\\[0.2cm]
& g^{\rm{(1b)}}_{T}(x) \Big |^{\epsilon_{\UV}}_{m_{q}}  =  \frac{\alpha_s C_F}{4\pi} \, \frac{1 + x}{1-x} \, \bigg ( {\cal P}_{\UV} + \ln \frac{\mu_{\UV}^2}{(1-x)^2 m_q^2} \bigg ) \, ,
\\[0.2cm]
& g^{\rm{(1b)}}_{T}(x) \Big |^{\epsilon_{\UV}}_{\epsilon_{\IR}}  =  \frac{\alpha_s C_F}{4\pi} \, \frac{1 + x}{1-x} \, \bigg ( {\cal P}_{\UV} - {\cal P}_{\IR} + \ln \frac{\mu_{\UV}^2}{\mu_{\IR}^2} \bigg ) \, ,
\end{dcases}
\end{eqnarray}
while in the cut-off scheme we find
\begin{eqnarray}
g^{\rm{(1b)}}_{T} (x) \Big |^{\Lambda_{\UV}}=
\begin{dcases}
& g^{\rm{(1b)}}_{T}(x) \Big |^{\Lambda_{\UV}}_{m_{g}}  =  \frac{\alpha_s C_F}{4\pi} \, \frac{1 + x}{1-x} \, \bigg ( \ln \frac{\Lambda_{\UV}^2}{x m_g^2} \bigg ) \, , 
\\[0.2cm]
& g^{\rm{(1b)}}_{T}(x) \Big |^{\Lambda_{\UV}}_{m_{q}}  =  \frac{\alpha_s C_F}{4\pi} \, \frac{1 + x}{1-x} \, \bigg ( \ln \frac{\Lambda_{\UV}^2}{(1-x)^2 m_q^2} \bigg ) \, ,
\end{dcases}
\end{eqnarray}
for the two IR regulators. Just as in the case of $g_{1}(x)$, the diagram in Fig.~($\rm{1c}$) gives the same result as that of Fig.~($\rm{1c}$), except for an overall sign. Moreover, the $x$-integrals of the contributions from Figs.~(1b) and~(1c) provide the very same results as diagrams~(2b) and~(2c), except for an overall sign. In the case of the cut-off scheme, our results for the UV-divergent parts of $g_{1}(x)$ ($h_{1}(x)$) and $g_{T}(x)$ ($h_{L}(x)$) are in agreement with the results of Refs.~\cite{Kundu:2001pk, Burkardt:2001iy}, where similar calculations were provided in the light-cone gauge.\footnote{In Ref.~\cite{Kundu:2001pk}, which employed a light-front Hamiltonian approach, a $\delta(x)$ term was missed for $h_{L}(x)$.}

We are now in a position to check the BC sum rule. 
The total result for $g_{1}(x)$ through one loop reads
\begin{eqnarray}
g_{1} (x) &=& \delta(1-x) + g^{\rm{(1a)}}_{1}(x) + g^{\rm{(1b)}}_{1}(x) + g^{\rm{(1c)}}_{1}(x) + \delta(1-x) \bigg ( \dfrac{\partial \Sigma (p)}{\partial \slashed{p}} + g^{\rm{(2b)}}_{1} + g^{\rm{(2c)}}_{1} \bigg ) \, ,
\end{eqnarray}
where the first term represents the (trivial) tree-level contribution. 
Upon taking the $x$-integral of the above equation, we see that Fig.~($\rm{1b}$) (Fig.~($\rm{1c}$)) cancels the contribution from Fig.~($\rm{2b}$) (Fig.~($\rm{2c}$)), such that
\begin{eqnarray}
\int^{1}_{0} dx \, g_{1} (x) &=& 1 + \int^{1}_{0} dx \, g^{\rm{(1a)}}_{1}(x) + \dfrac{\partial \Sigma (p)}{\partial \slashed{p}} \, .
\label{e:sum_rule_1a_2a}
\end{eqnarray}
(To understand the aforementioned point on cancellation, see the paragraphs after the Eqs.~(\ref{e:g1_1b_DR_QTM}) and~(\ref{e:2a_DR_QTM}).) 
This argument holds true for $g_{T}(x)$ as well. Since the contribution of the quark self-energy diagram is the same for both $g_{1}(x)$ and $g_{T}(x)$, we immediately see that, as far as the check of the sum rules are concerned, it suffices to consider the contribution from Fig.~($\rm{1a}$). 
In the following sections we will therefore provide the results for Fig.~($\rm{1a}$) only.

We begin our analysis in the instance that one does DR for the UV. We find that in the case of $m_{g} \neq 0$, the BC sum rule is satisfied provided one handles the IR divergence related to the zero-mode contributions with $m_{q} \neq 0$. Specifically, we have
\begin{eqnarray}
\int^{1}_{0} dx \, g^{\rm{(1a)}}_{1} (x)\Big |^{\epsilon_{\UV}}_{m_{g}} &=& \dfrac{\alpha_{s}C_{F}}{2\pi} \bigg ( \dfrac{1}{2} {\cal P}_{\UV} + \ln \dfrac{\mu_{\UV}}{m_{g}} - \dfrac{1}{4} \bigg ) \nonumber \\[0.2cm] 
&=& \int^{1}_{0} dx \, g^{\rm{(1a)}}_{T \rm{(s)}} (x) \Big |^{\epsilon_{\UV}}_{m_{q}} + \int^{1}_{0} dx \, g^{\rm{(1a)}}_{T \rm{(c)}} (x) \Big |^{\epsilon_{\UV}}_{m_{g}} \, .
\end{eqnarray}
On the other hand, if one applies DR for the IR of the zero-mode contribution, the BC sum rule is violated. 
Put differently, the recipe of using different IR regulators for the canonical and the singular terms in twist-3 PDFs, in general, fails, which re-emphasizes that $m_g \neq 0$ for twist-3 PDFs is problematic. (In fact, this issue is more severe for $h_{L}(x)$ as we discuss in the next section.)
In previous studies~\cite{Bhattacharya:2020xlt,Bhattacharya:2020jfj} we had already mentioned that $m_{g}\neq 0$ is problematic for the $x$-dependent results of the twist-3 PDFs. We did, however, not abandon a nonzero gluon mass, but rather proposed to work with either $m_{q}$ or DR for the singular terms, as already mentioned above. This recipe worked well for the calculation of matching coefficients,\footnote{``Matching" is a perturbative procedure that connects the quasi-PDFs to the light-cone PDFs.
We refer to the works in Refs.~\cite{Ji:2013dva,Ma:2014jla,Radyushkin:2017cyf,Wang:2017qyg,Stewart:2017tvs,Izubuchi:2018srq,Bhattacharya:2020xlt,Bhattacharya:2020jfj,Braun:2021aon} for more details on matching.} in the sense that these coefficients did not show a regulator-dependence. 
However, in the context of the sum rules, we observe that this recipe fails. 

For $m_{q} \neq 0$, we find
\begin{eqnarray}
\int^{1}_{0} dx \, g^{\rm{(1a)}}_{1} (x)\Big |^{\epsilon_{\UV}}_{m_{q}} &=& \dfrac{\alpha_{s}C_{F}}{2\pi} \bigg ( \dfrac{1}{2} {\cal P}_{\UV} + \ln \dfrac{\mu_{\UV}}{m_{q}} + \dfrac{2}{\beta} + 1 \bigg ) \nonumber \\[0.2cm] 
&=& \int^{1}_{0} dx \, g^{\rm{(1a)}}_{T \rm{(s)}} (x) \Big |^{\epsilon_{\UV}}_{m_{q}} + \int^{1}_{0} dx \, g^{\rm{(1a)}}_{T \rm{(c)}} (x) \Big |^{\epsilon_{\UV}}_{m_{q}} \, ,
\end{eqnarray}
where, $\beta <0$ denotes the $x=1$ pole present in diagram $\rm{(1a)}$.\footnote{Note that Eq.~(\ref{e:gT_mq_QTM}) has a term $\sim 1/1-x$. In order to carry out it's integral, we make the replacement $1/1-x \rightarrow 1/(1-x)^{1+\beta}$, with $\beta<0$, leading to singularities proportional to $1/\beta$.
Such a situation appears only for $m_{q} \neq 0$.
The singularities get cancelled when combining the diagrams in Fig.~(1a) and Fig.~(2a).}
Clearly, one has to include the contribution from the $\delta(x)$ term in order to satisfy the BC sum rule~\cite{Burkardt:2001iy,Aslan:2018tff}. 
When working with DR for the IR, we find that the BC sum rule is satisfied, although, as mentioned before, the $\delta(x)$ term drops out from $g_{T}(x)$,
\begin{eqnarray}
\int^{1}_{0} dx \, g^{\rm{(1a)}}_{1} (x)\Big |^{\epsilon_{\UV}}_{\epsilon_{\IR}} &=& \dfrac{\alpha_{s}C_{F}}{2\pi} \bigg ( \dfrac{1}{2} ({\cal P}_{\UV} - {\cal P}_{\IR} ) + \dfrac{1}{2} \ln \dfrac{\mu^{2}_{\UV}}{\mu^{2}_{\IR}}  \bigg ) \nonumber \\[0.2cm] 
&=& \int^{1}_{0} dx \, g^{\rm{(1a)}}_{T \rm{(c)}} (x) \Big |^{\epsilon_{\UV}}_{\epsilon_{\IR}} \, .
\label{e:g1_gT_DR_IR}
\end{eqnarray}
We note in passing that our results for $g_1$ and $g_T$ in the QTM allow us to make a comparison to the calculation of the structure function $g_2^{\rm s.f.}$ for deep-inelastic scattering off a quark target~\cite{Mertig:1993kq, Altarelli:1994dj}.
Specifically, the UV-divergent terms can be compared to terms in $g_2^{\rm s.f.}(x,Q^2)$ that are proportional to $\ln(Q^2/m_q^2)$, and we find complete agreement. Furthermore, we refer to Appendix~\ref{a:interesting_point_DR_IR} for a subtle point about DR for IR when $(\epsilon_{\IR}, \epsilon_{\UV})$ are held finite.

In the cut-off scheme, one can basically read off from the results that the UV divergent parts of $g_{1}(x)$ and $g_{T}(x)$ satisfy the BC sum rule --- compare also Refs.~\cite{Kundu:2001pk, Burkardt:2001iy}.
However, for the first time, we observe a violation of the BC sum rule for finite terms when using a cut-off regulator.
For $m_{g} \neq 0$, we find
\begin{eqnarray}
\int^{1}_{0} dx \, g^{\rm{(1a)}}_{1} (x) \Big |^{\Lambda_{\UV}}_{m_{g}} &=& \dfrac{\alpha_{s}C_{F}}{2\pi} \bigg ( \ln \dfrac{\Lambda_{\UV}}{m_{g}} + \dfrac{1}{4} \bigg ) \, ,
\\[0.2cm]
\int^{1}_{0} dx \, g^{\rm{(1a)}}_{T} (x) \Big |^{\Lambda_{\UV}}_{m_{g}} 
&=& \dfrac{\alpha_{s} C_{F}}{2\pi} \bigg ( \ln \dfrac{\Lambda_{\UV}}{m_{g}} + \dfrac{1}{4} + \dfrac{1}{2} \bigg ) \, ,
\end{eqnarray}
while for $m_{q} \neq 0$, we find
\begin{eqnarray}
\int^{1}_{0} dx \, g^{\rm{(1a)}}_{1} (x) \Big |^{\Lambda_{\UV}}_{m_{q}} &=& \dfrac{\alpha_{s}C_{F}}{2\pi} \bigg ( \ln \dfrac{\Lambda_{\UV}}{m_{q}} + \dfrac{2}{\beta} + \dfrac{3}{2}  \bigg ) \, ,
\\[0.2cm]
\int^{1}_{0} dx \, g^{\rm{(1a)}}_{T} (x) \Big |^{\Lambda_{\UV}}_{m_{q}} &=& \dfrac{\alpha_{s} C_{F}}{2\pi} \bigg ( \ln \dfrac{\Lambda_{\UV}}{m_{q}} + \dfrac{2}{\beta} + \dfrac{3}{2} + \dfrac{1}{2} \bigg ) \, .
\end{eqnarray}
It is worthwhile to pause and contemplate on: why is there a difference in the finite terms for the two UV schemes and, in particular, why is there a violation in the cut-off scheme? The underlying reason is rather simple. For cut-off and DR to give the same result, one must have the same prefactor in front of the UV-divergent integrals. If not, then the finite terms will depend on scheme. For example, consider the integrals\footnote{By ``divergent terms" in the two schemes, we mean ${\cal P}_{\UV} + \ln \mu^{2}_{\UV}/Q^{2}$ in DR, which translates to $\ln \Lambda^{2}_{\UV}/Q^{2}$ for a cut-off. By ``finite terms" we mean terms other than the aforementioned ones.}
\begin{eqnarray}
\mu^{2\epsilon} \, (1-\epsilon) \, \int^{\infty}_{0} \, \dfrac{d^{n - 2}k_{\perp}}{(2\pi)^{n - 2}} \, \dfrac{k^{2}_{\perp}}{\big ( k^{2}_{\perp} + Q^{2} \big )^{2}} & \approx & \dfrac{1}{4\pi} \bigg ( {\cal P}_{\UV} + \ln \dfrac{\mu^{2}_{\UV}}{Q^{2}} - 2 \bigg ) \, , 
\\[0.2cm]
\int^{\Lambda_{\UV}}_{0} \, \dfrac{d^{2}k_{\perp}}{(2\pi)^{2}} \, \dfrac{k^{2}_{\perp}}{\big ( k^{2}_{\perp} + Q^{2} \big )^{2}} & \approx & \dfrac{1}{4\pi} \bigg ( \ln \dfrac{\Lambda^{2}_{\UV}}{Q^{2}} - 1 \bigg ) \nonumber \, , 
\end{eqnarray}
with $Q^{2} > 0$. We see that the UV-divergent factors in the two schemes coincide. The difference in the finite factor can be attributed to an extra $(-\epsilon)$ term in DR. This is exactly the reason why we find different finite factors in the results for $g^{\rm{(1a)}}_{1} (x)$ in the two UV schemes, whereas the result for the canonical part of $g_{T}(x)$ remains unchanged. 
We now pose a hypothetical situation: what if one consistently carries out the algebra of the Dirac matrices in 4 dimensions, and then switches on DR or cut-off only at the time of carrying out the $k_{\perp}$ integrals? It is straightforward to check that this approach leads to the same final result in the two UV schemes, including the finite factors. 
But switching on DR right from the start, that is, keeping properly the factors in $\epsilon$, respects rotational invariance which underlies the BC sum rule.
This explains why the sum rule holds for DR and fails for a cut-off. 

Could we have avoided such a caveat with cut-off schemes? 
If our goal was to ``only" calculate the moments/integrals of the PDFs, we could have right from the start applied a cut-off to all 4 components of $k$, in the same spirit as one does in textbooks on quantum field theory.
We will outline such an analysis in the context of the h-sum rule in the next section. (The non-trivialities that stem from such an analysis can be appreciated more in the case of h-sum rule, which is the reason why we choose to highlight this case. The qualitative outcome of this study is however the same for both sum rules.)

\subsubsection{h-sum rule}
\label{ss:h_accident}
We first take up the calculations for $h_{1}(x)$. The diagram in  Fig.~($\rm{1a}$) contributes to $h_{1}(x)$ as 
\begin{eqnarray}
h^{\rm{(1a)}}_{1} (x) &=& - \dfrac{g^{2}_{s} C_{F} \mu^{2\epsilon}}{2\pi} \, (1-x) \int \dfrac{d^{n-2}k_{\perp}}{(2\pi)^{n-2}} \, \dfrac{ \bigg ( -\dfrac{\epsilon^{2}}{1-\epsilon} \bigg )  k^{2}_{\perp} + \epsilon (1+x^{2}) m^{2}_{q} + 2(1-\epsilon) x m^{2}_{q}}{ \big ( k^{2}_{\perp} + (1-x)^{2} m^{2}_{q} + x m^{2}_{g} \big ) ^{2} } \, .
\label{e:h1_QTM_general}
\end{eqnarray}
The final results in the DR scheme are
\begin{eqnarray}
h^{\rm{(1a)}}_{1} (x) \Big |^{\epsilon_{\UV}}=
\begin{dcases}
& h^{\rm{(1a)}}_{1} (x) \Big |^{\epsilon_{\UV}}_{m_{g}} = 0 \, ,
\\[0.2cm]
& h^{\rm{(1a)}}_{1} (x) \Big |^{\epsilon_{\UV}}_{m_q} = - \dfrac{\alpha_{s} C_{F}}{2\pi} \, \bigg ( \dfrac{2x}{1-x} \bigg ) \, ,
\\[0.2cm]
& h^{\rm{(1a)}}_{1} (x) \Big |^{\epsilon_{\UV}}_{\epsilon_{\IR}} = 0 \, ,
\end{dcases}
\label{e:h1_DR}
\end{eqnarray}
and in the cut-off scheme we get
\begin{eqnarray}
h^{\rm{(1a)}}_{1} (x) \Big |^{\Lambda_{\UV}}=
\begin{dcases}
& h^{\rm{(1a)}}_{1} (x) \Big |^{\Lambda_{\UV}}_{m_{g}} = 0 \, ,
\\[0.2cm]
& h^{\rm{(1a)}}_{1} (x) \Big |^{\Lambda_{\UV}}_{m_q} = - \dfrac{\alpha_{s} C_{F}}{2\pi} \, \bigg ( \dfrac{2x}{1-x} \bigg ) \, .
\end{dcases}
\label{e:h1_cut-off}
\end{eqnarray}
The above results suggest that $h^{\rm{(1a)}}_{1}(x)$ is UV-finite.

We now turn to $h_{L}(x)$. 
Before the $k_\perp$ integration is carried out, we find for the singular and canonical parts 
\begin{eqnarray}
h^{\rm{(1a)}}_{L \rm{(s)}} (x) &=& - \dfrac{g^{2}_{s} C_{F} \mu^{2\epsilon}}{2\pi} \, \delta (x) \int \dfrac{d^{n-2}k_{\perp}}{(2\pi)^{n-2}} \, \dfrac{1- \epsilon}{(k^{2}_{\perp} + m^{2}_{q})}  \, , \nonumber \\[0.2cm]
h^{\rm{(1a)}}_{L \rm{(c)}} (x) &=& \dfrac{g^{2}_{s} C_{F} \mu^{2\epsilon}}{2\pi} \int \dfrac{d^{n-2}k_{\perp}}{(2\pi)^{n-2}} \, \dfrac{x k^{2}_{\perp} + (1-2\epsilon) (1-x) k^{2}_{\perp}  - (1-x^{2}) m^{2}_{q} + x m^{2}_{g} + (1- \epsilon ) (1-x) m^{2}_{g}}{ \big ( k^{2}_{\perp} + (1-x)^{2} m^{2}_{q} + x m^{2}_{g} \big ) ^{2} } \, .
\end{eqnarray}
For $m_{g} \neq 0$, the singular part has two results~\cite{Bhattacharya:2020jfj}
\begin{eqnarray}
h^{\rm{(1a)}}_{L \rm{(s)}} (x) \Big |^{\epsilon_{\UV}} =
\begin{dcases}
& h^{\rm (1a)}_{L \rm{(s)}} (x) \Big |^{\epsilon_{\UV}}_{m_{q}} = - \dfrac{\alpha_{s} C_{F}}{2\pi} \, \delta (x) \bigg ( {\cal P}_{\UV} + \ln \dfrac{\mu^{2}_{\UV}}{m^{2}_{q}}-1 \bigg ) \, ,
\\[0.2cm]
& h^{\rm{(1a)}}_{L \rm{(s)}} (x) \Big |^{\epsilon_{\UV}}_{\epsilon_{\IR}}= - \dfrac{\alpha_{s} C_{F}}{2\pi} \, \delta (x) \bigg ( {\cal P}_{\UV} - {\cal P}_{\IR} + \ln \dfrac{\mu^{2}_{\UV}}{\mu^{2}_{\IR}} \bigg ) \, . 
\end{dcases}
\label{e:hL_sing_DR_QTM}
\end{eqnarray}
The result for the canonical part with $m_{g} \neq 0$ is~\cite{Bhattacharya:2020jfj}
\begin{eqnarray}
h^{\rm{(1a)}}_{L \rm{(c)}} (x) \Big |^{\epsilon_{\UV}}_{m_{g}} &=& \dfrac{\alpha_{s} C_{F}}{2\pi} \bigg ( {\cal P}_{\UV} + \ln \dfrac{\mu^{2}_{\UV}}{x m^{2}_{g}} +  \dfrac{(1-x)(1-2x)}{x} \bigg ) \, .
\end{eqnarray}
The results with $m_{q} \neq 0$ and DR for the IR are~\cite{Bhattacharya:2020jfj}
\begin{eqnarray}
h^{\rm{(1a)}}_{L} (x) \Big |^{\epsilon_{\UV}}_{m_{q}} &=& h^{\rm{(1a)}}_{L \rm{(s)}} (x) \Big |^{\epsilon_{\UV}}_{m_{q}} + h^{\rm{(1a)}}_{L \rm{(c)}} (x) \Big |^{\epsilon_{\UV}}_{m_{q}} \nonumber \\[0.2cm]
&=& - \dfrac{\alpha_{s} C_{F}}{2\pi} \, \delta (x) \bigg ( {\cal P}_{\UV} + \ln \dfrac{\mu^{2}_{\UV}}{m^{2}_{q}}-1 \bigg ) \nonumber + \dfrac{\alpha_{s} C_{F}}{2\pi} \bigg ( {\cal P}_{\UV} + \ln \dfrac{\mu^{2}_{\UV}}{(1-x)^{2}m^{2}_{q}} - \dfrac{2}{1-x} -2(1-x) \bigg ) \,,
\label{e:hL_full_DR_QTM}
\\[0.2cm]
h^{\rm{(1a)}}_{L} (x) \Big |^{\epsilon_{\UV}}_{\epsilon_{\IR}}  &=& h^{\rm{(1a)}}_{L \rm{(s)}} (x) \Big |^{\epsilon_{\UV}}_{\epsilon_{\IR}} + h^{\rm{(1a)}}_{L \rm{(c)}} (x) \Big |^{\epsilon_{\UV}}_{\epsilon_{\IR}} \nonumber \\[0.2cm]
&=& - \dfrac{\alpha_{s} C_{F}}{2\pi} \, \delta (x) \bigg ( {\cal P}_{\UV} - {\cal P}_{\IR}  + \ln \dfrac{\mu^{2}_{\UV}}{\mu^{2}_{\IR}} \bigg ) + \dfrac{\alpha_{s} C_{F}}{2\pi} \, \bigg ( {\cal P}_{\UV} - {\cal P}_{\IR}  + \ln \dfrac{\mu^{2}_{\UV}}{\mu^{2}_{\IR}}  \bigg ) \, .
\end{eqnarray}
Note that the prefactors of the $\delta(x)$ terms have an IR pole.
In the cut-off scheme, the full result for $h_{L}(x)$ with $m_{g} \neq 0$ is
\begin{eqnarray}
h^{\rm{(1a)}}_{L} (x) \Big |^{\Lambda_{\UV}}_{m_{g}} &=& h^{\rm{(1a)}}_{L \rm{(s)}} (x) \Big |^{\Lambda_{\UV}} + h^{\rm{(1a)}}_{L \rm{(c)}} (x) \Big |^{\Lambda_{\UV}}_{m_{g}} \nonumber \\[0.2cm]
&=& - \dfrac{\alpha_{s} C_{F}}{2\pi} \, \delta (x) \bigg ( \ln \dfrac{\Lambda^{2}_{\UV}}{m^{2}_{q}} \bigg ) + \dfrac{\alpha_{s}C_{F}}{2\pi} \bigg ( \ln \dfrac{\Lambda^{2}_{\UV}}{x m^{2}_{g}} + \dfrac{1-x}{x} \bigg ) \, .
\end{eqnarray}
With $m_{q} \neq 0$, we get
\begin{eqnarray}
h^{\rm{(1a)}}_{L} (x) \Big |^{\Lambda_{\UV}}_{m_{q}} &=& h^{\rm{(1a)}}_{L \rm{(s)}} (x) \Big |^{\Lambda_{\UV}}_{m_{q}} + h^{\rm{(1a)}}_{L \rm{(c)}} (x) \Big |^{\Lambda_{\UV}}_{m_{q}} \nonumber \\[0.2cm]
&=& - \dfrac{\alpha_{s} C_{F}}{2\pi} \, \delta (x) \bigg ( \ln \dfrac{\Lambda^{2}_{\UV}}{m^{2}_{q}} \bigg ) +\dfrac{\alpha_{s} C_{F}}{2\pi} \bigg ( \ln \dfrac{\Lambda^{2}_{\UV}}{(1-x)^{2}m^{2}_{q}} - \dfrac{2}{1-x} \bigg ) \, .
\label{e:hL_full_cut-off_QTM}
\end{eqnarray}

We are now ready to check the h-sum rule. The h-sum rule is violated for $m_{g} \neq 0$, in both DR and cut-off schemes, because the IR poles associated with the $\delta(x)$ terms in $h_{L}(x)$ contribute either $\ln (m_{q})$ or $1/\epsilon_{\IR}$, both of which are clearly absent in $h_{1}(x)$. For $m_{q} \neq 0$, and with DR for the UV, we find,
\begin{eqnarray}
\int^{1}_{0} dx \, h^{\rm{(1a)}}_{1} (x)\Big |^{\epsilon_{\UV}}_{m_{q}} &=& \dfrac{\alpha_{s}C_{F}}{2\pi} \bigg ( \dfrac{2}{\beta} + 2\bigg ) \nonumber \\[0.2cm] 
&=& \int^{1}_{0} dx \, h^{\rm{(1a)}}_{L \rm{(s)}} (x) \Big |^{\epsilon_{\UV}}_{m_{q}} + \int^{1}_{0} dx \, h^{\rm{(1a)}}_{L \rm{(c)}} (x) \Big |^{\epsilon_{\UV}}_{m_{q}} \, ,
\end{eqnarray}
where, $\beta$ denotes the $x=1$ pole present in diagram~($\rm{1a}$). 
Therefore, the h-sum rule holds provided one takes the $\delta(x)$ contribution into account. Similar studies in the past have also advanced the necessity of including the zero-mode contributions for the validity of the sum rules~\cite{Burkardt:2001iy,Aslan:2018tff,Aslan:2020zik}. It is interesting to discuss the above result. Recall that $h^{\rm{(1a)}}_{1} (x)$ is UV-finite. In this context, we note that the integrals of $h^{\rm{(1a)}}_{1}(x)$ and $h^{\rm{(1a)}}_{L}(x)$ agree because the UV poles from the (integral of the) singular and the canonical terms exactly cancel. Also, the $\ln(\mu_{\UV}/m_{q})$ terms, present in $h^{\rm{(1a)}}_{L}(x)$, cancel between these two terms. It is straightforward to verify that the h-sum rule holds when one does DR for both UV and IR:
\begin{eqnarray}
\int^{1}_{0} dx \, h^{\rm{(1a)}}_{1} (x)\Big |^{\epsilon_{\UV}}_{\epsilon_{\IR}} &=& 0 \nonumber \\[0.2cm] 
&=& \int^{1}_{0} dx \, h^{\rm{(1a)}}_{L \rm{(s)}} (x) \Big |^{\epsilon_{\UV}}_{\epsilon_{\IR}} + \int^{1}_{0} dx \, h^{\rm{(1a)}}_{L \rm{(c)}} (x) \Big |^{\epsilon_{\UV}}_{\epsilon_{\IR}} \, .
\end{eqnarray}

We now proceed to check the validity of the h-sum rule in the cut-off scheme. It is remarkable that the h-sum rule continues to hold even in the cut-off scheme when one works with $m_{q} \neq 0$:
\begin{eqnarray}
\int^{1}_{0} dx \, h^{\rm{(1a)}}_{1} (x)\Big |^{\Lambda_{\UV}}_{m_{q}} &=& \dfrac{\alpha_{s}C_{F}}{2\pi} \bigg ( \dfrac{2}{\beta} + 2\bigg ) \nonumber \\[0.2cm] 
&=& \int^{1}_{0} dx \, h^{\rm{(1a)}}_{L \rm{(s)}} (x) \Big |^{\Lambda_{\UV}}_{m_{q}} + \int^{1}_{0} dx \, h^{\rm{(1a)}}_{L \rm{(c)}} (x) \Big |^{\Lambda_{\UV}}_{m_{q}} \, .
\end{eqnarray}
Let us now examine why the h-sum rule continues to hold in the cut-off scheme, since the BC sum rule does not and since both sum rules are based on rotational invariance.
First of all, note that, just as in the case of DR, the $\ln(\Lambda_{\UV}/m_{q})$ terms, present in $h^{\rm{(1a)}}_{L}(x)$, cancel between the singular and the canonical terms. 
Now, notice that the ``extra" finite factors of $\delta(x)$ and $-2(1-x)$ present in the DR scheme (see the first expression in Eq.~(\ref{e:hL_full_DR_QTM}), and compare with Eq.~(\ref{e:hL_full_cut-off_QTM})) integrate to zero. 
Therefore, the absence of these terms in the cut-off scheme do not cause an issue for the h-sum rule.
We therefore conclude that the h-sum rule accidentally holds in the QTM for the cut-off scheme.
This conclusion is also supported by the fact that the h-sum rule is violated in the YM for a cut-off, as we discuss below in more detail.

We now want to discuss the application of DR and cut-off to all components of $k$, and consequently its impact on sum rules. 
We first go through the basic steps for the DR scheme. 
Our starting point for the calculation of $h^{\rm{(1a)}}_{1}$ is
\begin{eqnarray}
\int \dfrac{dk^{+}}{p^{+}} s^{i}_{\perp} \, h^{\rm{(1a)}}_{1} ( k^{+}) &=& - \dfrac{i g^{2}_{s} C_{F} \mu^{2\epsilon}}{4 p^{+}} \, \int \dfrac{d^{n}k}{(2\pi)^{n}} \int^{1}_{0} dy \dfrac{2(1-y)}{(k^{2} - Q^{2})^{3}} N_{h1} (k) \, ,
\label{e:h1_DR_4k}
\end{eqnarray}
where 
\begin{eqnarray}
N_{h1} (k) &=& 4(n-4) s^{i}_{\perp} p^{+} k^{2} - 8 (n-4) s^{i}_{\perp} k^{+} (k \cdot p) - 8 (n-4) p^{+} k^{i}_{\perp} (k \cdot s) \nonumber \\[0.2cm]
&& \phantom{This text is invisible.} + \,  4(n-4) s^{i}_{\perp} p^{+} m^{2}_{q} (y^{2}-1) + 8 s^{i}_{\perp} p^{+} m^{2}_{q} \big ( (n-2)y - (n-4) y^{2} \big ) \, ,
\end{eqnarray}
$Q^{2} = y m^{2}_{g} + (1-y)^{2} m^{2}_{q}$, and $y$ is the Feynman parameter. By using
\begin{eqnarray}
\int \dfrac{d^{n}k}{(2\pi)^{n}} \dfrac{k^{\mu} k^{\nu}}{(k^{2}-Q^{2})^{3}} &=& \bigg ( \dfrac{i}{4} Q^{2} \dfrac{(\pi)^{n/2}}{(2\pi)^{n}} \dfrac{\Gamma (2-n/2)}{(Q^{2})^{3-n/2}} \bigg ) g^{\mu \nu} \, ,
\end{eqnarray}
and 
\begin{eqnarray}
\int \dfrac{d^{n}k}{(2\pi)^{n}} \dfrac{1}{(k^{2}-Q^{2})^{3}} &=&  - \dfrac{i}{2} \dfrac{(\pi)^{n/2}}{(2\pi)^{n}} \dfrac{\Gamma (3-n/2)}{(Q^{2})^{3-n/2}} \, ,
\end{eqnarray}
we find that the first three terms in $N_{h1} (k)$ cancel one another, and the forth term is proportional to $\epsilon_{\UV}$. Therefore, the final result is given entirely by the last term in $N_{h1} (k)$:
\begin{eqnarray}
\int \dfrac{dk^{+}}{p^{+}} h^{\rm{(1a)}}_{1} (k^{+}) \Big |^{\epsilon_{\UV}}_{(m_{g}, m_{q})} &=& - \dfrac{\alpha_{s}C_{F}}{2\pi} \int^{1}_{0} dy \dfrac{2y(1-y)m^{2}_{q}}{Q^{2}} \, .
\end{eqnarray}
By taking the limits $m_{q} \rightarrow 0$ or $m_{g} \rightarrow 0$, we get back our results with DR applied to the transverse dimensions.
Our starting point for $h^{\rm{(1a)}}_{L}$ is,
\begin{eqnarray}
\int \dfrac{dk^{+}}{p^{+}} \dfrac{m_{q} \lambda}{p^{+}} \, h^{\rm{(1a)}}_{L} ( k^{+}) &=& - \dfrac{i g^{2}_{s} C_{F} \mu^{2\epsilon}}{4 p^{+}} \, \int \dfrac{d^{n}k}{(2\pi)^{n}} \int^{1}_{0} dy \dfrac{2(1-y)}{(k^{2} - Q^{2})^{3}} N_{hL} (k) \, ,
\label{e:hL_DR_4k}
\end{eqnarray}
where 
\begin{eqnarray}
N_{hL} (k) &=& 4(n-4) \lambda m_{q} k^{2} + 8 (n-4) s^{-} k^{+} (k \cdot p) - 8 (n-4) s^{+} k^{-} (k \cdot p) - 8 (n-4) p^{-} k^{+} (k \cdot s) + 8 (n-4) p^{+} k^{-} (k \cdot s) \nonumber \\[0.2cm]
&+& 8(n-2) \lambda m^{3}_{q} y - 4(n-4) \lambda m^{3}_{q} (1+y^{2}) \, .
\end{eqnarray}
The second, third, forth and the fifth terms in $N_{hL} (k)$ add up to cancel exactly the contribution from the first term in $N_{hL} (k)$. The seventh term is proportional to $\epsilon_{\UV}$, and it is the sixth term only that contributes to $h_{L}(x)$,
\begin{eqnarray}
\int \dfrac{dk^{+}}{p^{+}} h^{\rm{(1a)}}_{L} (k^{+}) \Big |^{\epsilon_{\UV}}_{(m_{g}, m_{q})} &=& - \dfrac{\alpha_{s}C_{F}}{2\pi} \int^{1}_{0} dy \dfrac{2y(1-y)m^{2}_{q}}{Q^{2}} \, .
\label{e:non_commute_hL_mg}
\end{eqnarray}
This means that the h-sum rule is satisfied. 
Few comments are in order: 
First, there is no need for a separate discussion of the zero-modes in this approach. 
Since the sum rule is satisfied, the contribution of the zero-modes is automatically included in such an analysis. 
Second, for $m_{g} \neq 0$, it is clear from Eq.~(\ref{e:non_commute_hL_mg}) that this approach does not give the same final result for $\int h_{L}$, when compared to the case where we first extract the $x$-dependence and then calculate the moment; but, $\int h_{1}$ agrees.
Third, in this approach, we observe that there is no problem in using a nonzero gluon mass as an IR regulator. 
This is different from the scenario when we first extract the $x$-dependent results, and then calculate their moments.

We now turn to the cut-off scheme. To evaluate the integrals, we first perform a Wick rotation, which allows us to carry out the integral in 
Euclidean space instead of 
Minkowski space.
By using
\begin{eqnarray}
\int \dfrac{d^{4}k}{(2\pi)^{4}} \dfrac{k^{2}}{(k^{2} - Q^{2})^{3}} &=& \dfrac{i}{32 \pi^{2}} \bigg ( 2 \ln \dfrac{\Lambda^{2}_{\UV}}{ Q^{2}} -3 \bigg ) \, , \\[0.2cm]
\int \dfrac{d^{4}k}{(2\pi)^{4}} \dfrac{1}{(k^{2} - Q^{2})^{3}} &=& - \dfrac{i}{32 \pi^{2}} \dfrac{1}{Q^{2}} \, , \\[0.2cm]
\quad k^{\mu} k^{\nu} \rightarrow \dfrac{1}{4} g^{\mu \nu} k^{2} \, , \quad \int \dfrac{d^{4}k}{(2\pi)^{4}} \dfrac{k^{\mu} (k \cdot a)}{(k^{2} - Q^{2})^{3}} &=& \dfrac{a^{\mu}}{4} \bigg [ \dfrac{i}{32 \pi^{2}} \bigg ( 2 \ln \dfrac{\Lambda^{2}_{\UV}}{ Q^{2}} -3 \bigg ) \bigg ] \quad a = (p,s) \, ,
\end{eqnarray}
we find
\begin{eqnarray}
\int \dfrac{dk^{+}}{p^{+}} h^{\rm{(1a)}}_{1} (k^{+}) \Big |^{\Lambda_{\UV}}_{(m_{g}, m_{q})} &=& - \dfrac{\alpha_{s}C_{F}}{2\pi} \int^{1}_{0} dy \dfrac{2y(1-y)m^{2}_{q}}{Q^{2}} \, , \\[0.2cm]
&=& \int \dfrac{dk^{+}}{p^{+}} h^{\rm{(1a)}}_{L} (k^{+}) \Big |^{\Lambda_{\UV}}_{(m_{g}, m_{q})} \, .
\end{eqnarray}
Therefore, we see that the h-sum rule is satisfied when cut-off is applied to all 4 components of $k$. This result is not surprising, because obviously without a bias for any specific direction, the rotational invariance is no longer broken. 
Once again, with a cut-off, $m_g \neq 0$ does not pose an issue as an IR regulator.

\subsection{Results in Yukawa Model}
The YM describes the pointlike interaction between fermions and a scalar field.
The diagrams in Fig.~($\rm{1a}$) and Fig.~($\rm{2a}$), with the obvious replacement of the gluon propagator by the propagator of the scalar, are the only ones that contribute to the PDFs in this model.
\subsubsection{BC sum rule}
Fig.~($\rm{1a}$) contributes to $g_{1}(x)$ as
\begin{eqnarray}
g^{\rm{(1a)}}_{1}(x) &=& \frac{g^{2}_{\rm Y} \mu^{2\epsilon}}{2(2\pi)} (1-x) \int \dfrac{d^{n-2}{k}_{\perp}}{(2\pi)^{n-2}} \, \frac{ \big ( -k^{2}_{\perp} + (1 + x )^{2} \, m^{2}_{q} \big )}{ \big ( k^{2}_{\perp} + (1-x)^{2} m^{2}_{q} + x m^{2}_{s} \big ) ^{2} } \, ,
\label{e:g1_SDM}
\end{eqnarray}
where $g_{\rm Y}$ is the counterpart of $g_{s}$, and $m_{s}$ is the mass of the scalar particle. 
Applying DR for the UV, we obtain the following results for $g_{1}(x)$ in three different IR schemes,
\begin{eqnarray}
g^{\rm{(1a)}}_{1} (x) \Big |^{\epsilon_{\UV}}=
\begin{dcases}
& g^{\rm{(1a)}}_{1} (x) \Big |^{\epsilon_{\UV}}_{m_{s}} = -\dfrac{\alpha_{\rm Y}}{4\pi} (1-x) \bigg ( {\cal P}_{\UV} + \ln \dfrac{\mu^{2}_{\UV}}{x m^{2}_{s}} - 1 \bigg )  \, ,
\\[0.2cm]
& g^{\rm{(1a)}}_{1} (x) \Big |^{\epsilon_{\UV}}_{m_q} = - \dfrac{\alpha_{\rm Y}}{4\pi} (1-x) \bigg ( {\cal P}_{\UV} + \ln \dfrac{\mu^{2}_{\UV}}{(1-x)^{2} m^{2}_{q}} - \dfrac{2(1+x^{2})}{(1-x)^{2}} \bigg ) \, ,
\\[0.2cm]
& g^{\rm{(1a)}}_{1} (x) \Big |^{\epsilon_{\UV}}_{\epsilon_{\IR}} = - \dfrac{\alpha_{\rm Y}}{4\pi} (1-x) \bigg ( {\cal P}_{\UV} - {\cal P}_{\IR}  + \ln \dfrac{\mu^{2}_{\UV}}{\mu^{2}_{\IR}} \bigg ) \, ,
\end{dcases}
\label{e:g1_DR}
\end{eqnarray}
where $\alpha_{\rm Y}$ is the counterpart of $\alpha_{s}$. 
Applying a cut-off for the UV, we get
\begin{eqnarray}
g^{\rm{(1a)}}_{1} (x) \Big |^{\Lambda_{\UV}}=
\begin{dcases}
& g^{\rm{(1a)}}_{1} (x) \Big |^{\Lambda_{\UV}}_{m_{s}} = - \dfrac{\alpha_{\rm Y}}{4\pi} (1-x) \bigg ( \ln \dfrac{\Lambda^{2}_{\UV}}{x m^{2}_{s}} - 1 \bigg ) \, ,
\\[0.2cm]
& g^{\rm{(1a)}}_{1} (x) \Big |^{\Lambda_{\UV}}_{m_q} = - \dfrac{\alpha_{\rm Y}}{4\pi} (1-x) \bigg ( \ln \dfrac{\Lambda^{2}_{\UV}}{(1-x)^{2}m^{2}_{q}} - \dfrac{2(1+x^{2})}{(1-x)^{2}} \bigg ) \, .
\end{dcases}
\label{e:g1_cut-off}
\end{eqnarray}

For the sake of completeness, we mention the results for the fermion self-energy diagram in YM. In DR, we obtain
\begin{eqnarray}
\dfrac{\partial \Sigma (p)}{\partial \slashed{p}} \Big |^{\epsilon_{\UV}} =
\begin{dcases}
& \dfrac{\partial \Sigma (p)}{\partial \slashed{p}} \Big |^{\epsilon_{\UV}}_{m_{s}} =
-\dfrac{\alpha_{\rm Y}}{4 \pi} \int_{0}^{1} dy \, \bigg ( y \, {\cal P}_{\UV} + y \ln \dfrac{\mu^{2}_{\UV}}{y m^{2}_{s}} \bigg ) \, , 
\\[0.2cm]
& \dfrac{\partial \Sigma (p)}{\partial \slashed{p}} \Big |^{\epsilon_{\UV}}_{m_{q}} =
-\dfrac{\alpha_{\rm Y}}{4 \pi} \int_{0}^{1} dy \,  \bigg ( (1-y) \, {\cal P}_{\UV} + (1-y) \ln \dfrac{\mu_{\UV}^{2}}{(1-y)^2 m_q^{2}} + \dfrac{4y}{1-y} \bigg ) \, ,
\\[0.2cm]
& \dfrac{\partial \Sigma (p)}{\partial \slashed{p}} \Big |^{\epsilon_{\UV}}_{\epsilon_{\IR}}  =
-\dfrac{\alpha_{\rm Y}}{4 \pi} \int_{0}^{1} dy \, y  \bigg (  {\cal P}_{\UV}- {\cal P}_{\IR}  + \ln \dfrac{\mu_{\UV}^{2}}{\mu_{\IR}^{2} } \bigg ) \, ,
\end{dcases}
\end{eqnarray}
while for a cut-off we find
\begin{eqnarray}
\dfrac{\partial \Sigma (p)}{\partial \slashed{p}} \Big |^{\Lambda_{\UV}} =
\begin{dcases}
& \dfrac{\partial \Sigma (p)}{\partial \slashed{p}} \Big |^{\Lambda_{\UV}}_{m_{s}} =
-\dfrac{\alpha_{\rm Y}}{4 \pi} \int_{0}^{1} dy \, \bigg ( y \ln \dfrac{\Lambda^{2}_{\UV}}{y m^{2}_{s}} \bigg ) \, ,
\label{e:...}
\\[0.2cm]
& \dfrac{\partial \Sigma (p)}{\partial \slashed{p}} \Big |^{\Lambda_{\UV}}_{m_{q}} =
-\dfrac{\alpha_{\rm Y}}{4 \pi} \int_{0}^{1} dy \, \bigg ( (1-y) \ln \dfrac{\Lambda^{2}_{\UV}}{(1-y)^{2} m^{2}_{q}} + \dfrac{4y}{1-y} \bigg ) \, .
\end{dcases}
\end{eqnarray}

Turning now to $g_{T}(x)$ in YM, we can once again split the contribution from Fig.~($\rm{1a}$) into a singular and a canonical part. As a first step, one obtains the following expressions for the singular and canonical parts of $g_{T}(x)$, 
\begin{eqnarray}
g^{\rm{(1a)}}_{T \rm{(s)}} (x) &=&  \frac{g^{2}_{\rm Y} \mu^{2\epsilon}}{2(2\pi)} \, \delta (x) \int \dfrac{d^{n-2}{k}_{\perp}}{(2\pi)^{n-2}} \, \dfrac{1}{(k^{2}_{\perp} + m^{2}_{q})} \, , \nonumber \\[0.2cm]
g^{\rm{(1a)}}_{T \rm{(c)}} (x) &=& - \frac{g^{2}_{\rm Y} \mu^{2\epsilon}}{2(2\pi)} \, \int \dfrac{d^{n-2}{k}_{\perp}}{(2\pi)^{n-2}} \, \dfrac{ 2 k^{2}_{\perp} - (1-x) \dfrac{k^{2}_{\perp}}{1-\epsilon} - 2 (1-x^{2}) m^{2}_{q} + (1+x) m^{2}_{s}}{ \big ( k^{2}_{\perp} + (1-x)^{2} m^{2}_{q} + x m^{2}_{s} \big ) ^{2} } \, .
\label{e:g1_SDM} 
\end{eqnarray}
Working with $m_{s} \neq 0$ leads to the following two results for the singular parts,
\begin{eqnarray}
g^{\rm{(1a)}}_{T \rm{(s)}} (x) \Big |^{\epsilon_{\UV}} =
\begin{dcases}
& g^{\rm (1a)}_{T \rm{(s)}} (x) \Big |^{\epsilon_{\UV}}_{m_{q}} =  \dfrac{\alpha_{\rm Y}}{4\pi} \, \delta (x) \bigg ( {\cal P}_{\UV} + \ln \dfrac{\mu^{2}_{\UV}}{m^{2}_{q}}\bigg ) \, ,
\\[0.2cm]
& g^{\rm{(1a)}}_{T \rm{(s)}} (x) \Big |^{\epsilon_{\UV}}_{\epsilon_{\IR}} =  \dfrac{\alpha_{\rm Y}}{4\pi} \, \delta (x) \bigg ( {\cal P}_{\UV} - {\cal P}_{\IR} + \ln \dfrac{\mu^{2}_{\UV}}{\mu^{2}_{\IR}} \bigg ) \, . 
\end{dcases}
\label{e:gT_sing_DR}
\end{eqnarray}
For the canonical part with $m_{s} \neq 0$, we get
\begin{eqnarray}
g^{\rm{(1a)}}_{T \rm{(c)}} (x) \Big |^{\epsilon_{\UV}}_{m_{s}} &=& - \dfrac{\alpha_{\rm Y}}{4\pi} \bigg ( (1+x) \, {\cal P}_{\UV} + (1+x) \ln \frac{\mu_{\UV}^2}{x m_s^2} + \dfrac{1-x}{x} \bigg ) \, .
\end{eqnarray}
Finally, with $m_{q} \neq 0$ and DR for the IR, we find
\begin{eqnarray}
g^{\rm{(1a)}}_{T} (x) \Big |^{\epsilon_{\UV}}_{m_{q}} &=& g^{\rm{(1a)}}_{T \rm{(s)}} (x) \Big |^{\epsilon_{\UV}}_{m_{q}} + g^{\rm{(1a)}}_{T \rm{(c)}} (x) \Big |^{\epsilon_{\UV}}_{m_{q}} \nonumber \\[0.2cm]
&=& \dfrac{\alpha_{\rm Y}}{4\pi} \, \delta (x) \bigg ( {\cal P}_{\UV} + \ln \dfrac{\mu^{2}_{\UV}}{m^{2}_{q}}\bigg ) - \dfrac{\alpha_{\rm Y}}{4\pi}  \bigg ( (1+x) \, {\cal P}_{\UV} +
 (1+x) \ln \frac{\mu_{\UV}^2}{(1-x)^2 m_q^2} - \frac{4}{1 - x} \bigg )\,,
\label{e:gT_full_DR_YM}
\\[0.2cm]
g^{\rm{(1a)}}_{T} (x) \Big |^{\epsilon_{\UV}}_{\epsilon_{\IR}}  &=& g^{\rm{(1a)}}_{T \rm{(s)}} (x) \Big |^{\epsilon_{\UV}}_{\epsilon_{\IR}} + g^{\rm{(1a)}}_{T \rm{(c)}} (x) \Big |^{\epsilon_{\UV}}_{\epsilon_{\IR}} \nonumber \\[0.2cm]
&=& \dfrac{\alpha_{\rm Y}}{4\pi} \, \delta (x) \bigg ( {\cal P}_{\UV} - {\cal P}_{\IR} + \ln \dfrac{\mu^{2}_{\UV}}{\mu^{2}_{\IR}} \bigg ) - \dfrac{\alpha_{\rm Y}}{4\pi} (1+x) \bigg (  {\cal P}_{\UV} - {\cal P}_{\IR}  + \ln \frac{\mu_{\UV}^2}{\mu_{\IR}^2}  \bigg ) \, .
\end{eqnarray}
For the singular part in the cut-off scheme, we find 
\begin{eqnarray}
g^{\rm (1a)}_{T \rm{(s)}} (x) \Big |^{\Lambda_{\UV}}_{m_{q}} = \dfrac{\alpha_{\rm Y}}{4\pi} \, \delta(x) \, \ln \dfrac{\Lambda^{2}_{\UV}}{m^{2}_{q}} \, .
\label{e:gT_sing_cut-off}
\end{eqnarray}
The full result for $g_{T}(x)$ with $m_{s} \neq 0$ is
\begin{eqnarray}
g^{\rm{(1a)}}_{T} (x) \Big |^{\Lambda_{\UV}}_{m_{s}} &=& g^{\rm{(1a)}}_{T \rm{(s)}} (x) \Big |^{\Lambda_{\UV}} + g^{\rm{(1a)}}_{T \rm{(c)}} (x) \Big |^{\Lambda_{\UV}}_{m_{s}} \nonumber \\[0.2cm]
&=& \dfrac{\alpha_{\rm Y}}{4\pi} \, \delta(x) \, \ln \dfrac{\Lambda^{2}_{\UV}}{m^{2}_{q}} - \dfrac{\alpha_{\rm Y}}{4\pi} \bigg ( (1+x) \ln \dfrac{\Lambda^{2}_{\UV}}{x m^{2}_{s}} + (1-x) + \dfrac{1-x}{x} \bigg )\, .
\end{eqnarray}
With $m_{q} \neq 0$, we get
\begin{eqnarray}
g^{\rm{(1a)}}_{T} (x) \Big |^{\Lambda_{\UV}}_{m_{q}} &=& g^{\rm{(1a)}}_{T \rm{(s)}} (x) \Big |^{\Lambda_{\UV}}_{m_{q}} + g^{\rm{(1a)}}_{T \rm{(c)}} (x) \Big |^{\Lambda_{\UV}}_{m_{q}} \nonumber \\[0.2cm]
&=& \dfrac{\alpha_{\rm Y}}{4\pi} \, \delta(x) \, \ln \dfrac{\Lambda^{2}_{\UV}}{m^{2}_{q}} - \dfrac{\alpha_{\rm Y}}{4\pi} \bigg ( (1+x) \ln \dfrac{\Lambda^{2}_{\UV}}{(1-x)^{2}m^{2}_{q}} -\dfrac{4}{1-x} + (1-x) \bigg ) \, .
\label{e:gT_cano_cut-off}
\end{eqnarray}

We now proceed to check whether or not the BC sum rule holds in the YM. First of all, notice that, in contrast to the QTM, the zero-mode contributions survive in the YM. Also, different from the QTM, the prefactors of these contributions are IR-divergent in the two UV schemes. A consequence of this is that there is a violation of the BC sum rule when using $m_{s} \neq 0$ in both UV schemes. Furthermore, working with $m_{s} \neq 0$ leads to $1/x$ poles as $x \rightarrow 0$ in the canonical part of $g_{T}(x)$. Hence, the lowest moment of $g_{T}(x)$ is not defined in the YM with $m_{s} \neq 0$. In fact, divergent terms like $1/x$ are typically observed for $m_{s/g} \neq 0$. In the QTM, these terms can be seen in the canonical parts of $h_{L}(x)$~\cite{Bhattacharya:2020jfj}. When either $m_{q}$ or DR is used for the IR, it is straightforward to verify that the BC sum rule holds when DR is applied for the UV and when the zero-mode contributions are taken into account. Specifically, we find
\begin{eqnarray}
\int^{1}_{0} dx \, g^{\rm{(1a)}}_{1} (x)\Big |^{\epsilon_{\UV}}_{m_{q}} &=& \dfrac{\alpha_{\rm Y}}{4\pi} \bigg ( - \dfrac{1}{2} {\cal P}_{\UV} - \ln \dfrac{\mu_{\UV}}{m_{q}} - \dfrac{4}{\beta} - \dfrac{7}{2} \bigg ) \nonumber \\[0.2cm] 
&=& \int^{1}_{0} dx \, g^{\rm{(1a)}}_{T \rm{(s)}} (x) \Big |^{\epsilon_{\UV}}_{m_{q}} + \int^{1}_{0} dx \, g^{\rm{(1a)}}_{T \rm{(c)}} (x) \Big |^{\epsilon_{\UV}}_{m_{q}} \, ,
\end{eqnarray}
and
\begin{eqnarray}
\int^{1}_{0} dx \, g^{\rm{(1a)}}_{1} (x)\Big |^{\epsilon_{\UV}}_{\epsilon_{\IR}} &=& \dfrac{\alpha_{\rm Y}}{4\pi} \bigg ( - \dfrac{1}{2} ( {\cal P}_{\UV} - {\cal P}_{\IR} ) - \dfrac{1}{2} \ln \dfrac{\mu^{2}_{\UV}}{\mu^{2}_{\IR}} \bigg ) \nonumber \\[0.2cm] 
&=& \int^{1}_{0} dx \, g^{\rm{(1a)}}_{T \rm{(s)}} (x) \Big |^{\epsilon_{\UV}}_{\epsilon_{\IR}} + \int^{1}_{0} dx \, g^{\rm{(1a)}}_{T \rm{(c)}} (x) \Big |^{\epsilon_{\UV}}_{\epsilon_{\IR}} \, ,
\end{eqnarray}
where $\beta$ denotes the pole at $x=1$ present in diagram~(1a).

When cut-off is switched on for the UV, the BC sum rule continues to hold for the UV divergent parts of $g_{1}(x)$ and $g_{T}(x)$. Note that the finite factors for $g_{1}(x)$ in the two UV schemes are exactly the same. On the other hand, they change for the canonical part of $g_{T}(x)$. The source of this change is the scheme-dependence of the prefactor for the UV divergent term in Eq.~(\ref{e:g1_SDM}) (see the second term in the canonical part). Due to the absence of a similar ``compensating" change elsewhere in Eq.~(\ref{e:g1_SDM}), the BC sum rule is violated for the finite parts. With $m_{q} \neq 0$, we find,
\begin{eqnarray}
\int^{1}_{0} dx \, g^{\rm{(1a)}}_{1} (x) \Big |^{\Lambda_{\UV}}_{m_{q}} &=& \dfrac{\alpha_{\rm Y}}{4\pi} \bigg ( - \ln \dfrac{\Lambda_{\UV}}{m_{q}} - \dfrac{4}{\beta} - \dfrac{7}{2} \bigg ) \, ,
\\[0.2cm]
\int^{1}_{0} dx \, g^{\rm{(1a)}}_{T} (x) \Big |^{\Lambda_{\UV}}_{m_{q}} &=& \int^{1}_{0} dx \, g^{\rm{(1a)}}_{T \rm{(s)}} (x) \Big |^{\Lambda_{\UV}}_{m_{q}} + \int^{1}_{0} dx \, g^{\rm{(1a)}}_{T \rm{(c)}} (x) \Big |^{\Lambda_{\UV}}_{m_{q}} 
= \dfrac{\alpha_{\rm Y}}{4\pi} \bigg ( - \ln \dfrac{\Lambda_{\UV}}{m_{q}} - \dfrac{4}{\beta} - \dfrac{7}{2} - \dfrac{1}{2}\bigg ) \, .
\end{eqnarray}
Most of the above findings are in agreement with what we see in the QTM. The case of $m_{g} \neq 0$ is however distinct for the two models. The difference can be traced back to the IR-finiteness of the prefactors of the zero-mode terms in QTM. 

\subsubsection{h-sum rule}
Fig.~($\rm{1a}$) contributes to $h_{1}(x)$ as
\begin{eqnarray}
h^{\rm{(1a)}}_{1} (x) &=& \frac{g^{2}_{\rm Y} \mu^{2\epsilon}}{2(2\pi)} \, (1-x) \int \dfrac{d^{n-2}{k}_{\perp}}{(2\pi)^{n-2}} \, \frac{\bigg ( 1 - \dfrac{1}{1-\epsilon} \bigg ) k^{2}_{\perp}  + (1 + x)^{2} \, m^{2}_{q}}{\big ( k^{2}_{\perp} + (1-x)^{2} m^{2}_{q} + x m^{2}_{s} \big ) ^{2} } \, .
\label{e:h1_SDM}
\end{eqnarray}
Using DR for the $k_{\perp}$ integrals, we obtain
\begin{eqnarray}
h^{\rm{(1a)}}_{1} (x) \Big |^{\epsilon_{\UV}}=
\begin{dcases}
& h^{\rm{(1a)}}_{1} (x) \Big |^{\epsilon_{\UV}}_{m_{s}} = - \dfrac{\alpha_{\rm Y}}{4\pi} (1-x) \, ,
\\[0.2cm]
& h^{\rm{(1a)}}_{1} (x) \Big |^{\epsilon_{\UV}}_{m_q} =  \dfrac{\alpha_{\rm Y}}{4\pi} \, \bigg ( \dfrac{(1+x)^{2}}{1-x} -(1-x) \bigg ) \, ,
\\[0.2cm]
& h^{\rm{(1a)}}_{1} (x) \Big |^{\epsilon_{\UV}}_{\epsilon_{\IR}} = 0 \, .
\end{dcases}
\label{e:h1_DR}
\end{eqnarray}
Using a cut-off for the $k_{\perp}$ integrals, we find
\begin{eqnarray}
h^{\rm{(1a)}}_{1} (x) \Big |^{\Lambda_{\UV}}=
\begin{dcases}
& h^{\rm{(1a)}}_{1} (x) \Big |^{\Lambda_{\UV}}_{m_{s}} = 0 \, ,
\\[0.2cm]
& h^{\rm{(1a)}}_{1} (x) \Big |^{\Lambda_{\UV}}_{m_q} =  \dfrac{\alpha_{\rm Y}}{4\pi} \, \bigg ( \dfrac{(1+x)^{2}}{1-x} \bigg ) \, .
\end{dcases}
\label{e:h1_cut-off}
\end{eqnarray}
Just as in the QTM, the contribution from the diagram in Fig.~($\rm{1a}$) to $h_{1}(x)$ is UV-finite.

Moving on to $h_{L}(x)$, we obtain the following general expressions for the singular and canonical parts,
\begin{eqnarray}
h^{\rm{(1a)}}_{L \rm{(s)}} (x) &=&  \frac{g^{2}_{\rm Y} \mu^{2\epsilon}}{2(2\pi)} \, \delta (x) \int \dfrac{d^{n-2}{k}_{\perp}}{(2\pi)^{n-2}} \, \dfrac{1}{(k^{2}_{\perp} + m^{2}_{q})} \,,  \nonumber \\[0.2cm]
h^{\rm{(1a)}}_{L \rm{(c)}} (x) &=& - \frac{g^{2}_{\rm Y} \mu^{2\epsilon}}{2(2\pi)} \, \int \dfrac{d^{n-2}{k}_{\perp}}{(2\pi)^{n-2}} \, \dfrac{ 2 x k^{2}_{\perp} - 2 (1-x^{2}) m^{2}_{q} + (1+x) m^{2}_{s}}{ \big ( k^{2}_{\perp} + (1-x)^{2} m^{2}_{q} + x m^{2}_{s} \big ) ^{2} } \, .
\label{e:hL_SDM} 
\end{eqnarray}
As discussed, the singular part for $m_{s} \neq 0$ has two results,
\begin{eqnarray}
h^{\rm{(1a)}}_{L \rm{(s)}} (x) \Big |^{\epsilon_{\UV}} =
\begin{dcases}
& h^{\rm (1a)}_{L \rm{(s)}} (x) \Big |^{\epsilon_{\UV}}_{m_{q}} =  \dfrac{\alpha_{\rm Y}}{4\pi} \, \delta (x) \bigg ( {\cal P}_{\UV} + \ln \dfrac{\mu^{2}_{\UV}}{m^{2}_{q}}\bigg ) \, ,
\\[0.2cm]
& h^{\rm{(1a)}}_{L \rm{(s)}} (x) \Big |^{\epsilon_{\UV}}_{\epsilon_{\IR}} =  \dfrac{\alpha_{\rm Y}}{4\pi} \, \delta (x) \bigg ( {\cal P}_{\UV} - {\cal P}_{\IR} + \ln \dfrac{\mu^{2}_{\UV}}{\mu^{2}_{\IR}} \bigg ) \, .
\end{dcases}
\label{e:hL_sing_DR}
\end{eqnarray}
With $m_{s} \neq 0$, we find for the canonical part
\begin{eqnarray}
h^{\rm{(1a)}}_{L \rm{(c)}} (x) \Big |^{\epsilon_{\UV}}_{m_{s}} &=&
- \dfrac{\alpha_{\rm Y}}{4\pi} \bigg ( 2 x \, {\cal P}_{\UV} + 2 x \, \ln \dfrac{\mu^{2}_{\UV}}{x m^{2}_{s}} + \dfrac{(1-x)(1+2x)}{x} \bigg ) \, .
\end{eqnarray}
With $m_{q} \neq 0$ and DR for the IR, the full results for $h_{L}(x)$ read
\begin{eqnarray}
h^{\rm{(1a)}}_{L} (x) \Big |^{\epsilon_{\UV}}_{m_{q}} &=& h^{\rm{(1a)}}_{L \rm{(s)}} (x) \Big |^{\epsilon_{\UV}}_{m_{q}} + h^{\rm{(1a)}}_{L \rm{(c)}} (x) \Big |^{\epsilon_{\UV}}_{m_{q}} \nonumber \\[0.2cm]
&=& \dfrac{\alpha_{\rm Y}}{4\pi} \, \delta (x) \bigg ( {\cal P}_{\UV} + \ln \dfrac{\mu^{2}_{\UV}}{m^{2}_{q}}\bigg ) - \dfrac{\alpha_{\rm Y}}{4\pi} \bigg ( 2 x \, {\cal P}_{\UV} + 2 x \,
 \ln \frac{\mu_{\UV}^2}{(1-x)^2 m_q^2} - \frac{2(-x^{2}+2x+1)}{1 - x} \bigg ) \,,
\\[0.2cm]
h^{\rm{(1a)}}_{L} (x) \Big |^{\epsilon_{\UV}}_{\epsilon_{\IR}}  &=& h^{\rm{(1a)}}_{L \rm{(s)}} (x) \Big |^{\epsilon_{\UV}}_{\epsilon_{\IR}} + h^{\rm{(1a)}}_{L \rm{(c)}} (x) \Big |^{\epsilon_{\UV}}_{\epsilon_{\IR}} \nonumber \\[0.2cm]
&=& \dfrac{\alpha_{\rm Y}}{4\pi} \, \delta (x) \bigg ( {\cal P}_{\UV} - {\cal P}_{\IR} + \ln \dfrac{\mu^{2}_{\UV}}{\mu^{2}_{\IR}} \bigg ) - \dfrac{\alpha_{\rm Y}}{4\pi} \, 2 x \, \bigg (  {\cal P}_{\UV} - {\cal P}_{\IR}  + \ln \frac{\mu_{\UV}^2}{\mu_{\IR}^2}  \bigg ) \, .
\end{eqnarray}
The full result for $h_{L}(x)$ in the cut-off scheme with $m_{s} \neq 0$ is
\begin{eqnarray}
h^{\rm{(1a)}}_{L} (x) \Big |^{\Lambda_{\UV}}_{m_{s}} &=& h^{\rm{(1a)}}_{L \rm{(s)}} (x) \Big |^{\Lambda_{\UV}} + h^{\rm{(1a)}}_{L \rm{(c)}} (x) \Big |^{\Lambda_{\UV}}_{m_{s}} \nonumber \\[0.2cm]
&=& \dfrac{\alpha_{\rm Y}}{4\pi} \, \delta(x) \, \ln \dfrac{\Lambda^{2}_{\UV}}{m^{2}_{q}} - \dfrac{\alpha_{\rm Y}}{4\pi} \bigg ( 2 x \ln \dfrac{\Lambda^{2}_{\UV}}{x m^{2}_{s}} + \dfrac{(1-x)(1+2x)}{x} \bigg ) \, .
\end{eqnarray}
Finally, with $m_{q} \neq 0$ we find
\begin{eqnarray}
h^{\rm{(1a)}}_{L} (x) \Big |^{\Lambda_{\UV}}_{m_{q}} &=& h^{\rm{(1a)}}_{L \rm{(s)}} (x) \Big |^{\Lambda_{\UV}}_{m_{q}} + h^{\rm{(1a)}}_{L \rm{(c)}} (x) \Big |^{\Lambda_{\UV}}_{m_{q}} \nonumber \\[0.2cm]
&=& \dfrac{\alpha_{\rm Y}}{4\pi} \, \delta(x) \, \ln \dfrac{\Lambda^{2}_{\UV}}{m^{2}_{q}} - \dfrac{\alpha_{\rm Y}}{4\pi} \bigg ( 2 x \, \ln \dfrac{\Lambda^{2}_{\UV}}{(1-x)^{2}m^{2}_{q}} - \dfrac{2(- x^{2} + 2x + 1)}{1-x}  \bigg ) \, .
\end{eqnarray}

We now turn to the calculation of the h-sum rule. We find that, irrespective of the choice of the UV scheme, the $\delta (x)$ terms in $h_L(x)$ are accompanied by prefactors that exhibit IR divergence. The h-sum rule is therefore not valid when working with $m_{s} \neq 0$. Furthermore, with $m_{s} \neq 0$, there are terms like $1/x$, which make the $x$-integrals diverge anyway. These complications appear in the QTM as well. When $m_{q}$ or DR is employed for the IR, the h-sum rule holds provided one applies DR for the UV. The corresponding results are
\begin{eqnarray}
\int^{1}_{0} dx \, h^{\rm{(1a)}}_{1} (x)\Big |^{\epsilon_{\UV}}_{m_{q}} &=& \dfrac{\alpha_{\rm Y}}{4\pi} \bigg ( - \dfrac{4}{\beta} - 4 \bigg ) \nonumber \\[0.2cm] 
&=& \int^{1}_{0} dx \, h^{\rm{(1a)}}_{L \rm{(s)}} (x) \Big |^{\epsilon_{\UV}}_{m_{q}} + \int^{1}_{0} dx \, h^{\rm{(1a)}}_{L \rm{(c)}} (x) \Big |^{\epsilon_{\UV}}_{m_{q}} \, ,
\end{eqnarray}
where $\beta$ reflects the $x=1$ pole, and
\begin{eqnarray}
\int^{1}_{0} dx \, h^{\rm{(1a)}}_{1} (x)\Big |^{\epsilon_{\UV}}_{\epsilon_{\IR}} &=& 0 \nonumber \\[0.2cm] 
&=& \int^{1}_{0} dx \, h^{\rm{(1a)}}_{L \rm{(s)}} (x) \Big |^{\epsilon_{\UV}}_{\epsilon_{\IR}} + \int^{1}_{0} dx \, h^{\rm{(1a)}}_{L \rm{(c)}} (x) \Big |^{\epsilon_{\UV}}_{\epsilon_{\IR}} \, .
\end{eqnarray}
As evident from the above results, it is mandatory to include the zero-mode contribution for the validity of the h-sum rule. In fact, these zero-mode contributions cancel the UV poles and the $\ln(m_{q})$ terms present in the canonical terms, such that $\int dx \, h^{\rm{(1a)}}_{L} (x)$ is UV-finite. 

Perhaps the most interesting finding in the Yukawa model is that the UV-finite parts in the cut-off scheme violate the h-sum rule. With $m_{q} \neq 0$, we find
\begin{eqnarray}
\int^{1}_{0} dx \, h^{\rm{(1a)}}_{1} (x) \Big |^{\Lambda_{\UV}}_{m_{q}} &=& \dfrac{\alpha_{\rm Y}}{4\pi} \bigg ( - \dfrac{4}{\beta} - \dfrac{7}{2} \bigg ) \, ,
\\[0.2cm]
\int^{1}_{0} dx \, h^{\rm{(1a)}}_{L} (x) \Big |^{\Lambda_{\UV}}_{m_{q}} &=& \int^{1}_{0} dx \, h^{\rm{(1a)}}_{L \rm{(s)}} (x) \Big |^{\Lambda_{\UV}}_{m_{q}} + \int^{1}_{0} dx \, h^{\rm{(1a)}}_{L \rm{(c)}} (x) \Big |^{\Lambda_{\UV}}_{m_{q}} 
= \dfrac{\alpha_{\rm Y}}{4\pi} \bigg ( - \dfrac{4}{\beta} - \dfrac{7}{2} - \dfrac{1}{2}\bigg ) \, .
\end{eqnarray}
The first term in the expression for $h_{1}(x)$ (see Eq.~(\ref{e:h1_SDM})) makes all the difference. Clearly, this is the term that gives rise to a different finite factor in the cut-off scheme. On the other hand, the result for $h_{L}(x)$ remains unchanged.
Obviously, in such a situation the sum rule couldn't have been valid simultaneously in the two UV schemes, and we indeed find a violation in the cut-off scheme. 
Let us mention that by performing calculations in the YM, we have a strong support to the picture that the h-sum rule is by no means ``superior'' to the BC sum rule. The fact that the h-sum rule holds in QTM can therefore be regarded as an ``accident''.
Generally, our work shows that, contrary to what is frequently assumed in the literature, it is not sufficient to check the sum rules for only the UV-divergent parts of the perturbative corrections. 
In fact, similar to what we are reporting here, there may well exist other cases of violation of relations that are based on Lorentz invariance in schemes that break rotational invariance.
We close this section by mentioning that the sum rules hold in the YM if the cut-off is applied in a rationally invariant manner. This analysis can be preformed in exactly the same manner as what we have shown in Sec.~\ref{ss:h_accident}. The (non-trivial) observation of the sum rule holding with $m_{g} \neq 0$ as pointed out in the QTM, applies for YM as well.

\section{Analytical results for the quasi-PDFs}
\label{s:main_results_quasi}
This section provides analytical results for the quasi-PDFs ($g_{1, \rm{Q}}(x), \, $ $g_{T, \rm{Q}}(x)$), and ($h_{1, \rm{Q}}(x), \, $ $h_{L, \rm{Q}}(x)$) in the QTM and the YM. Once again, our focus will be on the diagram in Fig.~(1a), which is sufficient for the check of the sum rules --- see Eq.~(\ref{e:sum_rule_1a_2a}) and the associated discussion. 

\subsection{Results in Quark Target Model}
The correlator for the quasi-PDFs for Fig.~(1a) is given by 
\begin{eqnarray}
\Phi_{\rm{Q}}^{[\Gamma]} (x; p^{3}) &=& -\dfrac{i \, g^{2}_{s} C_{F} \mu^{2\epsilon}}{2 (2\pi)^{2}} \int \dfrac{d^{n-2}k_{\perp}}{(2\pi)^{n-2}} \, \int dk^{0} \, \dfrac{\bar{u}(p) \, \big ( \gamma^{\mu} \, (\slashed{k} + m_{q}) \, \Gamma \, (\slashed{k} + m_{q}) \, \gamma_{\mu} \big ) \, u(p) }{D_{\rm{PDF}}} \, .
\end{eqnarray} 
The results for the quasi-PDFs can in general be cast in the form
\begin{eqnarray}
q_{\rm Q}(x; p^3) &=& q_{\rm {Q (s)}}(x; p^{3}) + q_{\rm {Q (c)}}(x; p^{3}) \nonumber \\[0.2cm]
&=& -\dfrac{g^{2}_{s} C_{F} \mu^{2\epsilon}}{2(2\pi)} \int \dfrac{d^{n-2} k_\perp}{(2\pi)^{n-2}} \, N_{q \rm{(s)}}  - \frac{i \, g^2_{s} C_{F} \mu^{2\epsilon}}{(2\pi)^2} \int \dfrac{d^{n-2} k_{\perp}}{(2\pi)^{n-2}} \, \int dk^{0}  \, \frac{N_{q \rm{(c)}}}{D_{\rm PDF}} \, . 
\label{e:qPDF_general1}
\end{eqnarray}
Here, $q_{\rm {Q (s)}}$ is the singular term, which is relevant for $g_{T, \rm{Q}}(x)$ and  $h_{L, \rm{Q}}(x)$ only, while $q_{\rm {Q (c)}}$ is the canonical term. 
The numerators for the specific PDFs are
\begin{eqnarray}
N_{g1 \rm{(c)}}  &=&  \dfrac{2}{\delta_{0}} \bigg \{ (1-\epsilon) \delta_{0} p^{3} (k^{0})^{2} + 2 k^{0} \big ( \epsilon \, m^{2}_{q} - (1-\epsilon) x p^{2}_{3} \big ) - (1-\epsilon) \delta_{0} p^{3} \big ( k^{2}_{\perp} - x^{2} p^{2}_{3} + m^{2}_{q} \big ) \bigg \} \, ,
\label{e:num_f10}
\\[0.2cm]
N_{gT \rm{(s)}} &=&  \dfrac{\epsilon \, p^{3}}{\big ( k^{2}_{\perp} + x^{2} p^{2}_{3} + m^{2}_{q} \big )^{3/2}}  \, , 
\\[0.2cm]
N_{gT (\rm{c})} &=& 2 p^{3} \bigg ( (k^{0})^{2} - x^{2} p^{2}_{3} + m^{2}_{q} -  \epsilon m^{2}_{g} \bigg ) \, ,
\\[0.2cm]
N_{h1 \rm{(c)}}  &=& \dfrac{2}{\delta_{0}} \bigg \{ \epsilon \delta_{0} p^{3} (k^{0})^{2} + 2 k^{0} \big ( (1-\epsilon) m^{2}_{q} - \epsilon x p^{2}_{3} \big ) +  \epsilon \bigg ( 1  - \dfrac{1}{1-\epsilon} \bigg ) \delta_{0} p^{3} k^{2}_{\perp} + \epsilon \delta_{0} p^{3} \big ( x^{2} p^{2}_{3} + m^{2}_{q} \big ) \bigg \} \, , 
\\[0.2cm]
N_{hL (\rm{s})} &=& \dfrac{(1- \epsilon) \, p^{3}}{\big ( k^{2}_{\perp} + x^{2} p^{2}_{3} + m^{2}_{q} \big )^{3/2}} \, ,  \\[0.2cm]
N_{hL (\rm{c})} &=& 2 p^{3} \bigg ( (k^{0})^{2} - x^{2} p^{2}_{3} + (-1 + 2\epsilon) k^{2}_{\perp} + m^{2}_{q} - (1 - \epsilon) m^{2}_{g} \bigg ) \, ,
\label{e:..}
\end{eqnarray}
and the denominator is given as
\begin{equation}
D_{\rm PDF} = \big ( k^{2} - m^{2}_{q} + i \varepsilon \big )^2 \, \big ( (P - k)^{2} - m^{2}_{g} + i \varepsilon \big ) \, .
\label{e:den_PDF}
\end{equation}
In these expressions, $\delta_{0} = \sqrt{1 + m^{2}_{q}/p^{2}_{3}}$. We perform the $k^{0}$ integral by means of contour integration. The $k^{0}$-poles in the complex plane are given by
\begin{eqnarray}
k^{0}_{1\pm} = k^{0}_{2\pm} &=& \pm \sqrt{x^{2} p^{2}_{3} + k^{2}_{\perp} + m^{2}_{q} -i\varepsilon} \,, 
\\[0.2cm]
k^{0}_{3\pm} &=& \delta_{0} p^{3} \pm \sqrt{(1-x)^{2} p^{2}_{3} + k^{2}_{\perp}+m^{2}_{g}-i\varepsilon} \, .
\end{eqnarray}
Since the two quark propagators are identical, one has double poles arising from the quark lines. Choosing to close the contour in the upper half plane, we pick up contributions from the (single) pole $k_{3-}^0$ and the double pole $k_{1-}^0 = k_{2-}^0$.
In the specific case of $g_{1,{\rm Q}}$, the result after the $k^0$-integration reads
\begin{eqnarray}
g_{1,{\rm Q}}(x; p^3) & = &  \frac{g^{2}_{s} \, C_{F} \mu^{2\epsilon}}{2\pi} \int  \dfrac{d^{n-2} k_\perp}{(2\pi)^{n-2}} \bigg[
\frac{N_{g1}(k_{3-}^0)}{(k_{3-}^0 - k_{1+}^0)^2 \, (k_{3-}^0 - k_{1-}^0)^2 \, (k_{3-}^0 - k_{3+}^0)}
\nonumber 
\\[0.1cm]
& & + \, \frac{N_{g1}'(k_{1-}^0)}{(k_{1-}^0 - k_{1+}^0)^2 \, (k_{1-}^0 - k_{3+}^0) \, (k_{1-}^0 - k_{3-}^0)}
- \frac{2 \, N_{g1}(k_{1-}^0)}{(k_{1-}^0 - k_{1+}^0)^3 \, (k_{1-}^0 - k_{3+}^0) \, (k_{1-}^0 - k_{3-}^0)}
\nonumber 
\\[0.1cm]
& & - \, \frac{N_{g1}(k_{1-}^0)}{(k_{1-}^0 - k_{1+}^0)^2 \, (k_{1-}^0 - k_{3+}^0)^2 \, (k_{1-}^0 - k_{3-}^0)}
- \frac{N_{g1}(k_{1-}^0)}{(k_{1-}^0 - k_{1+}^0)^2 \, (k_{1-}^0 - k_{3+}^0) \, (k_{1-}^0 - k_{3-}^0)^2}
\bigg] \,,
\label{e:qPDF_general2}
\end{eqnarray}
where $N_{g1}' \equiv \frac{d}{d k^0} N_{g1}$.
There is no need to carry out the $k_{\perp}$ integral analytically. By keeping $p^{3}$ finite, we will be evaluating the $k_{\perp}$ integral numerically using DR and a cut-off. Note that the above form for the quasi-PDFs holds true for any $x$. 
We want to mention at this point that for the twist-2 light-cone PDFs one can use the cut-graph method,
which amounts to putting the gluon on-shell.
However, this is not the case for the quasi-PDFs. Although the first term in the above equation corresponds exactly to the cut-graph method, all of the other terms provide finite contribution to the quasi-PDFs in all regions of $x$. Specifically, for the twist-2 case, we pointed out that one cannot recover the light-cone PDFs from quasi-PDFs for $x<0$, even in the limit $p^{3} \rightarrow \infty$, if quasi-PDFs are calculated in the cut-graph approach~\cite{Bhattacharya:2018zxi}. The problem with using the cut-graph method is even more serious at twist-3. Calculation of light-cone and quasi-PDFs in a cut-graph method excludes the contribution from the zero-modes, which by now we know is crucial to satisfy the sum rules~\cite{Burkardt:2001iy, Aslan:2020zik}.

\subsection{Results in Yukawa Model}
\label{s:qPDF_YM}
For quasi-PDFs in the YM, the general structure of the singular term is
\begin{eqnarray}
q_{\rm {Q (s)}}(x; p^{3}) &=& \dfrac{g^{2}_{\rm Y} \mu^{2\epsilon}}{4(2\pi)} \int \dfrac{d^{n-2} k_\perp}{(2\pi)^{n-2}} \, N_{q \rm{(s)}}   \, , 
\end{eqnarray}
while the canonical terms are given by Eqs.~(\ref{e:qPDF_general1}) and~(\ref{e:qPDF_general2}), but with an overall sign, and of course without the color factors as well as the replacement of the coupling constant. 
The numerators for the different PDFs are given by
\begin{eqnarray}
N_{g1 \rm{(c)}} &=& \dfrac{p^{3}}{\delta_{0}} \left \{ \delta_{0} (k^{0})^{2} - \frac{2k^{0}}{p^{3}} \Big( xp^{2}_{3} - m^{2}_{q} \Big) +  \delta_{0} \Big(- k^{2}_{\perp}+x^{2} p^{2}_{3} + m^{2}_{q} \Big) \right \} \, ,
\\[0.2cm]
N_{gT \rm{(s)}} &=&  \dfrac{ p^{3}}{\big ( k^{2}_{\perp} + x^{2} p^{2}_{3} + m^{2}_{q} \big )^{3/2}}  \, ,  \\[0.2cm]
N_{gT (\rm{c})} &=& \dfrac{p^{3}}{m_{q}} \bigg ( 2 m_{q} (k^{0})^{2} - 2 m_{q} k^{2}_{\perp} + m_{q} \dfrac{k^{2}_{\perp}}{1-\epsilon} - 2m_{q} x^{2} p^{2}_{3} + 2 m^{3}_{q} - m_{q} m^{2}_{s} \bigg ) \, ,
\\[0.2cm]
N_{h1 \rm{(c)}} &=& \dfrac{p^{3}}{\delta_{0}} \left \{ \delta_{0} (k^{0})^{2}-\frac{2k^{0}}{p^{3}}\Big(x p^{2}_{3} - m^{2}_{q} \Big) + \bigg ( 1 - \dfrac{1}{1-\epsilon} \bigg ) \delta_{0} k^{2}_{\perp}  + \delta_{0} \Big(x^{2}p^{2}_{3} + m^{2}_{q}\Big) \right \} \, , 
\\[0.2cm]
N_{hL \rm{(s)}} &=& \dfrac{ p^{3}}{\big ( k^{2}_{\perp} + x^{2} p^{2}_{3} + m^{2}_{q} \big )^{3/2}} \, , \\[0.2cm]
N_{hL (\rm{c})} &=& \dfrac{p^{3}}{m_{q}} \bigg ( 2 m_{q} (k^{0})^{2} - 2 m_{q} x^{2} p^{2}_{3} + 2 m^{3}_{q} - m_{q} m^{2}_{s} \bigg ) \, ,
\end{eqnarray}
The caveats regarding working with cut-graph methods, mentioned in the context of the QTM, also apply for the YM.

\subsection{Analytical proof of sum rules for quasi-PDFs in Quark Target Model}
\label{s:analytical_proof_moments_qPDFs}
In this section, we show that the UV-divergent parts of quasi-PDFs satisfy the BC-type sum rules. (For the sake of analytical simplicity, we are limiting ourselves to the UV-divergent parts only. However, in the next section, we provide numerical results for the full results of the quasi-PDFs.) We provide a sample calculation in the QTM, in the instance that one works with $m_{q} \neq 0$ as an IR regulator.
We provide the most important steps involved in this check with $g_{T, \rm{Q}}$ and $h_{L, \rm{Q}}$ as examples. 
We begin with $g_{T, \rm{Q}}$ which is calculated as
\begin{eqnarray}
\dfrac{m_{q}s^{i}_{\perp}}{p^{3}}\,g^{\rm{(1a)}}_{T, \rm{Q}} (x) &=& -\dfrac{i g^{2}_{s} C_{F} \mu^{2\epsilon}}{4} \int^{\infty}_{-\infty} \dfrac{d^{n}k}{(2\pi)^{n}} \dfrac{{\rm Tr} \big [ (\slashed{p}+ m_{q}) (1+\gamma_{5}\slashed{s}) \, \gamma^{\mu} \, (\slashed{k}+m_{q}) \, \gamma^{i}_{\perp}\gamma_{5} \, (\slashed{k}+m_{q}) \, \gamma_{\mu} \big ]}{(k^{2} -m^{2}_{q}+ i\varepsilon)^{2} ((p-k)^{2}-m^{2}_{g} + i\varepsilon)} \, \delta \bigg ( x- \dfrac{k^{3}}{p^{3}} \bigg ) \dfrac{1}{p^{3}} \nonumber \\[0.2cm]
&=& g^{\rm{(1a)}}_{T, \rm{Q (s)}} (x) + g^{\rm{(1a)}}_{T, \rm{Q (c)}} (x) \, , 
\end{eqnarray}
with
\begin{eqnarray}
g^{\rm{(1a)}}_{T, \rm{Q (s)}} (x) &=& - \dfrac{\alpha_{s}C_{F}}{2\pi} \dfrac{\epsilon_{\UV} \, p^{3}}{\sqrt{x^{2}p^{2}_{3} + m^{2}_{q}}} 
\label{gT_sing_app}
\, , \\[0.2cm]
g^{\rm{(1a)}}_{T, \rm{Q (c)}} (x) &=& \dfrac{\alpha_{s}C_{F}}{2\pi} \int^{1}_{0} dy \, \bigg \{ \dfrac{(1-y)}{2 Q} + \dfrac{(1-y) (x^{2}-y^{2})}{2 Q^{3}} - \dfrac{(1-y)(1+y^{2}) \rho_{q}}{2 Q^{3}} + \dfrac{\epsilon_{\UV} \, (1-y)}{Q}\bigg \} \, , 
\label{gT_cano_app}
\end{eqnarray}
where,
\begin{eqnarray}
\rho_{q} &=& \dfrac{m^{2}_{q}}{p^{2}_{3}} \, , \\[0.2cm]
 Q^{2} &=& (y-x)^{2} + (1-y)^{2}\rho_{q} \, .
\end{eqnarray}
In order to arrive at these expressions, we have taken a slightly different route compared to what we have presented in the previous sections. 
As long as we want to limit ourselves to the UV-divergent terms, it is better to invoke the following steps for the sake of analytical simplicity: combine the quark and gluon propagators via Feynman parameterization, then perform $\int dk^{0}$ via contour integral, and then, keeping $p^{3}$ finite, integrate over transverse parton momenta. These steps yield Eq.~(\ref{gT_sing_app}) and Eq.~(\ref{gT_cano_app}). In the end, we integrate over the Feynman parameter $y$. Analyzing the large-$x$ behavior of the resulting expression, we get
\begin{eqnarray}
g_{T, \rm{Q(c)}}^{\rm{(1a)}} (x)  & = & \frac{\alpha_s C_F}{2\pi}
\begin{dcases}
\phantom{+} \dfrac{1}{2x} 
& \quad  x \rightarrow +\infty \\[0.2cm]
- \dfrac{1}{2x}
& \quad  x \rightarrow -\infty  \, .
\end{dcases}
\end{eqnarray}
DR for the $x$-integral in $1-2\epsilon$ dimensions yields~\footnote{Note that we ``shift" DR for the UV-divergences to the $x$-integrals, just for the sake of analytical simplicity. More discussion regarding applying DR for the $k_\perp$-integrals or $x$-integrals can be found in the paragraph after Eq.~(\ref{e:h_p3_independent}).},
\begin{eqnarray}
\int dx \, g_{T, \rm{Q(c)}}^{\rm{(1a)}} (x)  & = & \frac{\alpha_s C_F}{2\pi}
\begin{dcases}
\dfrac{1}{4 \epsilon_{\UV}} 
& \quad  x \rightarrow +\infty  \\[0.2cm]
\dfrac{1}{4 \epsilon_{\UV}} 
& \quad  x \rightarrow -\infty  \, ,
\end{dcases}
\end{eqnarray}
while, $\int g_{T, \rm{Q(s)}}^{\rm{(1a)}}$ is UV-finite.
Hence,
\begin{eqnarray}
\int dx \, g_{T}^{\rm{(1a)}} (x) = \int dx \, g_{T, \rm{Q}}^{\rm{(1a)}} (x) = \dfrac{\alpha_{s}C_{F}}{2\pi} \, \bigg ( \dfrac{1}{2\epsilon_{\UV}} \bigg ) \, .
\end{eqnarray}
It is straightforward to repeat the corresponding steps for $g_{1, \rm{Q}}$, for which we obtain,
\begin{eqnarray}
g_{1, \rm{Q}}^{\rm{(1a)}} (x)  & = & \frac{\alpha_s C_F}{2\pi}
\begin{dcases}
\phantom{+} \dfrac{1}{2x} 
& \quad  x \rightarrow +\infty  \\[0.2cm]
- \dfrac{1}{2x}
& \quad  x \rightarrow -\infty  \, ,
\end{dcases}
\end{eqnarray}
and hence,
\begin{eqnarray}
\int dx \, g_{1, \rm{Q}}^{\rm{(1a)}} (x)  & = & \frac{\alpha_s C_F}{2\pi}
\begin{dcases}
\dfrac{1}{4 \epsilon_{\UV}} 
& \quad  x \rightarrow +\infty  \\[0.2cm]
\dfrac{1}{4 \epsilon_{\UV}} 
& \quad  x \rightarrow -\infty  \, .
\end{dcases}
\end{eqnarray}
Therefore,
\begin{eqnarray}
\int dx \, g_{1}^{\rm{(1a)}} (x) = \int dx \, g_{1, \rm{Q}}^{\rm{(1a)}} (x) = \dfrac{\alpha_{s}C_{F}}{2\pi} \, \bigg ( \dfrac{1}{2\epsilon_{\UV}} \bigg ) \, .
\end{eqnarray}
This means
\begin{eqnarray}
\int dx \, g^{\rm{(1a)}}_{1, \rm{Q}} (x)  = \int dx \, g^{\rm{(1a)}}_{T, \rm{Q}} (x) \, ,
\end{eqnarray}
which establishes the BC sum rule for the UV-divergent terms. We emphasize that our focus here is on the asymptotic expressions for the quasi-PDFs since we are interested in verifying the sum rules for the UV-divergent terms only. Therefore, the symbol $\int dx$ in the above expressions should not be misinterpreted as an $x$-integral over the entire range of $x$. In Sec.~\ref{s:numerical_results} we will provide a numerical check of the sum rules by integrating over all $x$.

The equivalent of Eq.~(\ref{gT_sing_app}) and Eq.~(\ref{gT_cano_app}) for $h_{L, \rm{Q}}$ is
\begin{eqnarray}
h^{\rm{(1a)}}_{L, \rm{Q (s)}} (x) &=& - \dfrac{\alpha_{s}C_{F}}{2\pi} \dfrac{p^{3}}{\sqrt{x^{2}p^{2}_{3} + m^{2}_{q}}} 
\label{hL_sing_app}
\, , \\[0.2cm]
h^{\rm{(1a)}}_{L, \rm{Q (c)}} (x) &=& \dfrac{\alpha_{s}C_{F}}{2\pi} \int^{1}_{0} dy \, \bigg \{ \dfrac{(1-y)}{2 Q} + \dfrac{(1-y) (x^{2}-y^{2})}{2 Q^{3}} - \dfrac{(1-y)(1+y^{2}) \rho_{q}}{2 Q^{3}} + \dfrac{(1-2\epsilon_{\UV}) (1-y)}{Q} \bigg \} \, . 
\label{hL_cano_app}
\end{eqnarray}
Picking out the UV-divergent terms at the end points $x \rightarrow \pm \infty$, we obtain
\begin{eqnarray}
h_{L, \rm{Q(s)}}^{\rm{(1a)}} (x)  & = & \frac{\alpha_s C_F}{2\pi}
\begin{dcases}
- \dfrac{1}{x} 
& \quad  x \rightarrow +\infty  \\[0.2cm]
\phantom{+} \dfrac{1}{x}
& \quad  x \rightarrow -\infty  \, ,
\end{dcases}
\\[0.2cm]
h_{L, \rm{Q(c)}}^{\rm{(1a)}} (x)  & = & \frac{\alpha_s C_F}{2\pi}
\begin{dcases}
\phantom{+} \dfrac{1}{x} 
& \quad  x \rightarrow +\infty  \\[0.2cm]
- \dfrac{1}{x}
& \quad  x \rightarrow -\infty  \, .
\end{dcases}
\end{eqnarray}
This means
\begin{eqnarray}
\int dx \, h_{L}^{\rm{(1a)}} (x) = \int dx \, \big ( h_{L, \rm{Q(s)}}^{\rm{(1a)}} (x) + h_{L, \rm{Q(c)}}^{\rm{(1a)}} (x) \big ) = 0 \, .
\end{eqnarray}
A corresponding calculation for $h_{1 , \rm{Q}}$ yields
\begin{eqnarray}
h_{1, \rm{Q}}^{\rm{(1a)}} (x)  & = & \frac{\alpha_s C_F}{2\pi}
\begin{dcases}
\mathcal{O}(\epsilon_{\UV})
& \quad  x \rightarrow +\infty  \\[0.2cm]
\mathcal{O}(\epsilon_{\UV})
& \quad  x \rightarrow -\infty  \, ,
\end{dcases}
\end{eqnarray}
and hence,
\begin{eqnarray}
\int dx \, h_{1}^{\rm{(1a)}} (x) = \int dx \, h_{1, \rm{Q}}^{\rm{(1a)}} (x) = 0 \, .
\end{eqnarray}
This means
\begin{eqnarray}
\int dx \, h^{\rm{(1a)}}_{1, \rm{Q}} (x)  = \int dx \, h^{\rm{(1a)}}_{L, \rm{Q}} (x) \, ,
\end{eqnarray}
which establishes the h-sum rule for the UV-divergent terms. It is straightforward to generalize this analysis to calculations in the YM.
In a similar fashion, one can check the validity of the sum rules for the UV-divergent parts in the cut-off scheme.

\section{Numerical Results for sum rules}
\label{s:numerical_results}
We now proceed to discuss numerical results for the sum rules. We will provide results in the QTM only, since the same analysis can be repeated in a straightforward manner in the YM.
We set the coupling constant $g_s$ and the color factor $C_{F}$ to unity because our numerical checks do not depend on the absolute values of these quantities. 
Our ``standard values" for the masses in the QTM are $m_{q} = 0.35$ GeV and $m_{g} = 0.1$ GeV. We will present our results for the sum rules with $\epsilon_{\UV} = (0.6, \, 0.8)$ for the DR scheme, and $\Lambda_{\UV} = (1, \, 4)$ GeV for the cut-off scheme. 
We emphasize that all the numerical results are for the exact expressions of the PDFs, that is, the expressions which are not expanded in the UV regulator.
Now, it is clear that conclusions from a model can only be considered robust if different values of model-parameters do not lead to qualitatively different results. In order to establish this, we will show some results with (extreme) values for the gluon mass, $m_{g} = 0.01$ GeV and $m_{g} = 0.7$ GeV.

\begin{table}[t]
\centering
\begin{tabular}{| c | c | c | c |}
\hline
\multicolumn{4}{|c|}{BC sum rule in QTM: DR for the UV} \\
\hline
Parameters and Moments of LC PDFs \qquad & $P^{3}$ (\rm GeV) \qquad &  $\int dx \, g_{1, \rm{Q}}(x)$  \qquad &  $\int dx \, g_{T, \rm{Q}}(x)$ \\
\hline 
\hline 
\multirow{3}{15em}{\begin{eqnarray*}
                  \epsilon_{\UV} &=& 0.8  \\
                   \int dx \, g_{1} (x) &=& -3.241 \\
                   \int dx \, g_{T} (x) &=& -3.241
                   \end{eqnarray*}}
 &  1  &  -3.241 &  -3.241 \\
\cline{2-4}
 &  2  &  -3.241 &  -3.241 \\
\cline{2-4}
 &  3  &  -3.238 &  -3.241 \\
\cline{2-4}
 &  4  &  -3.241 &  -3.241 \\
\hline
\hline 
\multirow{3}{15em}{\begin{eqnarray*}
                  \epsilon_{\UV} &=& 0.6 \\
                   \int dx \, g_{1} (x) &=& -0.8274 \\
                   \int dx \, g_{T} (x) &=& -0.8274
                   \end{eqnarray*}}
 &  1  &  -0.8275 &  -0.8274 \\
\cline{2-4}
 &  2  &  -0.8274 &  -0.8274 \\
\cline{2-4}
 &  3  &  -0.8275 &  -0.8274 \\
\cline{2-4}
 &  4  &  -0.8274 &  -0.8274 \\
\hline
\end{tabular}
\caption{All the numerical results have been obtained for $\mu = 1 \, {\rm GeV}$, $m_{q} = 0.35 \, {\rm GeV}$, and $m_{g} = 0.1 \, {\rm GeV}$. The BC sum rule is obeyed because the zero-mode contributions in $g_{T}(x)$ and $g_{T, \rm{Q}}(x)$ has been taken into account.}
\label{tab: QTM_BC_DR}
\end{table}

\begin{table}[b]
\centering
\begin{tabular}{| c | c | c | c |}
\hline
\multicolumn{4}{|c|}{BC sum rule in QTM: Cut-off for the UV} \\
\hline
Parameters and Moments of LC PDFs \qquad & $P^{3}$ (\rm GeV) \qquad &  $\int dx \, g_{1, \rm{Q}}(x)$  \qquad &  $\int dx \, g_{T, \rm{Q}}(x)$ \\
\hline 
\hline 
\multirow{3}{15em}{\begin{eqnarray*}
                  \Lambda_{\UV} &=& 1 \, {\rm{GeV}} \\
                   \int dx \, g_{1} (x) &=& -0.01082 \\
                   \int dx \, g_{T} (x) &=& -0.004864
                   \end{eqnarray*}}
 &  1  &  -0.01082 &  -0.004864 \\
\cline{2-4}
 &  2  &  -0.01082 &  -0.004864 \\
\cline{2-4}
 &  3  &  -0.01082 &  -0.004864 \\
\cline{2-4}
 &  4  &  -0.01082 &  -0.004864 \\
\hline
\hline 
\multirow{3}{15em}{\begin{eqnarray*}
                  \Lambda_{\UV} &=& 4 \, {\rm{GeV}} \\
                   \int dx \, g_{1} (x) &=& 0.005220 \\
                   \int dx \, g_{T} (x) &=& 0.01153
                   \end{eqnarray*}}
 &  1  &  0.005220 &  0.01153 \\
\cline{2-4}
 &  2  &  0.005220 &  0.01153 \\
\cline{2-4}
 &  3  &  0.005220 &  0.01153 \\
\cline{2-4}
 &  4  &  0.005220 &  0.01153 \\
\hline
\end{tabular}
\caption{All the numerical results have been obtained for $m_{q} = 0.35 \, {\rm GeV}$, and $m_{g} = 0.1 \, {\rm GeV}$. The BC sum rule is violated for both light-cone and quasi-PDFs in the cut-off scheme.}
\label{tab: QTM_BC_cut-off}
\end{table}

\begin{table}[t]
\centering
\begin{tabular}{| c | c | c | c |}
\hline
\multicolumn{4}{|c|}{BC sum rule in QTM: DR for the UV} \\
\hline
Parameters and Moments of LC PDFs \qquad & $P^{3}$ (\rm GeV) \qquad &  $\int dx \, g_{1, \rm{Q}}(x)$  \qquad &  $\int dx \, g_{T, \rm{Q}}(x)$ \\
\hline 
\hline 
\multirow{3}{22em}{\begin{eqnarray*}
                  \epsilon_{\UV} &=& 0.8 \quad 
                  (m_{q}, m_{g}) = (0.35, 0.7) \, {\rm{GeV}}
                  \end{eqnarray*}
                  \begin{eqnarray*}
                   \int dx \, g_{1} (x) &=& -0.05117 \\
                   \int dx \, g_{T} (x) &=& -0.05117
                   \end{eqnarray*}}
 &  1  &  -0.05117 &  -0.05117 \\
\cline{2-4}
 &  2  &  -0.05117 &  -0.05117 \\
\cline{2-4}
 &  3  &  -0.05118 &  -0.05117 \\
\cline{2-4}
 &  4  &  -0.05117 &  -0.05117 \\
\hline
\hline 
\multirow{3}{22em}{\begin{eqnarray*}
                  \epsilon_{\UV} &=& 0.8 \quad 
                  (m_{q}, m_{g}) = (0.35, 0.01) \, {\rm{GeV}}
                  \end{eqnarray*}
                  \begin{eqnarray*}
                   \int dx \, g_{1} (x) &=& -169.99 \\
                   \int dx \, g_{T} (x) &=& -169.95
                   \end{eqnarray*}}
 &  1  &  -169.99 &  -169.99 \\
\cline{2-4}
 &  2  &  -169.94 &  -169.99 \\
\cline{2-4}
 &  3  &  -169.97 &  -169.98 \\
\cline{2-4}
 &  4  &  -169.66 &  -169.89 \\
\hline
\end{tabular}
\caption{BC sum rule with variation of gluon mass in DR scheme. Note that $\mu = 1 \, {\rm GeV}$.}
\label{tab:BC_mg_DR}
\end{table}

Table~\ref{tab: QTM_BC_DR} shows results for the BC sum rule in the DR scheme. These numerical results confirm that the sum rule holds for both the light-cone and quasi-PDFs for finite values of the DR parameter $\epsilon_{\UV}$. Specifically, one not only has
\begin{eqnarray}
\int^{\infty}_{-\infty} dx \, g_{1, \rm{Q}} (x; p^{3}) \Big |^{\epsilon_{\UV}} = \int^{\infty}_{-\infty} dx \, g_{T, \rm{Q}} (x; p^{3}) \Big |^{\epsilon_{\UV}} \, , 
\end{eqnarray}
which is the BC sum rule for quasi-PDFs, but one also has
\begin{eqnarray}
\int^{\infty}_{-\infty} dx \, g_{1, \rm{Q}} (x; p^{3}) \Big |^{\epsilon_{\UV}} = \int^{1}_{0} dx \, g_{1} (x) \Big |^{\epsilon_{\UV}} \, , \qquad \int^{\infty}_{-\infty} dx \, g_{T, \rm{Q}} (x; p^{3}) \Big |^{\epsilon_{\UV}} = \int^{1}_{0} dx \, g_{T} (x) \Big |^{\epsilon_{\UV}} \, .
\label{e:BC_p3_independent}
\end{eqnarray}
As also shown in a model-independent manner in Sec.~\ref{s:definitions}, Eq.~(\ref{e:BC_p3_independent}) confirms that the (explicit) $p^{3}$-dependence of the quasi-PDFs drops out upon taking their lowest moment. Results in Table~\ref{tab: QTM_BC_DR} reflect that, for a complete match of the moments between light-cone and quasi-PDFs, one has to take the zero-mode contributions into account. Interestingly, the moments of the zero-mode terms exactly match between the light-cone PDF $g_{T}$ and the quasi-PDF $g_{T, \rm{Q}}$. (We have illustrated this point analytically in the context of $h_{L, \rm{Q}}$ towards the end of this section.) Similarly, we find an exact match in the moments of the canonical terms between the two distributions. To sum up, we infer
\begin{eqnarray}
\int^{\infty}_{-\infty} dx \, g_{T, \rm{Q (s)}} (x; p^{3}) \Big |^{\epsilon_{\UV}} = \int^{1}_{0} dx \, g_{T \rm{(s)}} (x) \Big |^{\epsilon_{\UV}} \, , \qquad \int^{\infty}_{-\infty} dx \, g_{T, \rm{Q (c)}} (x; p^{3}) \Big |^{\epsilon_{\UV}} = \int^{1}_{0} dx \, g_{T \rm(c)} (x) \Big |^{\epsilon_{\UV}} \, .
\end{eqnarray}
Calculation of moments can therefore be considered to be an independent check of our analytical results.
We find these results to be very encouraging because they have been obtained for the most general situation when one has all the partonic masses in the picture, $m_{q} \neq 0$ and $m_{g} \neq 0$. 

Table~\ref{tab: QTM_BC_cut-off} demonstrates the violation of the BC sum rule for both light-cone and quasi-PDFs when a cut-off is applied. Although,
\begin{eqnarray}
\int^{\infty}_{-\infty} dx \, g_{1, \rm{Q}} (x; p^{3}) \Big |^{\Lambda_{\UV}} \neq  \int^{\infty}_{-\infty} dx \, g_{T, \rm{Q}} (x; p^{3}) \Big |^{\Lambda_{\UV}} \, , \qquad 
\int^{1}_{0} dx \, g_{1} (x) \Big |^{\Lambda_{\UV}} \neq  \int^{1}_{0} dx \, g_{T} (x) \Big |^{\Lambda_{\UV}} \, ,
\end{eqnarray}
one still has,
\begin{eqnarray}
\int^{\infty}_{-\infty} dx \, g_{1, \rm{Q}} (x; p^{3}) \Big |^{\Lambda_{\UV}} = \int^{1}_{0} dx \, g_{1} (x) \Big |^{\Lambda_{\UV}} \, , \qquad \int^{\infty}_{-\infty} dx \, g_{T, \rm{Q}} (x; p^{3}) \Big |^{\Lambda_{\UV}} = \int^{1}_{0} dx \, g_{T} (x) \Big |^{\Lambda_{\UV}} \, .
\end{eqnarray}
The bottom line is that, whether or not the sum rules hold among different quasi-PDFs, the moment of the quasi-PDFs agrees with those of their light-cone counterparts. This is a general statement, and is true at least for the regulated results. 
We will return to this point at the end of this section.

One can repeat the same exercise to obtain results for the h-sum rule in the DR and cut-off schemes. Our numerical results confirm
\begin{eqnarray}
\int^{\infty}_{-\infty} dx \, h_{1, \rm{Q}} (x; p^{3}) \Big |^{\epsilon_{\UV}} = \int^{\infty}_{-\infty} dx \, h_{L, \rm{Q}} (x; p^{3}) \Big |^{\epsilon_{\UV}} \, , 
\end{eqnarray}
which is the h-sum rule for quasi-PDFs, and
\begin{eqnarray}
\int^{\infty}_{-\infty} dx \, h_{1, \rm{Q}} (x; p^{3}) \Big |^{\epsilon_{\UV}} = \int^{1}_{0} dx \, h_{1} (x) \Big |^{\epsilon_{\UV}} \, , \qquad \int^{\infty}_{-\infty} dx \, h_{L, \rm{Q}} (x; p^{3}) \Big |^{\epsilon_{\UV}} = \int^{1}_{0} dx \, h_{L} (x) \Big |^{\epsilon_{\UV}} \, .
\end{eqnarray}
More importantly, our numerical results reaffirm the accidental validity of the h-sum rule for the light-cone PDFs in the cut-off scheme. And, not surprisingly, this accident is exactly reproduced by their corresponding quasi-PDFs. We therefore infer the following relations, 
\begin{eqnarray}
\int^{\infty}_{-\infty} dx \, h_{1, \rm{Q}} (x; p^{3}) \Big |^{\Lambda_{\UV}} &=&  \int^{\infty}_{-\infty} dx \, h_{L, \rm{Q}} (x; p^{3}) \Big |^{\Lambda_{\UV}} \, , \qquad 
\int^{1}_{0} dx \, h_{1} (x) \Big |^{\Lambda_{\UV}} =  \int^{1}_{0} dx \, h_{L} (x) \Big |^{\Lambda_{\UV}} \, , \\[0.2cm]
\int^{\infty}_{-\infty} dx \, h_{1, \rm{Q}} (x; p^{3}) \Big |^{\Lambda_{\UV}} &=& \int^{1}_{0} dx \, h_{1} (x) \Big |^{\Lambda_{\UV}} \, , \qquad \int^{\infty}_{-\infty} dx \, h_{L, \rm{Q}} (x; p^{3}) \Big |^{\Lambda_{\UV}} = \int^{1}_{0} dx \, h_{L} (x) \Big |^{\Lambda_{\UV}} \, .
\label{e:h_p3_independent}
\end{eqnarray}

Table~\ref{tab:BC_mg_DR} confirms the robustness of the discussions we have had in the context of the BC sum rule under the variation of the gluon mass. We have also confirmed the robustness of our results in the cut-off scheme. Finally, all the discussions we have had in the context of the h-sum rule remains valid if the gluon mass is changed. To summarize, we find that the BC and the h sum rules hold in QTM when DR is employed. And, it is the h-sum rule only that remains accidentally valid in the cut-off scheme. 
We have repeated the same analysis in the YM. The general results in the YM are the same as in the QTM, with the exception that the h-sum rule is violated in the cut-off scheme. 
We have also verified that the results in the YM remain valid when changing the model parameters.

We conclude this section by discussing two general aspects related to the calculation of moments of quasi-PDFs: (1) the convergence of such moments, and (2) the role of a twist-expansion in the calculation of moments of quasi-PDFs.
Unlike in the case of light-cone PDFs, the $k_{\perp}$-integral 
for quasi-PDFs is finite.
However, when computing the lowest $x$-moment for quasi-PDFs, one encounters a singularity due to the $1/|x|$ behaviour of the quasi-PDFs as $|x| \to \infty$, as we saw above in Sec.~\ref{s:analytical_proof_moments_qPDFs}.
Instead, in our numerical calculations we have used DR and a cut-off to the $k_{\perp}$ integrals (despite the fact that those integrals are finite without such ``regulators''), but with the $x$-integrals extending to infinity, leading to finite lowest moments for the quasi-PDFs.
In the following we will make explicit how such a situation can arise.
To this end we consider the integral which appears in the singular part of the twist-3 quasi-PDFs, and employ both DR and a cut-off to the $k_\perp$ integral.
We find that with DR,
\begin{eqnarray}
\int^{\infty}_{0} dk_{\perp} \, \dfrac{p^{3} \, k^{1-2\epsilon}_{\perp}}{\big ( k^{2}_{\perp} + x^{2} p^{2}_{3} + m^{2}_{q} \big )^{3/2}} &=& \dfrac{p^{3}}{\sqrt{\pi}} \bigg [ \Gamma(1-\epsilon) \Gamma(\epsilon + 1/2) \big ( x^{2} p^{2}_{3} + m^{2}_{q} \big )^{- \frac{1}{2} - \epsilon} \bigg ] \nonumber \\[0.2cm]
& \xrightarrow{x \to \infty} & \dfrac{1}{|x|^{1+2\epsilon}} \, , 
\end{eqnarray}
where $\epsilon >0$, and with a cut-off,
\begin{eqnarray}
\int^{\Lambda}_{0} dk_{\perp} \, \dfrac{p^{3} \, k_{\perp}}{\big ( k^{2}_{\perp} + x^{2} p^{2}_{3} + m^{2}_{q} \big )^{3/2}} &=& p^{3} \left [ \dfrac{1}{\sqrt{x^{2}p^{2}_{3} + m^{2}_{q}}} - \dfrac{1}{\sqrt{x^{2}p^{2}_{3} + m^{2}_{q} + \Lambda^{2}}} \right ] \nonumber \\[0.2cm]
& \xrightarrow{x \to \infty} & \dfrac{\Lambda^{2}}{2p^{2}_{3}|x|^{3}}  \, . 
\end{eqnarray}
Therefore, by providing regulation for the $k_{\perp}$-integrals, we essentially alter the large-$x$ behavior of the quasi-PDFs. (In other words, by providing regulation for the $k_{\perp}$ integrals, we automatically provide regulation for the $x$ integrals.)
Specifically, with regulated $k_{\perp}$ integrals, the quasi-PDFs are forced to fall faster than $1/x$, and hence their lowest moments are well-defined.
This explains the finiteness of the moments in all the tables that has been presented so far.
(If we calculate $\int dx$ of the above expressions, and then take $\epsilon \approx 0$ or $\Lambda \approx \infty$, we immediately ``recover" the poles in the moments for the quasi-PDFs, which are in fact the same poles present in the light-cone PDFs. See the next paragraph for this point.)

Let us now proceed to the second point.
In matching-type calculations, one calculates moments for the quasi-PDFs after a twist-expansion in powers of $1/p^{3}$. As shown in Sec.~\ref{ss:model_independent_sum_rule_quasi}, we expect the moments to match between quasi and light-cone PDFs before any twist-expansion. More specifically, we expect an agreement in the moments for finite values of $p^{3}$. To demonstrate the subtleties involved in the calculation of moments after a twist expansion, we (again) take as an example case the singular terms for the quasi-PDFs, namely the one that appears for $h_{L, \rm{Q}}(x)$, which reads 
\begin{eqnarray}
h_{L, \rm{Q(s)}}(x) &=& -\dfrac{g^{2}_{s} C_{F}\mu^{2\epsilon}}{2(2\pi)} \int \dfrac{d^{n-2} k_\perp}{(2\pi)^{n-2}} \, \dfrac{(1- \epsilon) \, p^{3}}{\big ( k^{2}_{\perp} + x^{2} p^{2}_{3} + m^{2}_{q} \big )^{3/2}} \nonumber \\[0.2cm]
&=& g^{2}_{s} C_{F} \mu^{2\epsilon} p^{3} \left ( 2^{-3+2\epsilon} \pi^{-5/2 + \epsilon} (-1+\epsilon) \, \Gamma (1/2+\epsilon) (x^{2}p^{2}_{3} + m^{2}_{q})^{-1/2 - \epsilon} \right ) \, .
\label{e:analytical_moment_check_hL}
\end{eqnarray}
Eq.~(\ref{e:analytical_moment_check_hL}) is exact. 
Performing a twist expansion of this expression provides
\begin{eqnarray}
h_{L, \rm{Q(s)}} (x) & \approx & g^{2}_{s} C_{F} \mu^{2\epsilon} (x^{2} p^{2}_{3} + m^{2}_{q})^{-\epsilon} \bigg ( \dfrac{2^{-3+2\epsilon}  \pi^{-5/2 + \epsilon} \, (-1+\epsilon) \, \Gamma (1/2+\epsilon)}{x} \nonumber \\[0.2cm]
&& \phantom{This text is invisible ...} - \dfrac{2^{-4+2\epsilon}  \pi^{-5/2 + \epsilon} \, (-1+\epsilon)  \, \Gamma (1/2+\epsilon)}{x^{3}} \dfrac{m^{2}_{q}}{p^{2}_{3}} \, + \, \mathcal{O} \bigg ( \dfrac{1}{p^{3}_{3}} \bigg ) \bigg ) \, .
\label{e:divergence_x=0}
\end{eqnarray}
We immediately see that this expression cannot be integrated upon $x$, since the leading term and the higher-order terms have a pole at $x = 0$.
(However, there is no problem for the integral as $x= \pm \infty$.)
On the other hand, it is possible to calculate directly the $x$-integral of Eq.~(\ref{e:analytical_moment_check_hL}), without encountering any divergence at all. The resulting expression in the limit $\epsilon \to 0$ is 
\begin{eqnarray}
\int^{\infty}_{-\infty} dx \, h_{L, \rm{Q(s)}} (x) & \approx & \dfrac{g^{2}_{s} C_{F}}{8 \pi^{2}} \bigg ( - {\cal P}_{\UV} - \ln \dfrac{\mu^{2}_{\UV}}{m^{2}_{q}} + 1 \bigg ) \, ,
\end{eqnarray}
which exactly agrees with the moment of the light-cone $h_{L \rm{(s)}}(x)$ (see Eq.~(\ref{e:hL_sing_DR_QTM})). The situation is the same for the case of a cut-off. 
Specifically, after a twist-expansion, the singular part of $h_{L, \rm{Q}} (x)$ reads
\begin{eqnarray}
h_{L, \rm{Q(s)}} (x) &=& \dfrac{g^{2}_{s} C_{F}}{8\pi^{2}} p^{3} \left ( - \dfrac{1}{\sqrt{x^{2}p^{2}_{3} + m^{2}_{q}}} + \dfrac{1}{\sqrt{x^{2}p^{2}_{3} + m^{2}_{q} + \Lambda^{2}}} \right ) 
\label{e:analytical_moment_check_cut-off_hL} 
\, \\[0.2cm] 
& \approx & g^{2}_{s} C_{F}  \left ( - \dfrac{1}{16\pi^{2} x^{3}} \dfrac{\Lambda^{2}}{p^{2}_{3}} + \dfrac{3}{64 \pi^{2} x^{5}} \dfrac{(2 m^{2}_{q} \, \Lambda^{2} + \Lambda^{4})}{p^{4}_{3}} \, + \, \mathcal{O} \bigg ( \dfrac{1}{p^{6}_{3}} \bigg ) \right ) \, ,
\end{eqnarray}
which clearly reflects the non-integrability at $x=0$. However, it is (again) possible to calculate directly the $x$-integral of Eq.~(\ref{e:analytical_moment_check_cut-off_hL}). The result in the limit $\Lambda \to \infty$ agrees exactly with the moment of the light-cone $h_{L (\rm{s})}$ (see Eq.~(\ref{e:hL_full_cut-off_QTM})),
\begin{eqnarray}
\int^{\infty}_{-\infty} dx \, h_{L, \rm{Q(s)}} (x) & \approx & - \dfrac{g^{2}_{s} C_{F}}{8 \pi^{2}} \ln \dfrac{\Lambda^{2}}{m^{2}_{q}}\, .
\end{eqnarray}
The above analysis shows a non-trivial issue related to the non-commutativity of two limits: performing a twist-expansion, and the calculation of $\int dx $. We repeat that our numerical results for the moments have been calculated for finite values of $p^{3}$, that is, without any twist-expansion, and the regulated results are in complete agreement with the corresponding moments of the light-cone PDFs. Note that, in general, higher-twist terms may have divergences at the end points $x \rightarrow 0$ and $x \rightarrow 1$, and therefore in such cases, the moments will also be divergent. 
Specifically, these type of divergences are bound to arise from the canonical component of the quasi-PDFs. For example, $h_{L, \rm{Q}(c)}$ in QTM shows a behavior like,
\begin{align}
h_{L, \rm{Q}(c)}^{\rm{(1a)}} (x) & \approx 
\dfrac{\alpha_{s}C_{F}}{2\pi}
\begin{dcases}
\ln \frac{x}{x-1} 
& x > 1\\[0.2cm]
\ln \dfrac{4x p^{2}_{3}}{(1-x)m^{2}_{q}} - \dfrac{2}{1-x}
& 0 < x  <1 \\[0.2cm]
\ln \frac{x - 1}{x}
& x<0 
\end{dcases}
\\[0.2cm]
& - \dfrac{\alpha_{s}C_{F}}{8\pi} \, \dfrac{m^{2}_{q}}{p^{2}_{3}} \, \dfrac{1}{x^{2}(1-x)}
\begin{dcases}
\phantom{+} 2x^{2} - x - 3 - 2x^{2} (1-x) \ln \dfrac{x-1}{x} 
& x > 1  \\[0.2cm]
\phantom{+} 2x^{3} - x - 3 + 2x^{2} (1-x) \ln \dfrac{4x p^{2}_{3}}{(1-x)m^{2}_{q}}
&   0 < x < 1 \\[0.2cm]
- 2x^{2} + x + 3 + 2x^{2} (1-x) \ln \dfrac{x-1}{x} 
&   x < 0 
\end{dcases}
\nonumber
\\[0.3cm]
& + \mathcal{O} \bigg ( \dfrac{1}{p^{4}_{3}} \bigg ) \, .
\end{align}

\section{Sum rule involving the twist-3 PDF $e(x)$}
\label{s:results_e_eQ}
In this section, we shift our focus to the twist-3 PDF $e(x)$. 
Our goal is to address a particular sum rule which relates $e(x)$ to the target mass. 
We check this relation in both models.
We also provide the model results for the quasi-PDF $e_{\rm{Q}}(x;p^3)$ and confirm numerically that the light-cone PDF $e(x)$ and the quasi-PDF $e_{\rm{Q}}(x; p^3)$ have the exact same lowest moments, as it should be in a model-independent manner.
The light-cone PDF $e(x)$ and it's quasi-counterpart $e_{\rm Q} (x; p^{3})$ are defined as,
\begin{eqnarray}
\Phi^{[1]} = \dfrac{m_{q}}{p^{+}} \, e(x) \, , \qquad
\Phi^{[1]}_{\rm Q}  =\dfrac{m_{q}}{p^{3}} \, e_{\rm Q} (x; p^{3}) \, .
\label{e:def_e}
\end{eqnarray}
By repeating the steps shown in Sec.~\ref{ss:model_independent_sum_rule_quasi}, it is straight forward to check that the above definitions imply that,
\begin{eqnarray}
 \int dx \, e (x) = \int dx \, e_{\rm Q} (x; p^{3}) \,.
\end{eqnarray}
This implies that the lowest moment of $e_{\rm Q}$ is also related to target mass. (We repeat that in the QTM the target mass coincides with the quark mass.)

\subsection{Results in Quark Target Model}
\begin{table}[t]
\centering
\begin{tabular}{| c | c | c |}
\hline
\multicolumn{3}{|c|}{Moments of $e(x)$ and $e_{\rm{Q}}(x)$ in QTM: DR for the UV} \\
\hline
Parameters and Moments of LC PDFs \qquad & $P^{3}$ (\rm GeV) \qquad &  $\int dx \, e_{\rm{Q}}(x)$  \\
\hline 
\hline 
\multirow{3}{15em}{\begin{eqnarray*}
                  \epsilon_{\UV} &=& 0.8  \\
                   \int dx \, e (x) &=& -2.705
                   \end{eqnarray*}}
 &  1  &  -2.705  \\
\cline{2-3}
 &  2  &  -2.705  \\
\cline{2-3}
 &  3  &  -2.705  \\
\cline{2-3}
 &  4  &  -2.705  \\
\hline
\hline 
\multirow{3}{15em}{\begin{eqnarray*}
                  \epsilon_{\UV} &=& 0.6 \\
                   \int dx \, e (x) &=& -0.4687
                   \end{eqnarray*}}
 &  1  &  -0.4693  \\
\cline{2-3}
 &  2  &  -0.4693  \\
\cline{2-3}
 &  3  &  -0.4693  \\
\cline{2-3}
 &  4  &  -0.4693  \\
\hline
\end{tabular}
\caption{All the numerical results have been obtained for $\mu = 1 \, {\rm GeV}$, $m_{q} = 0.35 \, {\rm GeV}$, and $m_{g} = 0.1 \, {\rm GeV}$. }
\label{table:numerical_moment_e_QTM_DR}
\end{table}

It is known that loop corrections to the quark propagator can be summed up into a renormalized propagator as
\begin{eqnarray}
i G^{\rm{R}} (\slashed{p}) &=& \dfrac{i}{\slashed{p} - m_{\rm{R}} + \Sigma_{\rm{R}} (\slashed{p})} \, ,
\label{e:renorm_propagator}
\end{eqnarray}
where $\Sigma_{\rm{R}} (\slashed{p})$ denotes one-particle irreducible Feynman diagrams together with contributions from counter-terms. 
For the renormalized self energy one has
\begin{eqnarray}
\Sigma_{\rm{R}} (\slashed{p}) &=& \Sigma (\slashed{p}) + \delta_{2} \slashed{p} - (\delta_{2} + \delta_{\rm{m}}) m_{\rm{R}} \,,
\label{e:Sigma}
\end{eqnarray}
where $\Sigma (\slashed{p})$ is the result for the diagram in Fig.~($\rm{2a}$),
$\delta_{2}$ is the counter-term entering the wave-function renormalization factor $Z_{2} = 1+ \delta_{2}$, and $m_{q}$ is the bare mass of the quark given by $m_{q} = m_{\rm{R}} + \delta_{\rm m} \, m_{\rm{R}}$. 
By choosing the counter-terms as 
\begin{eqnarray}
\delta_{2} &=& \dfrac{\alpha_{s} C_{F}}{4\pi} {\cal P}_{\UV} \, ,\\[0.2cm] 
\delta_{\rm{m}} &=& \dfrac{3 \alpha_{s} C_{F}}{4\pi} {\cal P}_{\UV} \, ,
\end{eqnarray}
in $\MSb$ scheme, we obtain the following renormalized expression for the self-energy,
\begin{eqnarray}
\Sigma_{\rm{R}} (\slashed{p} = m_q) = \dfrac{\alpha_{s} C_{F}}{ 2 \pi} \bigg (  \dfrac{3}{2} m_{q} \ln \dfrac{\mu^{2}_{\UV}}{m^{2}_{\rm{R}}} + 2 m_{q} \bigg ) \, ,
\end{eqnarray}
at the pole $\slashed{p} = m_{q}$. 
With these results, the relation between the renormalized mass of the target and the bare quark mass reads
\begin{eqnarray}
m_{\rm{R}} \Big |^{\epsilon_{\UV}}  &=& m_{q} \bigg [ 1 + \dfrac{\alpha_{s} C_{F}}{2\pi} \bigg ( \dfrac{3}{2} \ln \dfrac{\mu^{2}_{\UV}}{m^{2}_{q}} + 2 \bigg ) \bigg ] \,.
\label{e:renorm_mass_QTM_DR}
\end{eqnarray}
In the cut-off scheme, we choose the counter-terms as 
\begin{eqnarray}
\delta_{2} &=& \dfrac{\alpha_{s} C_{F}}{4\pi} \ln \dfrac{\Lambda^{2}_{\UV}}{\mu^{2}} \, ,\\[0.2cm] 
\delta_{\rm{m}} &=& \dfrac{3 \alpha_{s} C_{F}}{4\pi} \ln \dfrac{\Lambda^{2}_{\UV}}{\mu^{2}} \, ,
\end{eqnarray}
where $\mu$ is an arbitrary scale introduced to render the counter-terms dimensionless. 
One can then show that the relation between the renormalized mass of the target and the bare quark mass is given by
\begin{eqnarray}
m_{\rm{R}} \Big |^{\Lambda_{\UV}} &=& m_{q} \bigg [ 1 + \dfrac{\alpha_{s} C_{F}}{2\pi} \bigg ( \dfrac{3}{2} \ln \dfrac{\mu^{2}}{m^{2}_{q}} + 1 \bigg ) \bigg ] \, .
\label{e:renorm_mass_QTM_cutoff}
\end{eqnarray}
There is a well-known sum rule that relates the lowest moment of $e(x)$ to the derivative of the renormalized target mass, 
\begin{eqnarray}
\int dx \, e(x) &=& \dfrac{\partial m_{\rm{R}}}{\partial m_{q}} \, .
\label{e:mass_sum_rule}
\end{eqnarray}
In order to verify Eq.~(\ref{e:mass_sum_rule}), 
we expect that $e(x)$ should be renormalized.
However, it is known that the renormalization of twist-3 PDFs is not diagonal, and a complete 
renormalization program will therefore require to involve quark-gluon-quark matrix elements,
which goes beyond the scope of our present work. 
Also, sum rules in general cannot be taken for granted after switching on renormalization. 
Throughout this work, our focus has been on the regulated PDFs. 
Therefore, below we will use a regulated mass to verify if $\int e$ obeys the relation in Eq.~(\ref{e:mass_sum_rule}).

In order to calculate a regulated mass, we repeat the same steps as above, but we do not make a subtraction of the counter-terms. In doing so, we arrive at
\begin{eqnarray}
m_{\rm{Reg.}} \Big |^{\epsilon_{\UV}} &=& m_{q} \bigg [ 1 + \dfrac{\alpha_{s} C_{F}}{2\pi} \bigg ( \dfrac{3}{2} {\cal P}_{\UV} + \dfrac{3}{2} \ln \dfrac{\mu^{2}_{\UV}}{m^{2}_{q}} + 2 \bigg ) \bigg ] \, , \\[0.2cm]
\therefore \dfrac{\partial{m_{\rm{Reg.}}}}{\partial{m_{q}}} \Big |^{\epsilon_{\UV}} &=& 1 + \dfrac{\alpha_{s} C_{F}}{2\pi} \bigg ( \dfrac{3}{2} {\cal P}_{\UV} + \dfrac{3}{2} \ln \dfrac{\mu^{2}_{\UV}}{m^{2}_{q}} -1 \bigg ) \, ,
\label{e:regulated_mass_QTM_DR}
\\[0.2cm]
m_{\rm{Reg.}} \Big |^{\Lambda_{\UV}} &=& m_{q} \bigg [ 1 + \dfrac{\alpha_{s} C_{F}}{2\pi} \bigg ( \dfrac{3}{2} \ln \dfrac{\Lambda^{2}_{\UV}}{m^{2}_{q}} + 1 \bigg ) \bigg ] \, , 
\\[0.2cm]
\therefore \dfrac{\partial{m_{\rm{Reg.}}}}{\partial{m_{q}}} \Big |^{\Lambda_{\UV}} &=& 1 + \dfrac{\alpha_{s} C_{F}}{2\pi} \bigg ( \dfrac{3}{2} \ln \dfrac{\Lambda^{2}_{\UV}}{m^{2}_{q}} - 2 \bigg ) \, .
\label{e:regulated_mass_QTM_cutoff}
\end{eqnarray}

We now turn to the results for $e(x)$ in the QTM. The starting expressions for the singular and canonical parts for $e(x)$ are
\begin{eqnarray}
e^{\rm{(1a)}}_{\rm{(s)}} (x) &=&  \dfrac{g^{2}_{s} C_{F} \mu^{2\epsilon}}{2\pi} \, \delta (x) \int \dfrac{d^{n-2}k_{\perp}}{(2\pi)^{n-2}}  \,  \dfrac{1-\epsilon}{(k^{2}_{\perp} + m^{2}_{q})}  \, , \nonumber \\[0.2cm]
e^{\rm{(1a)}}_{\rm{(c)}} (x) &=& \dfrac{g^{2}_{s} C_{F} \mu^{2\epsilon}}{2\pi} \int \dfrac{d^{n-2}k_{\perp}}{(2\pi)^{n-2}} \, \dfrac{k^{2}_{\perp} - (1-x^{2}) m^{2}_{q} + x m^{2}_{g} - (1-\epsilon) (1-x) m^{2}_{g}}{ \big ( k^{2}_{\perp} + (1-x)^{2} m^{2}_{q} + x m^{2}_{g} \big ) ^{2} }  \, .
\end{eqnarray}
From these equations, our final results for $e(x)$ with $m_{q} \neq 0$ reads~\cite{Bhattacharya:2020jfj}
\begin{eqnarray}
e^{\rm{(1a)}} (x) \Big |^{\epsilon_{\UV}}_{m_{q}}  &=& e^{\rm{(1a)}}_{\rm{(s)}} (x) \Big |^{\epsilon_{\UV}}_{m_{q}} + e^{\rm{(1a)}}_{\rm{(c)}} (x) \Big |^{\epsilon_{\UV}}_{m_{q}} \nonumber \\[0.2cm]
&=& \dfrac{\alpha_{s} C_{F}}{2\pi} \, \delta (x) \bigg ( {\cal P}_{\UV} + \ln \dfrac{\mu^{2}_{\UV}}{m^{2}_{q}}-1 \bigg )  + \dfrac{\alpha_{s} C_{F}}{2\pi} \bigg ( {\cal P}_{\UV} + \ln \dfrac{\mu^{2}_{\UV}}{(1-x)^{2}m^{2}_{q}} - \dfrac{2}{1-x} \bigg ) \, 
\label{e:e_QTM_DR}
\end{eqnarray}
in the DR scheme. Therefore, the lowest moment of $e(x)$ in DR is
\begin{eqnarray}
\int^{1}_{0} dx \, \delta(1-x) + \int \dfrac{dk^{+}}{p^{+}} \, e^{\rm{(1a)}} (k^{+}) \Big |^{\epsilon_{\UV}}_{m_{q}}  + \dfrac{\partial \Sigma}{\partial \slashed{p}} \Big |^{\epsilon_{\UV}}_{m_{q}} &=& 1 + \dfrac{\alpha_{s} C_{F}}{2\pi} \bigg ( \dfrac{3}{2} {\cal P}_{\UV} + \dfrac{3}{2} \ln \dfrac{\mu^{2}_{\UV}}{m^{2}_{q}} - 1 \bigg ) \, ,
\end{eqnarray}
which exactly agrees with the DR result in Eq.~(\ref{e:regulated_mass_QTM_DR}), and hence the sum rule in Eq.~(\ref{e:mass_sum_rule}) is satisfied. 
Note that this sum rule holds only if one takes the zero modes into account. This was already pointed out in Ref.~{\cite{Burkardt:2001iy}}.
In the cut-off scheme, with regularization applied to the transverse components only, our result for $e(x)$ is 
\begin{eqnarray}
e^{\rm{(1a)}} (x) \Big |^{\Lambda_{\UV}}_{m_{q}}  &=& e^{\rm{(1a)}}_{\rm{(s)}} (x) \Big |^{\Lambda_{\UV}}_{m_{q}} + e^{\rm{(1a)}}_{\rm{(c)}} (x) \Big |^{\Lambda_{\UV}}_{m_{q}} \nonumber \\[0.2cm]
&=& \dfrac{\alpha_{s} C_{F}}{2\pi} \, \delta (x) \ln \dfrac{\Lambda^{2}_{\UV}}{m^{2}_{q}} + \dfrac{\alpha_{s} C_{F}}{2\pi} \bigg ( \ln \dfrac{\Lambda^{2}_{\UV}}{(1-x)^{2} m^{2}_{q}} - \dfrac{2}{1-x} \bigg ) \, .
\label{e:e_QTM_cut-off}
\end{eqnarray}
One can check that the above results do not satisfy the sum rule. Now, by applying a regularization to all components of $k$, we obtain
\begin{eqnarray}
\int \dfrac{dk^{+}}{p^{+}} \, e^{\rm{(1a)}} (k^{+}) \Big |^{\Lambda_{\UV}}_{m_{q}} &=& \dfrac{\alpha_{s}C_{F}}{\pi} \int^{1}_{0} dy \bigg ( 2(1-y) \ln \dfrac{\Lambda^{2}_{\UV}}{(1-y)^{2}m^{2}_{q}} - 4(1-y) - \dfrac{y}{1-y} \bigg ) \, , \nonumber \\[0.2cm]
\dfrac{\partial \Sigma}{\partial \slashed{p}} \Big |^{\Lambda_{\UV}}_{m_{q}} &=& \dfrac{\alpha_{s}C_{F}}{2\pi} \int^{1}_{0} dy \bigg ( -y \ln \dfrac{\Lambda^{2}_{\UV}}{(1-y)^{2}m^{2}_{q}} + y + \dfrac{2y(2-y)}{1-y} \bigg ) \, , \nonumber \\[0.2cm] 
\int^{1}_{0} dx \, \delta(1-x) + \int \dfrac{dk^{+}}{p^{+}} \, e^{\rm{(1a)}} (k^{+}) \Big |^{\Lambda_{\UV}}_{m_{q}}  + \dfrac{\partial \Sigma}{\partial \slashed{p}} \Big |^{\Lambda_{\UV}}_{m_{q}} &=& 1 + \dfrac{\alpha_{s} C_{F}}{2\pi} \bigg ( \dfrac{3}{2} \ln \dfrac{\Lambda^{2}_{\UV}}{m^{2}_{q}} - 2 \bigg ) \, ,
\end{eqnarray}
which exactly matches with our result in Eq.~(\ref{e:regulated_mass_QTM_cutoff}). In the above equations, $y$ is the Feynman parameter, and we refer Sec.~\ref{ss:h_accident} for more details on calculations with cut-offs applied to all directions without bias.
References in the past have focused entirely on the UV-divergent terms for both $e$ and the target mass, see for example~\cite{Burkardt:2001iy,Mukherjee:2009uy}.  
Here, we see, for the first time, that there are (once again) caveats with regard to the UV-finite terms in the cut-off scheme.

For the sake of completeness, we also provide the results for $e(x)$ with nonzero gluon mass and DR as the IR regulators. The singular terms for $e(x)$ with $m_{g} \neq 0$ read~\cite{Bhattacharya:2020jfj}
\begin{eqnarray}
e^{\rm{(1a)}}_{\rm{(s)}} (x) \Big |^{\epsilon_{\UV}} =
\begin{dcases}
& e^{\rm{(1a)}}_{\rm{(s)}} (x) \Big |^{\epsilon_{\UV}}_{m_{q}} = \dfrac{\alpha_{s} C_{F}}{2\pi} \, \delta (x) \bigg ( {\cal P}_{\UV} + \ln \dfrac{\mu^{2}_{\UV}}{m^{2}_{q}}-1 \bigg ) \, ,
\\[0.2cm]
& e^{\rm{(1a)}}_{\rm{(s)}} (x) \Big |^{\epsilon_{\UV}}_{\epsilon_{\IR}} = \dfrac{\alpha_{s} C_{F}}{2\pi} \, \delta (x) \bigg ( {\cal P}_{\UV} - {\cal P}_{\IR} + \ln \dfrac{\mu^{2}_{\UV}}{\mu^{2}_{\IR}} \bigg ) \, ,
\end{dcases}
\label{e:e_sing}
\end{eqnarray}
and the canonical part reads~\cite{Bhattacharya:2020jfj}
\begin{eqnarray}
e^{\rm{(1a)}}_{\rm{(c)}} (x) \Big |^{\epsilon_{\UV}}_{m_{g}} & = & 
\dfrac{\alpha_{s} C_{F}}{2\pi} \bigg ( {\cal P}_{\UV} + \ln \dfrac{\mu^{2}_{\UV}}{x m^{2}_{g}} -  \dfrac{1-x}{x} \bigg ) \, .
\label{e:e_cano}
\end{eqnarray}
When DR is applied for the IR, we obtain~\cite{Bhattacharya:2020jfj}
\begin{eqnarray}
e^{\rm{(1a)}} (x) \Big |^{\epsilon_{\UV}}_{\epsilon_{\IR}}  &=& e^{\rm{(1a)}}_{\rm{(s)}} (x) \Big |^{\epsilon_{\UV}}_{\epsilon_{\IR}} + e^{\rm{(1a)}}_{\rm{(c)}} (x) \Big |^{\epsilon_{\UV}}_{\epsilon_{\IR}} \nonumber \\[0.2cm]
&=& \dfrac{\alpha_{s} C_{F}}{2\pi} \, \delta (x) \bigg ( {\cal P}_{\UV} - {\cal P}_{\IR} + \ln \dfrac{\mu^{2}_{\UV}}{\mu^{2}_{\IR}} \bigg ) + \dfrac{\alpha_{s} C_{F}}{2\pi} \, \bigg (  {\cal P}_{\UV} - {\cal P}_{\IR} + \ln \dfrac{\mu^{2}_{\UV}}{\mu^{2}_{\IR}}  \bigg ) \, .
\end{eqnarray}
When a cut-off is applied (transverse direction), the result for $e(x)$ with $m_{g} \neq 0$ reads
\begin{eqnarray}
e^{\rm{(1a)}} (x) \Big |^{\Lambda_{\UV}}_{m_{g}}  &=& e^{\rm{(1a)}}_{\rm{(s)}} (x) \Big |^{\Lambda_{\UV}} + e^{\rm{(1a)}}_{\rm{(c)}} (x) \Big |^{\Lambda_{\UV}}_{m_{g}} \nonumber \\[0.2cm]
&=& \dfrac{\alpha_{s} C_{F}}{2\pi} \, \delta (x) \ln \dfrac{\Lambda^{2}_{\UV}}{m^{2}_{q}}  + \dfrac{\alpha_{s} C_{F}}{2\pi} \, \bigg (  \ln \dfrac{\Lambda^{2}_{\UV}}{x m^{2}_{g}} - \dfrac{1-x}{x}  \bigg ) \, .
\end{eqnarray}

The general structure of the result for quasi-PDF $e_{\rm{Q}}(x)$ is given by Eq.~(\ref{e:qPDF_general1}). The numerator for the singular and the canonical parts are given by
\begin{eqnarray}
N_{e (\rm{s})} &=& - \dfrac{(1- \epsilon) \, p^{3}}{\big ( k^{2}_{\perp} + x^{2} p^{2}_{3} + m^{2}_{q} \big )^{3/2}} \, ,  \\[0.2cm]
N_{e (\rm{c})} &=& 2 p^{3} \bigg ( (k^{0})^{2} - x^{2} p^{2}_{3} - k^{2}_{\perp} + m^{2}_{q} + (1 - \epsilon) m^{2}_{g} \bigg ) \, .
\end{eqnarray}
Table~\ref{table:numerical_moment_e_QTM_DR} confirms that the moment of $e_{\rm{Q}}(x)$ agrees exactly with that of $e(x)$ when DR is applied for the UV. This agreement holds true even in the cut-off scheme. 

\subsection{Results in Yukawa Model}
\label{ss:results_e_eQ_YM}
To derive the renormalized mass of the target in the YM, we follow the procedure outlined in the previous section. 
In the $\MSb$ scheme, the counter-terms are
\begin{eqnarray}
\delta_{2} &=& \dfrac{\alpha_{\rm Y}}{8\pi} {\cal P}_{\UV} \, ,\\[0.2cm] 
\delta_{\rm{m}} &=& - \dfrac{3 \alpha_{\rm Y}}{8\pi} {\cal P}_{\UV} \, ,
\end{eqnarray}
and we arrive at the following relation between the renormalized mass of the target and the bare quark mass,
\begin{eqnarray}
m_{\rm{R}} \Big |^{\epsilon_{\UV}} &=& m_{q} \bigg [ 1 + \dfrac{\alpha_{\rm Y}}{4\pi} \bigg ( -\dfrac{3}{2} \ln \dfrac{\mu^{2}_{\UV}}{m^{2}_{q}} - \dfrac{7}{2} \bigg ) \bigg ] \, .
\end{eqnarray}
In the cut-off scheme, we choose the counter-terms as
\begin{eqnarray}
\delta_{2} &=& \dfrac{\alpha_{\rm Y}}{8\pi} \ln \dfrac{\Lambda^{2}_{\UV}}{\mu^{2}} \, ,\\[0.2cm] 
\delta_{\rm{m}} &=& - \dfrac{3 \alpha_{\rm Y}}{8\pi}  \ln \dfrac{\Lambda^{2}_{\UV}}{\mu^{2}} \, ,
\end{eqnarray}
and the relation between the two masses reads
\begin{eqnarray}
m_{\rm{R}} \Big |^{\Lambda_{\UV}} &=& m_{q} \bigg [ 1 + \dfrac{\alpha_{\rm Y}}{4\pi} \bigg ( -\dfrac{3}{2} \ln \dfrac{\mu^{2}}{m^{2}_{q}} - 2 \bigg ) \bigg ] \, .
\end{eqnarray}
Once again, by repeating the above steps, and not subtracting the counter-terms, we arrive at the following expressions for the regulated mass and it's derivative,
\begin{eqnarray}
m_{\rm{Reg.}} \Big |^{\epsilon_{\UV}} &=& m_{q} \bigg [ 1 + \dfrac{\alpha_{\rm Y}}{4\pi} \bigg ( - \dfrac{3}{2} {\cal P}_{\UV} - \dfrac{3}{2} \ln \dfrac{\mu^{2}_{\UV}}{m^{2}_{q}} - \dfrac{7}{2} \bigg ) \bigg ] \, , \\[0.2cm]
\therefore \dfrac{\partial{m_{\rm{Reg.}}}}{\partial{m_{q}}} \Big |^{\epsilon_{\UV}} &=& 1 + \dfrac{\alpha_{\rm Y}}{4\pi} \bigg ( - \dfrac{3}{2} {\cal P}_{\UV} - \dfrac{3}{2} \ln \dfrac{\mu^{2}_{\UV}}{m^{2}_{q}} - \dfrac{1}{2} \bigg ) \, ,
\label{e:regulated_mass_YM_DR}
\\[0.2cm]
m_{\rm{Reg.}} \Big |^{\Lambda_{\UV}} &=& m_{q} \bigg [ 1 + \dfrac{\alpha_{\rm Y}}{4\pi} \bigg ( - \dfrac{3}{2} \ln \dfrac{\Lambda^{2}_{\UV}}{m^{2}_{q}} -2 \bigg ) \bigg ] \, , 
\\[0.2cm]
\therefore \dfrac{\partial{m_{\rm{Reg.}}}}{\partial{m_{q}}} \Big |^{\Lambda_{\UV}} &=& 1 + \dfrac{\alpha_{\rm Y}}{4\pi} \bigg ( - \dfrac{3}{2} \ln \dfrac{\Lambda^{2}_{\UV}}{m^{2}_{q}} + 1 \bigg ) \, .
\label{e:regulated_mass_YM_cutoff}
\end{eqnarray}

Turning now to the results for $e(x)$, the starting expressions for the singular and canonical terms are 
\begin{eqnarray}
e^{\rm{(1a)}}_{\rm{(s)}} (x) &=&  \dfrac{g^{2}_{\rm Y} \mu^{2\epsilon}}{2(2\pi)} \, \delta (x) \int \dfrac{d^{n-2}k_{\perp}}{(2\pi)^{n-2}} \, \dfrac{1}{(k^{2}_{\perp} + m^{2}_{q})} \, , \nonumber \\[0.2cm]
e^{\rm{(1a)}}_{\rm{(c)}} (x) &=& - \dfrac{g^{2}_{\rm Y} \mu^{2\epsilon}}{2(2\pi)} \int \dfrac{d^{n-2}k_{\perp}}{(2\pi)^{n-2}} \, \dfrac{2 k^{2}_{\perp} - 2 (1-x^{2}) m^{2}_{q} + (1+x) m^{2}_{s}}{ \big ( k^{2}_{\perp} + (1-x)^{2} m^{2}_{q} + x m^{2}_{s} \big ) ^{2} } \, .
\end{eqnarray}
The result for $e(x)$ in the DR scheme is
\begin{eqnarray}
e^{\rm{(1a)}} (x) \Big |^{\epsilon_{\UV}}_{m_{q}}  &=& e^{\rm{(1a)}}_{\rm{(s)}} (x) \Big |^{\epsilon_{\UV}}_{m_{q}} + e^{\rm{(1a)}}_{\rm{(c)}} (x) \Big |^{\epsilon_{\UV}}_{m_{q}} \nonumber \\[0.2cm]
&=& \dfrac{\alpha_{\rm Y}}{4\pi} \, \delta (x) \bigg ( {\cal P}_{\UV} + \ln \dfrac{\mu^{2}_{\UV}}{m^{2}_{q}} \bigg )  + \dfrac{\alpha_{\rm Y}}{4\pi} \bigg ( -2 \, {\cal P}_{\UV} - 2 \ln \dfrac{\mu^{2}_{\UV}}{(1-x)^{2}m^{2}_{q}} + \dfrac{4}{1-x} \bigg ) \, ,
\label{e:e_YM_DR}
\end{eqnarray}
and we find that the lowest moment of $e(x)$ is
\begin{eqnarray}
\int^{1}_{0} dx \, \delta(1-x) + \int \dfrac{dk^{+}}{p^{+}} \, e^{\rm{(1a)}} (k^{+}) \Big |^{\epsilon_{\UV}}_{m_{q}}  + \dfrac{\partial \Sigma}{\partial \slashed{p}} \Big |^{\epsilon_{\UV}}_{m_{q}} &=& 1 + \dfrac{\alpha_{\rm Y}}{4\pi} \bigg ( - \dfrac{3}{2} {\cal P}_{\UV} - \dfrac{3}{2} \ln \dfrac{\mu^{2}_{\UV}}{m^{2}_{q}}  - \dfrac{1}{2} \bigg ) \, ,
\end{eqnarray}
which is in agreement with Eq.~(\ref{e:regulated_mass_YM_DR}). Hence the sum rule in Eq.~(\ref{e:mass_sum_rule}) is exactly satisfied in the DR scheme. The result for $e(x)$ in the cut-off scheme, with regularization applied to the transverse components, is
\begin{eqnarray}
e^{\rm{(1a)}} (x) \Big |^{\Lambda_{\UV}}_{m_{q}}  &=& e^{\rm{(1a)}}_{\rm{(s)}} (x) \Big |^{\Lambda_{\UV}}_{m_{q}} + e^{\rm{(1a)}}_{\rm{(c)}} (x) \Big |^{\Lambda_{\UV}}_{m_{q}} \nonumber \\[0.2cm]
&=& \dfrac{\alpha_{\rm Y}}{4\pi} \, \delta (x) \ln \dfrac{\Lambda^{2}_{\UV}}{m^{2}_{q}} + \dfrac{\alpha_{\rm Y}}{4\pi} \bigg ( -2 \ln \dfrac{\Lambda^{2}_{\UV}}{(1-x)^{2} m^{2}_{q}} + \dfrac{4}{1-x} \bigg ) \, .
\label{e:e_YM_cut-off}
\end{eqnarray}
Once again, we find that the sum rule is violated with the above results, but is satisfied provided $\int e$ and $\frac{\partial \Sigma}{\partial \slashed{p}}$ are consistently calculated with a regularization to all components of $k$. 
The results are
\begin{eqnarray}
\int \dfrac{dk^{+}}{p^{+}} \, e^{\rm{(1a)}} (k^{+}) \Big |^{\Lambda_{\UV}}_{m_{q}} &=& \dfrac{\alpha_{\rm Y}}{4\pi} \int^{1}_{0} dy \bigg ( - 2(1-y) \ln \dfrac{\Lambda^{2}_{\UV}}{(1-y)^{2}m^{2}_{q}} + 3(1-y) + \dfrac{(1+y)^{2}}{1-y} \bigg ) \, , \nonumber \\[0.2cm]
\dfrac{\partial \Sigma}{\partial \slashed{p}} \Big |^{\Lambda_{\UV}}_{m_{q}} &=& \dfrac{\alpha_{\rm Y}}{4\pi} \int^{1}_{0} dy \bigg ( -y \ln \dfrac{\Lambda^{2}_{\UV}}{(1-y)^{2}m^{2}_{q}} + y - \dfrac{2y(1+y)}{1-y} \bigg ) \, , \nonumber \\[0.2cm] 
\int^{1}_{0} dx \, \delta(1-x) + \int \dfrac{dk^{+}}{p^{+}} \, e^{\rm{(1a)}} (k^{+}) \Big |^{\Lambda_{\UV}}_{m_{q}}  + \dfrac{\partial \Sigma}{\partial \slashed{p}} \Big |^{\Lambda_{\UV}}_{m_{q}} &=& 1 + \dfrac{\alpha_{\rm Y}}{4\pi} \bigg ( - \dfrac{3}{2} \ln \dfrac{\Lambda^{2}_{\UV}}{m^{2}_{q}} + 1 \bigg ) \, ,
\end{eqnarray}
which exactly matches with our result in Eq.~(\ref{e:regulated_mass_YM_cutoff}). 
 
For $m_{s} \neq 0$, the singular part of $e(x)$ has two results,
\begin{eqnarray}
e^{\rm{(1a)}}_{\rm{(s)}} (x) \Big |^{\epsilon_{\UV}} =
\begin{dcases}
& e^{\rm (1a)}_{\rm{(s)}} (x) \Big |^{\epsilon_{\UV}}_{m_{q}} =  \dfrac{\alpha_{\rm Y}}{4\pi} \, \delta (x) \bigg ( {\cal P}_{\UV} + \ln \dfrac{\mu^{2}_{\UV}}{m^{2}_{q}} \bigg ) \, ,
\\[0.2cm]
& e^{\rm{(1a)}}_{\rm{(s)}} (x) \Big |^{\epsilon_{\UV}}_{\epsilon_{\IR}}=  \dfrac{\alpha_{\rm Y}}{4\pi} \, \delta (x) \bigg ( {\cal P}_{\UV} - {\cal P}_{\IR} + \ln \dfrac{\mu^{2}_{\UV}}{\mu^{2}_{\IR}} \bigg ) \, , 
\end{dcases}
\end{eqnarray}
while the canonical part is given by
\begin{eqnarray}
e^{\rm{(1a)}}_{\rm{(c)}} (x) \Big |^{\epsilon_{\UV}}_{m_{s}} &=& - \dfrac{\alpha_{\rm Y}}{4\pi} \bigg ( 2 \, {\cal P}_{\UV} + 2 \ln \dfrac{\mu^{2}_{\UV}}{x m^{2}_{s}} +  \dfrac{1-x}{x} \bigg ) \, .
\end{eqnarray}
When DR is applied for the IR, we get
\begin{eqnarray}
e^{\rm{(1a)}} (x) \Big |^{\epsilon_{\UV}}_{\epsilon_{\IR}}  &=& e^{\rm{(1a)}}_{\rm{(s)}} (x) \Big |^{\epsilon_{\UV}}_{\epsilon_{\IR}} + e^{\rm{(1a)}}_{\rm{(c)}} (x) \Big |^{\epsilon_{\UV}}_{\epsilon_{\IR}} \nonumber \\[0.2cm]
&=& \dfrac{\alpha_{\rm Y}}{4\pi} \, \delta (x) \bigg ( {\cal P}_{\UV} - {\cal P}_{\IR} + \ln \dfrac{\mu^{2}_{\UV}}{\mu^{2}_{\IR}} \bigg ) - \dfrac{2 \alpha_{\rm Y}}{4\pi} \, \bigg (  {\cal P}_{\UV} - {\cal P}_{\IR} + \ln \dfrac{\mu^{2}_{\UV}}{\mu^{2}_{\IR}}  \bigg ) \, .
\end{eqnarray}
Finally, when a cut-off is applied to the transverse direction, the result for $e(x)$ with $m_{s} \neq 0$ reads
\begin{eqnarray}
e^{\rm{(1a)}} (x) \Big |^{\Lambda_{\UV}}_{m_{s}}  &=& e^{\rm{(1a)}}_{\rm{(s)}} (x) \Big |^{\Lambda_{\UV}} + e^{\rm{(1a)}}_{\rm{(c)}} (x) \Big |^{\Lambda_{\UV}}_{m_{s}} \nonumber \\[0.2cm]
&=& \dfrac{\alpha_{\rm Y}}{4\pi} \, \delta (x) \ln \dfrac{\Lambda^{2}_{\UV}}{m^{2}_{q}}  - \dfrac{\alpha_{\rm Y}}{4\pi} \, \bigg ( 2 \ln \dfrac{\Lambda^{2}_{\UV}}{x m^{2}_{s}} + \dfrac{1-x}{x}  \bigg ) \, .
\end{eqnarray}

We refer to Section~\ref{s:qPDF_YM} for the general structure of the quasi-PDFs in YM. The numerators for the quasi-PDF $e_{\rm{Q}}(x)$ are given by
\begin{eqnarray}
N_{e (\rm{s})} &=&  \dfrac{ p^{3}}{\big ( k^{2}_{\perp} + x^{2} p^{2}_{3} + m^{2}_{q} \big )^{3/2}}  \, , \\[0.2cm]
N_{e (\rm{c})} &=& \dfrac{p^{3}}{m_{q}} \bigg ( 2 m_{q} (k^{0})^{2} - 2 m_{q} k^{2}_{\perp} - 2 m_{q} x^{2} p^{2}_{3} + 2 m^{3}_{q} - m_{q} m^{2}_{s} \bigg ) \, .
\end{eqnarray}
We have confirmed numerically that our results for the moment of $e_{\rm{Q}}(x)$ matches exactly with that of $e(x)$.

\section{Summary}
\label{s:summary}
In this paper, we have revisited BC-type sum rules which relate the lowest moment of certain twist-2 and twist-3 PDFs. 
While those sum rules have long been known in the case of light-cone PDFs, we argue that they also hold for the corresponding quasi-PDFs.
We have also scrutinized the sum rules through model calculations.
Specifically, we have calculated the light-cone PDFs ($g_{1}(x), \, $ $g_{T}(x)$) and ($h_{1}(x), \, $ $h_{L}(x)$), and their quasi-PDF counterparts ($g_{1, \rm{Q}}(x), \, $ $g_{T, \rm{Q}}(x)$) and ($h_{1, \rm{Q}}(x), \, $ $h_{L, \rm{Q}}(x)$) in the QTM and the YM, to lowest order in perturbation theory. 
We have regulated the IR divergences in 3 schemes: non-zero gluon mass $m_{g} \neq 0$, non-zero quark mass $m_{q} \neq 0$, and DR. 
For the UV divergences, we have made use of 2 schemes: DR, and cut-off. 

Related previous model calculations have focused on the UV-divergent parts of (the perturbative corrections to) the PDFs. 
As such, several works in the past have shown that the BC-type sum rules are valid in cut-off schemes. 
Here, we have presented the full results for the PDFs at one-loop order, that is, we have calculated the UV-divergent and the UV-finite parts of the PDFs. 
We have shown that the BC-type sum rules hold for both the UV-divergent and the UV-finite terms when DR is employed for the UV. 
However, we have found that these sum rules are generally violated for the UV-finite terms when a cut-off is employed.
The only exception is the h-sum rule in the QTM, which ``accidentally" remains valid in the cut-off scheme. 
Violations of the sum rules can be expected in cut-off schemes because they break rotational/Lorentz invariance which is the reason why the BC-type sum rules exist in the first place. 
We have also shown that working with $m_{g} \neq 0$ at twist-3 can lead to a violation of the BC-type sum rules.
Furthermore, we have clarified two important issues related to the moments for quasi-PDFs --- the moments are finite if a regulator is applied to the $k_\perp$-integral (even though this integral is finite for quasi-PDFs), and the moments of quasi-PDFs diverge if calculated after a twist-expansion. 
Finally, we have calculated the light-cone PDF $e(x)$, and its corresponding quasi-PDF $e_{\rm{Q}}(x)$ in both the QTM and the YM. 
In particular, we have scrutinized the sum rule which relates the lowest moment of $e(x)$ to the target mass. 
We repeat that we have not considered renormalization, which could give rise to additional complications when trying to establish BC-type and related sum rules --- see, for instance, Ref.~\cite{Fatma_talk_GHP}.
Nonetheless, the physics pertaining to the regulated results, which we have presented in this work, are fundamental for our concepts.

It is quite likely that there exist more instances in which sum rules or other relations that are rooted in the Lorentz invariance are spoiled in cut-off schemes.
One potential example are polynomiliaty relations for generalized parton distributions~\cite{Ji:1998pc}.
An important message of our work is that it is crucial to calculate the perturbative corrections for the various partonic functions beyond the UV-divergent parts.
And if such a calculation suggests a violation of a certain relation, one must check carefully the cause of the violation.

\begin{acknowledgments}
We express our gratitude to Krzysztof Cichy, Martha Constantinou, Aurora Scapellato, and Fernanda Steffens for collaboration on topics related to the present manuscript.
This work has been supported by the National Science Foundation under grant No. PHY-1812359, and by the U.S.~Department of Energy, Office of Science, Office of Nuclear Physics, within the framework of the TMD Topical Collaboration.
\end{acknowledgments}

\appendix
\section{An interesting point related to DR when applied for IR divergences}
\label{a:interesting_point_DR_IR}
Typically, when DR is applied for both IR and UV divergences, one introduces an arbitrary scale/cut-off, $\Lambda$, to set the boundaries between the IR and UV regions ($0 < \Lambda < \infty$). In the following, through the example of ($g_{1}, \, g_{T}$) in the QTM, we show that the BC sum rule is violated if the regularization is applied to the transverse dimensions and if ($\epsilon_{\IR}, \, \epsilon_{\UV}$) are kept finite. The reason for this violation can be traced back to the effect of the cut-off $\Lambda$ which continues to hold if ($\epsilon_{\IR}, \, \epsilon_{\UV}$) are kept finite. As we shall show below, it is only after a Taylor expansion in powers of ($\epsilon_{\IR}, \, \epsilon_{\UV}$) $ \approx 0$, that the logarithms in $\Lambda$ drop out (at least for the dominant IR pole and the finite term, that is, $\mathcal{O}(\epsilon^{0}_{\IR}, \, \epsilon^{0}_{\UV})$), such that the sum rules are exactly satisfied. This is the case that we have discussed at stretch throughout our manuscript.  

We first calculate $g_{1}(x)$:
\begin{eqnarray}
g^{\rm{(1a)}}_{1}(x) &=& 2 \, \alpha_{s} C_{F} \, (1-x) \, \left \{ \mu^{2\epsilon} (1-\epsilon) \, \int \dfrac{d^{n-2}k_{\perp}}{(2\pi)^{n-2}}  \dfrac{1}{k^{2}_{\perp}} \right \} \nonumber \\[0.2cm]
&=& 2 \, \alpha_{s} C_{F} \, (1-x) \, \left \{ - \dfrac{(4\pi)^{-1+\epsilon_{\IR}} (-1+\epsilon_{\IR})  \bigg (\dfrac{\Lambda}{\mu_{\IR}} \bigg )^{-2\epsilon_{\IR}}}{\epsilon^{2}_{\IR} \Gamma (-\epsilon_{\IR})} + \dfrac{(4\pi)^{-1+\epsilon_{\UV}} (-1+\epsilon_{\UV})  \bigg (\dfrac{\Lambda}{\mu_{\UV}} \bigg )^{-2\epsilon_{\UV}}}{\epsilon^{2}_{\UV} \Gamma (-\epsilon_{\UV})} \right \} \, .
\end{eqnarray}
On the other hand, the singular and the canonical parts for $g_{T}(x)$ are
\begin{eqnarray}
g^{\rm{(1a)}}_{T (\rm{s})}(x) &=& - 2 \, \alpha_{s} C_{F} \, \delta(x) \, \left \{ \mu^{2\epsilon} \epsilon \, \int \dfrac{d^{n-2}k_{\perp}}{(2\pi)^{n-2}}  \dfrac{1}{k^{2}_{\perp}} \right \} \nonumber \\[0.2cm]
&=& 2 \, \alpha_{s} C_{F} \, \delta(x) \, \left \{  \dfrac{(4\pi)^{-1+\epsilon_{\IR}}  \bigg (\dfrac{\Lambda}{\mu_{\IR}} \bigg )^{-2\epsilon_{\IR}}}{ \Gamma (1 -\epsilon_{\IR})} - \dfrac{(4\pi)^{-1+\epsilon_{\UV}}  \bigg (\dfrac{\Lambda}{\mu_{\UV}} \bigg )^{-2\epsilon_{\UV}}}{ \Gamma (1 -\epsilon_{\UV})} \right \} \, , \\[0.2cm]
g^{\rm{(1a)}}_{T (\rm{c})}(x) &=& 2 \, \alpha_{s} C_{F} x \, \left \{ \mu^{2\epsilon}  \, \int \dfrac{d^{n-2}k_{\perp}}{(2\pi)^{n-2}}  \dfrac{1}{k^{2}_{\perp}} \right \} \nonumber \\[0.2cm]
&=& 2 \, \alpha_{s} C_{F} x \, \left \{ \dfrac{(4\pi)^{-1+\epsilon_{\IR}}  \bigg (\dfrac{\Lambda}{\mu_{\IR}} \bigg )^{-2\epsilon_{\IR}}}{\epsilon^{2}_{\IR} \Gamma (-\epsilon_{\IR})} - \dfrac{(4\pi)^{-1+\epsilon_{\UV}} \bigg (\dfrac{\Lambda}{\mu_{\UV}} \bigg )^{-2\epsilon_{\UV}}}{\epsilon^{2}_{\UV} \Gamma (-\epsilon_{\UV})} \right \} \, .
\end{eqnarray}
Therefore, 
\begin{eqnarray}
\int dx \, g_{1}^{\rm{(1a)}} (x) \neq \int dx \, g_{T}^{\rm{(1a)}} (x) \, .
\end{eqnarray}
However, in the limit of ($\epsilon_{\IR}, \, \epsilon_{\UV}$) $\rightarrow 0$, the $\Lambda$-dependence drops out:
\begin{eqnarray}
g_{1}^{\rm{(1a)}} (x) &=& 2 \, \alpha_{s} C_{F} \, (1-x) \, \left \{ \dfrac{1}{4\pi} \bigg ( \dfrac{1}{\epsilon_{\UV}} - \dfrac{1}{\epsilon_{\IR}}\bigg ) + \dfrac{1}{2\pi} \bigg ( \ln \dfrac{\Lambda}{\mu_{\IR}} - \ln \dfrac{\Lambda}{\mu_{\UV}} \bigg ) \right \} \, , \\[0.2cm]
g^{\rm{(1a)}}_{T (\rm{s})}(x) &=& 0 \, , \\[0.2cm]
g^{\rm{(1a)}}_{T (\rm{c})}(x) &=& 2 \, \alpha_{s} C_{F} \, x \, \left \{ \dfrac{1}{4\pi} \bigg ( \dfrac{1}{\epsilon_{\UV}} - \dfrac{1}{\epsilon_{\IR}}\bigg ) + \dfrac{1}{2\pi} \bigg ( \ln \dfrac{\Lambda}{\mu_{\IR}} - \ln \dfrac{\Lambda}{\mu_{\UV}} \bigg ) \right \} \, .
\end{eqnarray}
Clearly,
\begin{eqnarray}
\int dx \, g_{1}^{\rm{(1a)}} (x) = \int dx \, g_{T}^{\rm{(1a)}} (x) \, .
\end{eqnarray}
This point is very interesting, and the reasoning for such a result is rather simple: sum rules do not hold for finite values of ($\epsilon_{\IR}, \, \epsilon_{\UV}$) because of the effect of the cut-off $\Lambda$, which has been applied to the transverse dimensions to demarcate the IR and UV regions. Therefore, obviously, in the process of creating boundaries between IR and UV regions, we ended up breaking rotational invariance. We note in passing that due to this reason, we did not quote numerical results for the moments with finite ($\epsilon_{\IR}, \, \epsilon_{\UV}$), as we did in other instances with nonzero parton mass regulators.

It is also interesting to check if the above observation changes when DR, for both IR and UV, is applied to all components of the loop momenta. Such a case would require us to create boundaries between IR and UV by introducing $\Lambda$ on all components of the loop momenta. Our starting point for $g^{\rm{(1a)}}_{1}$ is
\begin{eqnarray}
\int \dfrac{dk^{+}}{p^{+}} \lambda \, g^{\rm{(1a)}}_{1} ( k^{+}) &=& - \dfrac{i g^{2} C_{F} \mu^{2\epsilon}}{4 p^{+}} \, \int \dfrac{d^{n}k}{(2\pi)^{n}} \int^{1}_{0} dy \dfrac{2(1-y)}{(k^{2} - Q^{2})^{3}} N_{g1} (k) \, ,
\label{e:g1_DR_4k}
\end{eqnarray}
where 
\begin{eqnarray}
N_{g1}(k) = 4\lambda p^{+} (n-2) k^{2} - 8 m_{q} (n-2) k^{+} (k \cdot s) \, ,
\end{eqnarray}
and $Q^{2} = 0$. Using
\begin{eqnarray}
\mu^{2\epsilon} (n-2) \int \dfrac{d^{n}k}{(2\pi)^{n}} \, \dfrac{1}{k^{4}} &=& \dfrac{i}{8 \pi^{2}} \bigg ( \dfrac{1}{\epsilon_{\UV}} - \dfrac{1}{\epsilon_{\IR}} + \ln \dfrac{\mu^{2}_{\UV}}{\mu^{2}_{\IR}} \bigg ) \, , \\[0.3cm]
\mu^{2\epsilon} (n-2) \int \dfrac{d^{n}k}{(2\pi)^{n}} \, \dfrac{k^{+}(k \cdot s)}{k^{6}} &=& \mu^{2\epsilon} s^{+} \dfrac{(n-2)}{n} \int \dfrac{d^{n}k}{(2\pi)^{n}} \, \dfrac{k^{2}}{k^{6}} = \dfrac{i s^{+} }{32 \pi^{2}} \bigg ( \dfrac{1}{\epsilon_{\UV}} - \dfrac{1}{\epsilon_{\IR}} + \ln \dfrac{\mu^{2}_{\UV}}{\mu^{2}_{\IR}} \bigg ) \, ,
\end{eqnarray}
we obtain
\begin{eqnarray}
\int \dfrac{dk^{+}}{p^{+}} g^{\rm{(1a)}}_{1} (k^{+}) \Big |^{\epsilon_{\UV}}_{\epsilon_{\IR}} &=&  \dfrac{\alpha_{s}C_{F}}{2\pi} \dfrac{1}{2} \bigg ( \dfrac{1}{\epsilon_{\UV}} - \dfrac{1}{\epsilon_{\IR}} + \ln \dfrac{\mu^{2}_{\UV}}{\mu^{2}_{\IR}} \bigg ) \, ,
\end{eqnarray}
which agrees with Eq.~(\ref{e:g1_gT_DR_IR}). Our starting point for $g^{\rm{(1a)}}_{T}$ is given by
\begin{eqnarray}
\int \dfrac{dk^{+}}{p^{+}} \dfrac{m_{q} s^{i}_{\perp}}{p^{+}} \, g^{\rm{(1a)}}_{T} (k^{+}) &=& - \dfrac{i g^{2} C_{F} \mu^{2\epsilon}}{4 p^{+}} \, \int \dfrac{d^{n}k}{(2\pi)^{n}} \int^{1}_{0} dy \dfrac{2(1-y)}{(k^{2} - Q^{2})^{3}} N_{gT} (k) \, ,
\label{e:gT_DR_4k}
\end{eqnarray}
where
\begin{eqnarray}
N_{gT} (k) = 4m_{q} s^{i}_{\perp} (n-2) k^{2} - 8 m_{q} (n-2) k^{i}_{\perp} (k \cdot s) \, .
\end{eqnarray}
We see that the structure of the individual terms in $N_{gT} (k)$ and $N_{g1} (k)$ exactly agree. Therefore, prior to carrying out the integrals explicitly, one can already see that the BC sum rule will be satisfied. A direct consequence of this term-by-term equivalence is that the sum rule continues to hold also for finite values of $(\epsilon_{\IR},\epsilon_{\UV})$. Specifically, we find the following equality,
\begin{eqnarray}
\int \dfrac{dk^{+}}{p^{+}} g^{\rm{(1a)}}_{1}(k^{+}) &=&  \dfrac{\alpha_{s} C_{F}}{2\pi} \, \bigg \{ (4\pi)^{\epsilon_{\IR}} \bigg (\dfrac{\Lambda}{\mu_{\IR}} \bigg )^{-2\epsilon_{\IR}} \bigg ( \dfrac{1}{\epsilon^{2}_{\IR} \Gamma(-\epsilon_{\IR})}  -  \dfrac{(-1+\epsilon_{\IR})}{\epsilon_{\IR} \Gamma(3-\epsilon_{\IR})} \bigg ) \nonumber \\[0.2cm]
&& \phantom{This text is invisible.} - (4\pi)^{\epsilon_{\UV}} \bigg (\dfrac{\Lambda}{\mu_{\UV}} \bigg )^{-2\epsilon_{\UV}} \bigg ( \dfrac{1}{\epsilon^{2}_{\UV} \Gamma(-\epsilon_{\UV})} -  \dfrac{(-1+\epsilon_{\UV})}{\epsilon_{\UV} \Gamma(3-\epsilon_{\UV})} \bigg ) \bigg \} 
\nonumber \\[0.2cm]
&=& \int \dfrac{dk^{+}}{p^{+}} g^{\rm{(1a)}}_{T}(k^{+}) \, ,
\end{eqnarray}
which holds for arbitrary values of the cut-off $\Lambda$. 
Ultimately, all of these observations arise from the very same situation, namely, whether or not we are applying a cut-off in a rotationally invariant manner. 
While our observation here is very important and fundamental, we believe it is not widely known and, in fact, we are not aware of a paper which discusses this point.


\end{document}